%% file: EHOW++.tex
\documentclass[12pt,DIV=16]{scrartcl}

\DeclareOldFontCommand{\rm}{\normalfont\rmfamily}{\mathrm}
\DeclareOldFontCommand{\sf}{\normalfont\sffamily}{\mathsf}
\DeclareOldFontCommand{\tt}{\normalfont\ttfamily}{\mathtt}
\DeclareOldFontCommand{\bf}{\normalfont\bfseries}{\mathbf}
\DeclareOldFontCommand{\it}{\normalfont\itshape}{\mathit}
\DeclareOldFontCommand{\sl}{\normalfont\slshape}{\@nomath\sl}
\DeclareOldFontCommand{\sc}{\normalfont\scshape}{\@nomath\sc}

\input{paperdef.tex}

\hypersetup{
  pdfauthor={E. Bagnaschi, H. Bahl, J. Ellis, J. Evans, T. Hahn, S. Heinemeyer,
W. Hollik, K. A. Olive, S. Pa{\ss}ehr, H. Rzehak, I. V. Sobolev,
G. Weiglein and J. Zheng}
  ,pdftitle={Supersymmetric Models in Light of
Improved Higgs Mass Calculations}
  ,pdfsubject={}
  ,pdfkeywords={}
}

\newcount\timecount
\newcount\hours \newcount\minutes  \newcount\temp \newcount\pmhours
\hours = \time
\divide\hours by 60
\temp = \hours
\multiply\temp by 60
\minutes = \time
\advance\minutes by -\temp
\def\hour{\the\hours}
\def\minute{\ifnum\minutes<10 0\the\minutes
            \else\the\minutes\fi}
\def\clock{
\ifnum\hours=0 12:\minute\ AM
\else\ifnum\hours<12 \hour:\minute\ AM
      \else\ifnum\hours=12 12:\minute\ PM
            \else\ifnum\hours>12
                 \pmhours=\hours
                 \advance\pmhours by -12
                 \the\pmhours:\minute\ PM
                 \fi
            \fi
      \fi
\fi
}

\def\monthname{\relax\ifcase\month 0/\or January\or February\or
   March\or April\or May\or June\or July\or August\or September\or
   October\or November\or December\else\number\month/\fi}

\def\bold#1{\setbox0=\hbox{$#1$}%
     \kern-.025em\copy0\kern-\wd0
     \kern.05em\copy0\kern-\wd0
     \kern-.025em\raise.0433em\box0 }

\graphicspath{{figs/}}

\begin{document}
\begin{titlepage}
\pagestyle{empty}
\begin{center}
  {\large {\bf Supersymmetric Models in Light of\\[.3em]
      Improved Higgs Mass Calculations}} \\
\end{center}
\begin{center}
\vskip 0.10in
{\bf E.~Bagnaschi}$^{1}$,
{\bf H.~Bahl}$^2$,
{\bf J.~Ellis}$^3$,
{\bf J.~Evans}$^4$,
{\bf T.~Hahn}$^2$,\\
{\bf S.~Heinemeyer}$^5$,
{\bf W.~Hollik}$^2$,
{\bf K.~A.~Olive}$^{6}$,
{\bf S.~Pa{\sz}ehr}$^7$,\\
{\bf H.~Rzehak}$^8$,
{\bf I.~V.~Sobolev}$^{9}$,
{\bf G.~Weiglein}$^{9}$
and
{\bf J.~Zheng}$^{10}$
\vskip 0.2in
{\footnotesize {\it
$^1${Paul Scherrer Institut, CH-5232 Villigen PSI, Switzerland} \\[1ex]
$^2${Max-Planck-Institut f{\"u}r Physik, F{\"o}hringer Ring 6, D-80805 Munich, Germany}\\[1ex]
$^3${Theoretical Particle Physics and Cosmology Group, Dept.\ of Physics,
King's College London,\\ London WC2R 2LS, UK;\\
National Institute of Chemical Physics and Biophysics, R\"avala 10, 10143 Tallinn, Estonia;\\
Theory Division, CERN, CH-1211 Geneva 23, Switzerland}\\[1ex]
$^4${School of Physics, KIAS, Seoul 130-722, Korea}\\[1ex]
$^5${Instituto de F\'isica Te\'orica, Universidad Aut\'onoma de Madrid Cantoblanco, 28049 Madrid, Spain;\\
Campus of International Excellence UAM+CSIC, Cantoblanco, 28049, Madrid, Spain;\\
Instituto de F{\'i}sica de Cantabria (CSIC-UC), E-39005 Santander, Spain}\\[1ex]
$^6${William\,I.\,Fine\,Theoretical\,Physics\,Institute,\,%
School of Physics and Astronomy,
Univ.\,of\,Minnesota,\,%
Minneapolis,\,MN\,55455,\,USA}\\[1ex]
$^7${Sorbonne Universit\'e, CNRS, Laboratoire de Physique Th\'eorique et Hautes \'Energies (LPTHE), UMR 7589,\\ 4 Place Jussieu, F--75252 Paris CEDEX 05, France}\\[1ex]
$^8${CP3-Origins, University of Southern Denmark, Odense, Denmark}\\[1ex]
$^9${DESY, Notkestr.\ 85, D-22607 Hamburg, Germany}\\[1ex]
$^{10}${Department of Physics, University of Tokyo, Bunkyo-ku, Tokyo 113-0033, Japan}}}\\
\vskip 0.2in
{\bf Abstract}
\end{center}
\baselineskip=16pt
\noindent

\small{We discuss the parameter spaces of supersymmetry (SUSY) scenarios
taking into account the improved Higgs-mass {prediction} provided
by \FHnew. Among other improvements, this {prediction}
incorporates three-loop renormalization-group effects and two-loop
threshold corrections, and can accommodate three separate mass scales:
$m_{\tilde q}$ (for squarks), $\mgluino$ (for gluinos) and $m_{\tilde
  \chi}$ (for electroweakinos). Furthermore, it contains an
  improved treatment of the \DRbar\ scalar top
parameters avoiding problems with the conversion to on-shell
parameters, that yields {more
  accurate results} for large SUSY-breaking scales. We first consider
the CMSSM, in which the soft SUSY-breaking parameters $m_0$ and
$m_{1/2}$ are universal at the GUT scale, and then sub-GUT models in
which universality is imposed at some lower scale.  In both cases, we
consider the constraints {from the Higgs-boson mass~$\Mh$} in the
bulk of the \mbox{$(m_0, m_{1/2})$}~plane and also along stop
coannihilation strips where sparticle masses may extend into the
multi-TeV range.  We then consider the minimal anomaly-mediated
SUSY-breaking (mAMSB) scenario, in which large sparticle masses are
generic. In all these scenarios the substantial {improvements}
between the calculations of~$\Mh$ in \FHnew\ and \FHold{, which was
  used in an earlier study,} change significantly the preferred
portions of the models' parameter spaces.  Finally, we consider the
pMSSM11, in which sparticle masses may be significantly smaller and we
find only small changes in the preferred regions of parameter space.}
\vfill
\begin{center}{\footnotesize
    CERN-TH/2018-185, DESY-18-182, PSI-PR-18-11, UMN--TH--3801/18,\\
    FTPI--MINN--18/18, IFT--UAM/CSIC--18-081, KIAS-P18095,\\
    KCL-PH-TH/2018-41, MPP-2018-239, CP3-Origins-2018-039~DNRF90}
\end{center}

\end{titlepage}
\baselineskip=16pt


\section{Introduction\label{sec:intro}}

Given the persistent absence of any signal in the searches for
supersymmetric particles at the~Large~Hadron~Collider~(LHC) and in
direct searches for supersymmetric dark matter~(DM), there is
strengthened emphasis on the information about the scale of
supersymmetry~(SUSY) that can be obtained indirectly from other
measurements. The Higgs-boson
discovery~\cite{Aad:2012tfa,Chatrchyan:2012xdj} at the~LHC opened a
new window with {the} SUSY Higgs-boson {mass as a precision
  observable}. The Minimal Supersymmetric Standard
Model~(MSSM)~\cite{Nilles:1983ge,Haber:1984rc} contains---in contrast
to the single Higgs doublet of the~Standard~Model~(SM)---two Higgs
doublets. In the $\cp$-conserving case this leads to a physical
spectrum consisting of two $\cp$-even neutral Higgs bosons,~$h$
and~$H$, one $\cp$-odd,~$A$, and two charged Higgs bosons,~$H^\pm$. At
the tree level the Higgs sector can be described, besides
the~SM~parameters, by two additional input parameters, conveniently
chosen to be the mass of the $\cp$-odd Higgs boson,~\MA, (or the mass
of the charged Higgs,~\MHp) and the ratio of the two vacuum
expectation values,~$\tb \equiv v_2/v_1$.  The light (or heavy)
neutral $\cp$-even MSSM Higgs boson can be interpreted as the signal
discovered at~$\simord 125 \gev$~\cite{Heinemeyer:2011aa}.

Prominent among the predicted quantities is the mass of the light
$\cp$-even Higgs boson,~$\Mh$, which can be calculated in terms of
the~SM~parameters and the soft SUSY-breaking parameters. As is well
known, tree-level calculations implied that~\mbox{$\Mh < \MZ$} in
the~MSSM, whereas one-loop calculations raised the possibility
that~\mbox{$\Mh > \MZ$}~\cite{Haber:1990aw,Ellis:1990nz,Okada:1990vk}.
The more complete multi-loop calculations of~$\Mh$ that have become
available subsequently (as summarized in Section~\ref{sec:Mh-theo})
can accommodate comfortably the measured value~\mbox{$\Mh \simeq
  125$}~GeV~\cite{Aad:2015zhl}, and the similarities of the measured
Higgs couplings to those in the~SM~\cite{Khachatryan:2016vau} are also
consistent with the~MSSM.

The question then arises whether these successes of the~MSSM can be
used to estimate reliably the masses of SUSY particles such as the
scalar top quarks (stops), with the corollary question what ranges of
their masses are compatible with the strengthening lower limits
{from the~LHC} on sparticle masses. Several of us studied these
questions in the context of data from~LHC~Run~1, using the
\FHold\ code~\cite{Buchmueller:2013psa}. A particular emphasis in that
analysis was to understand the impact of the combination of
fixed-order calculations of~$\Mh$ and results obtained in
an~Effective~Field~Theory~(EFT)~approach, which had recently been
accomplished at that time and allowed the resummation of large
logarithmic contributions, stabilizing the calculation of~$\Mh$ for
large stop mass scales~\cite{Hahn:2013ria}.

During LHC~Run~2 the ATLAS and CMS experiments have been pushing the
lower limits on the masses of some strongly-interacting sparticles
into the~\mbox{$1$--$2$}~TeV range. It is therefore of key importance
to have available calculations of the Higgs mass that are as accurate
as possible when one or more soft SUSY-breaking parameters are in the
multi-TeV range, and there may be a {rather} large hierarchy
between different supersymmetric mass scales.

Important steps in this direction have been taken since the release of
\FHold. Many of these advances in the {prediction} of~$\Mh$ that
are particularly important for sparticle masses in the multi-TeV range
are incorporated in the recent release of {\tt FeynHiggs}~{\tt
  2.14.1}. These include three-loop renormalization-group
equations~(RGEs) with electroweak effects, as well as corresponding
two-loop threshold corrections including the possibility of
non-degenerate stop mass parameters. Moreover, whereas only a single
SUSY-breaking scale was incorporated in \FHold, three distinct scales
can be accommodated in \FHnew. These are {the squark and gluino
  masses,}~$m_{\tilde q}$,~$\mgluino$, and a scale~$\mew$
characterizing the overall electroweakino mass scale, thus making the
connection to DM, assuming it to be given by the lightest
neutralino,~$\neu1$~\cite{Go1983,EHNOS}.  In addition, {problems
that  occur when combining an infinite tower of resummed logarithms with
  a fixed-order result where \DRbar\ input parameters of the scalar
  top sector have been converted into the corresponding parameters of
  the on-shell~(OS) renormalization scheme} can now be avoided by
performing the calculation directly in the \DRbar~scheme. Finally a
new, improved procedure for determining the poles of the Higgs-boson
propagator matrix has been introduced. Section~\ref{sec:Mh-status}
contains a review of \FHnew\ and its relations to other codes for
calculating~$\Mh$ in the~MSSM, and Section~\ref{sec:FH} makes a
specific comparison of \FHnew\ with \FHold.

In Section~\ref{sec:Mh-calc} of this paper we explore the significance
of these advances for a number of MSSM~scenarios with different
phenomenological features that are sensitive to different aspects of
\FHnew.\footnote{We have checked in various specific cases that
  further advances going beyond \FHnew\ that have become available
  very recently~{\cite{FH214}} do not have a significant impact on the
  numerical analyses presented in this paper.} The first of these is
the CMSSM \cite{cmssm,ELOS,eelnos,azar,eeloz}, in which the soft
SUSY-breaking scalar mass parameter~$m_0$ and gaugino mass
parameter~$m_{1/2}$ are each assumed to be universal at the
GUT~scale~\MGUT.

The second example is provided by `sub-GUT'~models in which this
universality is imposed at some scale~\mbox{$\Min \le
  \MGUT$}~\cite{sub-GUT,ELOS,eelnos,eeloz,MCsubGUT}.  The LHC~searches
impose severe constraints on these models, favoring parameter sets
along the stop coannihilation~\cite{stopco,eds,eoz,interplay,raza} and
focus-point strips~\cite{fp}. These extend out to multi-TeV sparticle
masses with stop masses~$m_{\tilde t_{1,2}}$ that are strongly
non-degenerate in general. Moreover, in the focus-point
case~\mbox{$\mneu1 \ll \mstop1$}, whereas these masses are very
similar along the stop coannihilation strip.

Thirdly, we consider the minimal anomaly-mediated
SUSY-breaking~(mAMSB)~model~\cite{anom,mAMSB}, in which sfermion
masses are typically several tens~of~TeV, whereas values
of~\mbox{$\mneu1 \simeq 1$}~TeV or~${\simeq}\,3$~TeV are preferred by
the DM density constraint. For a recent global analysis of this model
taking into account the constraints from Run~1 of the~LHC,
see~\cite{MCmAMSB}.

Finally, we consider a phenomenological
MSSM~scenario~\cite{pMSSM,pMSSM2} with $11$~free parameters specified
at the electroweak scale, as has recently been analyzed including
LHC~Run~2 data in~\cite{MCpMSSM11}. {\it A priori}, this scenario
would allow many possible mass hierarchies, as well as many
near-degeneracies between sparticle masses that could dilute the
classic missing-transverse-energy~($\ETslash$) signatures at the~LHC
and permit lighter sparticles than are allowed in the~CMSSM and
sub-GUT~models.

In each of these scenarios, our primary concern is the implications of
improvements in the \FHnew\ calculation of~$\Mh$ (compared to
previous, less sophisticated calculations) for the model parameter
space.


\section{Higgs Mass Calculations\label{sec:Mh-theo}}

The experimental accuracy of the measured mass of the observed Higgs
boson has already reached the level of a precision observable, with an
uncertainty of less than~$300\mev$~\cite{Aad:2015zhl}. This precision
should ideally be matched by the theoretical uncertainty in the
prediction of the SM-like Higgs-boson mass. In the following we
briefly review the status of Higgs-boson mass calculations in
the~MSSM. Particularly we focus on the implementation in the code \fh,
where we summarize the relevant progress over the last years,
emphasizing the differences w.r.t.\ \FHold, which was used
in~\citere{Buchmueller:2013psa}.


\subsection{Status of MSSM Higgs Mass Calculations\label{sec:Mh-status}}

The tree-level predictions for the Higgs-boson masses in the~MSSM
receive large higher-order corrections, which in the case of~$\Mh$ can
be of~\order{100\%}, see
\citeres{Heinemeyer:2004ms,Heinemeyer:2004gx,Djouadi:2005gj,
  Draper:2016pys} for reviews. Beyond the one-loop level, the dominant
two-loop corrections
of~$\order{\alt\als}$~\cite{Heinemeyer:1998jw,Heinemeyer:1998kz,Heinemeyer:1998np,Zhang:1998bm,Espinosa:1999zm,Degrassi:2001yf}
and~\order{\alt^2}~\cite{Espinosa:2000df,Brignole:2001jy} as well as
the corresponding corrections
of~$\order{\alb\als}$~\cite{Brignole:2002bz,Heinemeyer:2004xw}
and~\order{\alt\alb}~\cite{Dedes:2003km} have been known for more than
a decade (see also \citere{Heinemeyer:2007aq, Hollik:2014wea,
  Hollik:2014bua,Passehr:2017ufr} for the \cp-violating case---the
last reference going beyond the large-$\tan\beta$~limit employed by
\citere{Dedes:2003km}).\footnote{Here and in the following we
  use~\mbox{$\al_f = (y_f)^2/(4\pi)$}, where~$y_f$ denotes the fermion
  Yukawa coupling.} The $\tan\beta$-enhanced threshold corrections to
the bottom Yukawa {coupling} in
the~MSSM~\cite{Hempfling:1993kv,Hall:1993gn,Carena:1994bv,Carena:1999py}
are included in the resummation of leading contributions from the
bottom/scalar bottom
sector~\cite{Brignole:2002bz,Dedes:2003km,Heinemeyer:2004xw} (see
also~\cite{Noth:2008tw,Noth:2010jy} for corresponding
next-to-leading~order~(NLO) threshold
contributions). Momentum-dependent two-loop contributions have also
been
computed~\cite{Martin:2004kr,Borowka:2014wla,Degrassi:2014pfa,Borowka:2015ura,Borowka:2018anu}.

In the case of SUSY spectra with large mass hierarchies, the
fixed-order calculation of the Higgs-boson mass loses its predictive
power, due to the appearance of large logarithms of the ratio of the
mass scales appearing in the result. To obtain {an accurate}
prediction, the resummation of these logarithms is required. To
achieve this goal, the calculation of the Higgs mass has to be cast
into the language of Effective Field Theories~(EFTs).  In this
approach, the heavy degrees of freedom are integrated out at their
characteristic scale,~$M_S$, where they enter the matching conditions
for the couplings of the low-energy~EFT. The {RGEs} are then used
to relate the values of the couplings at~$M_S$ with those at the low
scale, which in the simplest cases is the electroweak scale, where
physical observables such as the Higgs mass are computed. In this way,
the logarithms of the ratio of the relevant mass scales are taken into
account to all orders, while, at the same time, power-suppressed terms
of~$\order{v^2/M_S^2}$ are neglected, unless higher-dimensional
effective operators are matched and included in the low-energy~EFT.

The EFT~approach was originally developed about~$25$~years
ago~\cite{Barbieri:1990ja, Espinosa:1991fc,Casas:1994us}.  It has
subsequently been used to compute the coefficients of the logarithmic
terms appearing in the computation of the Higgs mass at
one~\cite{Haber:1993an}, two~\cite{Carena:1995bx, Carena:1995wu,
  Haber:1996fp, Carena:2000dp} and three~\cite{Degrassi:2002fi,
  Martin:2007pg} loops.  However, due to the missing~${v^2/M_S^2}$
terms mentioned above, this approach was not competitive with a
traditional fixed-order computation in the case of {relatively light}
SUSY scenarios.

The situation has changed in the past few years, due to the renewed
interest in scenarios with heavy sparticles caused by the (so far)
negative outcomes of the direct searches at the~LHC. Moreover, our
knowledge of the matching condition for the Higgs quartic coupling in
case of the~SM as a low-energy~EFT now has been extended to all the
contributions controlled by the strong and by the third-generation
Yukawa couplings at
two~loops~\cite{Draper:2013oza,Bagnaschi:2014rsa,Vega:2015fna,Bagnaschi:2017xid}.
The codes {\tt MhEFT}~\cite{Draper:2013oza}, {\tt
  SUSYHD}~\cite{Vega:2015fna} and {\tt
  HSSUSY}~{\cite{BVW,Athron:2017fvs} implement these
computations, with the latter including all the available corrections.
The more {complicated} case of a low-energy~EFT containing two Higgs
doublets also has been studied in several contexts, and several codes
are available for this case: {\tt MhEFT}~\cite{Lee:2015uza} and
several generators~\cite{Bagnaschi:2015pwa,BVW} based on {\tt
  FlexibleSUSY}~\cite{Athron:2014yba}. Scenarios
with~\mbox{$\mgluino/{m_{\tilde{q}}} \gg 1$}, where~$\mgluino$
denotes the gluino mass and~{$m_{\tilde{q}}$} the scalar top mass
scale, are not yet included in any code: the corrections
by~$\log(\mgluino/{m_{\tilde{q}}})$ in this hierarchy {can
  presently not yet be resummed}. {These logarithms} could lead to
large effects for~\mbox{$\mgluino/{m_{\tilde{q}}} \gsim 4$}, a
possibility that we comment on later in our numerical analysis.

In order to provide a reliable {prediction} for the Higgs-boson
masses in {both low- and} high-scale MSSM~scenarios, the
resummation of the leading and subleading logarithms can be combined
with the fixed-order results in the~MSSM in the so-called
\mbox{``hybrid approach''}, thereby keeping track of the
power-suppressed terms that are neglected in a simple EFT~approach in
which the low-energy~EFT does not include higher-dimensional
operators~\footnote{{In} \citere{Bagnaschi:2017xid}
  dimension-$6$~operators were included to perform an estimation of
  these effects in a pure EFT~approach.}. The hybrid approach was first
implemented into the code
\fh~\cite{Heinemeyer:1998yj,Hahn:2009zz,Heinemeyer:1998np,Degrassi:2002fi,Frank:2006yh,Hahn:2013ria,Bahl:2016brp,Bahl:2017aev,feynhiggs-www,FH214}.
In the first version that {adopted this method}, \FHold, one light
Higgs doublet at the low scale was assumed, and the logarithms
originating in the top/scalar top sector were
resummed~\cite{Hahn:2013ria}. Further refinements have been presented
more recently in \citeres{Bahl:2016brp,Bahl:2017aev}~\footnote{At present,
  the bottom Yukawa effects at {next-to-NLO~(NNLO)} level in the EFT part
  of the Higgs-boson mass calculations are incorporated only in
  {\tt HSSUSY}~\cite{BVW,Athron:2017fvs}.}.  More recently, the hybrid
approach has been extended  
to support such spectra where a full Two-Higgs-Doublet-Model~(2HDM) is
required as the low-energy~EFT~\cite{Bahl:2018jom}. However the latter
are not implemented in the current public release of \fh\ and
therefore they are not used in the current paper, see the discussion
in \refse{sec:FH} for more details.

For completeness, we also mention here some further corrections that
are available in the literature. The full~\order{\al\als} corrections,
{including the complete momentum dependence at the two-loop
  level,} became recently available in \citere{Borowka:2018anu}.  A
(nearly) full two-loop {effective potential} calculation, including
also the leading three-loop corrections {up to
  next-to-leading-logarithm~(NLL)~level}, has also been
published~\cite{Martin:2002iu,Martin:2002wn,Martin:2004kr,Martin:2007pg},
but is not publicly available as a computer code. Another leading
three-loop calculation of~\order{\alt\als^2}, depending on various SUSY mass hierarchies, has been performed
in~\cite{Harlander:2008ju,Kant:2010tf}, {and} is included in the
code {\tt H3m} that is now available as a stand-alone code, {\tt
  Himalaya}~\cite{Harlander:2017kuc}.  Another approach to the
combination of logarithmic resummation with fixed-order results has
been presented in \citere{Athron:2016fuq} and included in {\tt
  FlexibleSUSY}. Subsequently it was also implemented in the {\tt
  SARAH}+{\tt SPheno}~\cite{Staub:2017jnp} framework. We also note
that \citere{Allanach:2018fif} has studied the issue of the comparison
of the theoretical uncertainties in {\tt SoftSUSY} vs.\ {\tt
  HSSUSY}. Finally, there is a recent
calculation~\cite{Harlander:2018yhj} that resums terms of leading
order in the top Yukawa coupling and~NNLO in the strong
coupling~$\alpha_s$, including the three-loop matching coefficient for
the quartic Higgs coupling of the~SM to the~MSSM between the~EFT and
the fixed-order expression for the Higgs mass, which is available in
an updated version of the {\tt
  Himalaya}~code~\cite{Harlander:2017kuc}. However, a {detailed
  numerical} comparison of \FHnew\ with other codes to calculate~$\Mh$
{is} beyond the scope of this paper.


\subsection{Comparison between \FHnew\ and \FHold\label{sec:FH}}

The main {advances} in \FHnew\ in comparison to
\FHold\ {are related to the EFT~part of the calculation.}  The
resummation of large logarithmic contributions in \FHold\ was
restricted to~$\mathcal{O}(\alpha_s,\alpha_t)$
leading-logarithmic~(LL) and~NLL~contributions. Since then,
electroweak~LL and~NLL~contributions as well
as~$\mathcal{O}(\alpha_s,\alpha_t)$ {next-to~NLL~(NNLL)}~contributions
have been included. This means, in particular, that the full
SM~two-loop~RGEs and partial three-loop~RGEs\footnote{The electroweak
  gauge couplings are {neglected} at the three-loop level.} are
used for evolving the couplings between the electroweak scale and the
SUSY scale~$M_\text{SUSY}$. At the SUSY~scale, full one-loop threshold
corrections and (non-degenerate) threshold corrections
of~$\mathcal{O}(\alpha_s\alpha_t,\alpha_t^2)$ are used for the
matching of the effective~SM to the full~MSSM, taken from
\citere{Bagnaschi:2014rsa} and from
\citeres{Vega:2015fna,Bagnaschi:2017xid}, respectively. Numerically,
the electroweak~LL and~NLL~contributions amount to an upward shift
of~$M_h$ of~$\simord 1 \gev$ for a SUSY~scale of
a~few~TeV. The~NNLL~contributions are numerically relevant only for
large stop mixing, shifting~$M_h$ downwards by~$\simord 1\gev$ for
positive~$X_t$ and upwards by~$\simord 1\gev$ for negative~$X_t$
(where the off-diagonal entry in the stop mass matrix for real
parameters is~$\mt\,X_t$).

For consistency with this logarithmic precision, one must choose
appropriate matching conditions with physical observables at the
electroweak scale.  This is relevant, in particular, for the
\MSbar\ top quark mass in the~SM. In \FHold, the corrections
of~$\mathcal{O}(\alpha_s,\alpha_t)$ in the mass were used. The 
inclusion of electroweak~LL and~NLL~resummation as well as~NNLL
of~$\mathcal{O}(\alpha_s,\alpha_t)$ implies the need to use instead
the~NNLO \MSbar\ top quark mass of the~SM, as done in \FHnew. This
{modification} not only implies changes for large SUSY scales but
also impacts significantly the prediction of~$M_h$ for low
SUSY~scales, as the shift in the top quark mass affects the
non-logarithmic terms {that are} relevant in this regime. The
combined electroweak one-loop as well as the two-loop corrections
amount to a downwards shift of the \MSbar\ top mass of the~SM
by~$\simord 3 \gev$. The effect on~$M_h$ is of similar size.

The EFT calculation in the new \FH\ version allows one to take into
account three different relevant scales. In addition to the SUSY
scale~$m_{\tilde{q}}$---which was the only scale in \FHold---an
electroweakino scale~$\mew$ and a gluino scale~$\mgluino$ are
available. They allow one to investigate scenarios with light
electroweakinos and/or gluinos. This corresponds to a tower of up to
three~EFTs (SM, SM with electroweakinos, SM with gluinos, SM with
electroweakinos and gluinos). {Besides the limitation
  that~$\mgluino/m_{\tilde{q}}$ should not be too large (see the
  discussion above}, all scales can be chosen independently from each
other, though the gluino~threshold has a negligible numerical
influence {in this case}. Also, the electroweakino threshold
becomes relevant only for a large hierarchy between the electroweakino
scale and the SUSY scale (\mbox{${\mew}/M_\text{SUSY} \lesssim
  1/10$}), leading to upward shifts of~$M_h$ of~$\simord 1\gev$.

The second main advance is a better handling of \DRbar\ input
parameters. The fixed-order calculation of \FH\ by default employs a
mixed OS/\DRbar~scheme for renormalization, in which the parameters of
the stop sector are fixed employing the OS~scheme. In \FHold, this was
the only available renormalization scheme. Therefore, a one-loop
conversion between the \DRbar\ and the OS~scheme was employed in the
case of \DRbar\ input parameters. {Whilst, for a fixed-order result,
  such a conversion leads to shifts that are beyond the calculated
  order, this is no longer the case if a fixed-order result is
  supplemented by a resummation of large logarithms.  As shown
  in~\cite{Bahl:2017aev}, the parameter conversion in this case
  induces additional logarithmic higher-order terms that can spoil the
  resummation}. As a solution for this issue, an optional
\DRbar\ renormalization of the stop sector is implemented in
\FHnew. This renders a conversion of the stop parameters
unnecessary. Note, however, that the \DRbar\ sbottom input parameters
are still converted to the OS~scheme. In particular for large
SUSY~scales, {employing directly the \DRbar~scheme for the stop
  sector parameters and avoiding the conversion to the OS~scheme}
affects the results significantly: \EG, for SUSY~scales of~$\simord 20
\tev$, shifts in~$\Mh$ of~$\simord 10 \gev$ were observed
{compared to the result based on the parameter conversion} with
the sign of the shift depending on the size of the stop mixing.  Also,
for low SUSY~scales of~$\simord 1 \tev$, the {prediction using the
  \DRbar~scheme of the stop sector parameters differs from that
  employing the conversion to the OS~scheme by a downward shift
  in~$M_h$ of~$\simord 1\gev$ in the case of large stop mixing. For
  SUSY~scales below~$1 \tev$, where the impact of higher-order
  logarithmic contributions is relatively small, the observed shift
  can be interpreted to a large extent as an indication of the
  possible size of unknown higher-order corrections.}

In addition to these improvements, also the Higgs pole determination
has been reworked. It was noted in~\cite{Bahl:2017aev} that there is a
cancellation between two-loop contributions from sub-loop
renormalization and terms arising through the pole determination. In
the fixed-order calculation, these terms are of higher order, which
are not controlled. In \FHold, the pole determination was performed
numerically employing the \DRbar~scheme for the Higgs field
renormalization. As a consequence of this procedure, the two-loop
contributions from sub-loop renormalization were not included at the
same order as the terms arising through the pole determination,
resulting in an incomplete cancellation. In \FHnew\ the pole
determination has been adapted in order to ensure a complete
cancellation.\footnote{In \FHnew, the Higgs poles are determined by
  expanding the Higgs propagator matrix around the one-loop solutions
  for the Higgs masses. Due to {instabilities} in this method
  close to crossing points, where two of the Higgs bosons change their
  role, in the most recent {\tt FeynHiggs}
  version~{\tt 2.14.3}~\cite{FH214} the 
  Higgs poles are again determined numerically. {In order to}
  avoid inducing higher-order terms that would cancel in a more
  complete calculation, the Higgs field renormalization is used to
  absorb these.  Since no crossing points appear in {the} scenarios investigated in this
  work, using {\tt FeynHiggs 2.14.3} instead
  of \FHnew\ would not lead to significant numerical differences.} The
numerical impact of this {improved} pole determination procedure
increases with rising~$M_\text{SUSY}$. For~$M_\text{SUSY}$ in the
multi-TeV range, it amounts to a downward shift of~$M_h$ of~$\simord 1
\gev$.

Finally, the handling of complex input parameters in \FH\ was
improved. In the fixed-order calculation, the corrections
of~$\mathcal{O}(\alpha_t^2)$ with full dependence on {the phases
  of complex parameters} were
implemented~\cite{Hollik:2014wea,Hollik:2014bua,Hollik:2015ema,Hahn:2015gaa}
(see also~\cite{Passehr:2017ufr}). In addition, an interpolation of
the EFT~calculation in the case of non-zero phases was
introduced. Numerically, this can lead to shifts of~$M_h$ of up
to~$3\gev$. As we do not discuss here the effects of {the phases
  of complex parameters, we do not provide further details} that can
be found in \citere{FH214}.

Summing up this discussion, we {generally} expect the prediction
of~$M_h$ of \FHnew\ to be lower {than} that of \FHold. In the case of
\DRbar\ input parameters, the large shifts {compared to the
  previous result that employed a conversion to the OS~scheme} for the
renormalization of the stop sector can, however, outweigh the other
effects and lead to an overall upward shift of~$M_h$.


\section{Calculations in Specific MSSM Scenarios\label{sec:Mh-calc}}

In this Section, we illustrate the implications of the improved
{prediction for}~$\Mh$ implemented in \FHnew\ in the context of
several specific MSSM~scenarios. The first of these is
the~CMSSM~\cite{cmssm,ELOS,eelnos,azar,eeloz}, in which the soft
supersymmetry-breaking scalar masses~$m_0$, the gaugino
masses~$m_{1/2}$ and the trilinear parameters are all constrained to
be universal at the GUT~scale. The second scenario we study is a class
of sub-GUT~models~\cite{sub-GUT,ELOS,eelnos,eeloz,MCsubGUT}, in which
these universality relations hold at some renormalization
scale~\mbox{$\Min < \MGUT$}, as occurs, \EG, in mirage-mediation
models~\cite{mirage}. We then discuss minimal anomaly-mediated
models~\cite{anom,mAMSB}, in which the scalar masses are typically
much greater than the gaugino masses.  For all of these models, we use
{\tt SSARD}~\cite{SSARD} to compute the particle mass spectrum and
relic density.  We note that the convention for $A$~terms used in {\tt
  SSARD} is opposite to that used in \FH.  Finally, we study a
phenomenological version of the~MSSM~\cite{pMSSM,pMSSM2} with
$11$~free parameters in the soft supersymmetry-breaking sector,
the~pMSSM11, allowing for many possible sparticle mass hierarchies. In
all cases we assume that the lightest supersymmetric particle~(LSP) is
the lightest neutralino~$\neu1$ and provides the full DM
density~\cite{Planck}.


\subsection{The Light Higgs-Boson Mass in the CMSSM\label{sec:cmssm}}

The four-dimensional parameter space of the~CMSSM that we consider
here includes a common input gaugino mass parameter,~$m_{1/2}$, a
common input soft SUSY-breaking scalar mass parameter,~$m_0$, and a
common trilinear soft SUSY-breaking parameter,~$A_0$, which are each
assumed to be universal at the scale~$\MGUT$ (defined as the
renormalization scale where the two electroweak gauge couplings are
equal), and the ratio of MSSM Higgs vevs,~$\tb$.  There is also a
discrete ambiguity in the sign of the Higgs mixing
parameter,~$\mu$. In the~CMSSM, renormalization group~(RG) effects
typically produce hierarchies of physical sparticle masses, \EG,
between gluinos and electroweakly-interacting gauginos and between
squarks and sleptons. The {limits from} LHC~searches for
sparticles generally require at least the strongly-interacting
sparticles to be relatively heavy. Accurate calculations of~$\Mh$ for
MSSM~spectra in the multi-TeV range require many of the improvements
made in \FHnew\ compared to \FHold.

Reconciling the cosmological dark matter density~\cite{Planck} of
the~LSP with the relatively heavy spectra that LHC~searches impose on
the~CMSSM typically requires specific relations between some of the
sparticle masses. One such example is the stop coannihilation strip,
and another is the focus-point region, which we discuss in the two
following subsections.


\subsubsection{Stop Coannihilation Strips in the CMSSM\label{sec:stop-coan}}

We first consider in some detail examples of stop coannihilation
strips. In this case the lighter stop mass~$m_{\tilde t_1}$ and the
mass of the~LSP,~$\mneu1$, must be quite degenerate. The relic density
constraint alone would allow them to weigh several~TeV but the allowed
range of mass scales is in general restricted by the measurement
of~$\Mh$, for more details see \citere{eeloz} (where \fh~{\tt 2.13.1}
was used). It is therefore very important that the MSSM~calculation
of~$\Mh$ along the stop coannihilation strip is optimized.

In Fig.~\ref{fig:m12vsm0tb5} we show four examples of \mbox{$(m_{1/2},
  m_0)$}~planes in the~CMSSM for~\mbox{$\tan \beta = 5$}. The upper
panels are for~\mbox{$A_0 = 3\, m_0$}, and the lower panels are
for~\mbox{$A_0 = - 4.2\, m_0$}, assuming that the Higgs mixing
parameter~\mbox{$\mu > 0$} (left panels) or~\mbox{$\mu < 0$} (right
panels). In each panel, the brick-red shaded regions are excluded
because they feature a charged~LSP, which is the~${\tilde \tau_1}$ in
the lower right regions and the~${\tilde t_1}$ in the upper left
regions. There are very narrow dark blue strips close to these
excluded regions where the LSP~contribution to the dark matter
density~\mbox{$\Omega_\chi\, h^2 < 0.2$}. This range is chosen for
clarity, as the range~\mbox{$\Omega_{\rm CDM}\, h^2 = 0.1193 \pm
  0.0014$} allowed by cosmology~\cite{Planck} would correspond to a
much thinner strip that would be completely invisible.  Even with the
extended range for the relic density, the line is barely
visible.\footnote{Even taking a range which
  allows~\mbox{$\Omega_\chi\, h^2 < 1$} would not make the line thick
  enough to be more visible on the scale of these figures.}  As we
discuss in more detail below, the coannihilation strips generally have
endpoints at very high masses, where the cross section becomes too
small to ensure the proper relic density, even when~\mbox{$m_\chi =
  m_{\tilde t_1}$}. The locations of these endpoints for the Planck
range of~\mbox{$\Omega_{\rm CDM}\, h^2$} are indicated by~{\bf
  X}~marks along the strips.  The panels feature contours of~$\Mh$
calculated using \FHnew\ (red solid lines) and \FHold\ (thin gray
dashed lines). The latter are truncated in regions of large stop
masses for~\mbox{$\tb = 5$}, \mbox{$A_0 = 3\, m_0$} and~\mbox{$\mu >
  0$}, where \FHold\ fails to return valid calculations of~$\Mh$.

\begin{figure}[hbtp!]
\centerline{
\includegraphics[height=8cm]{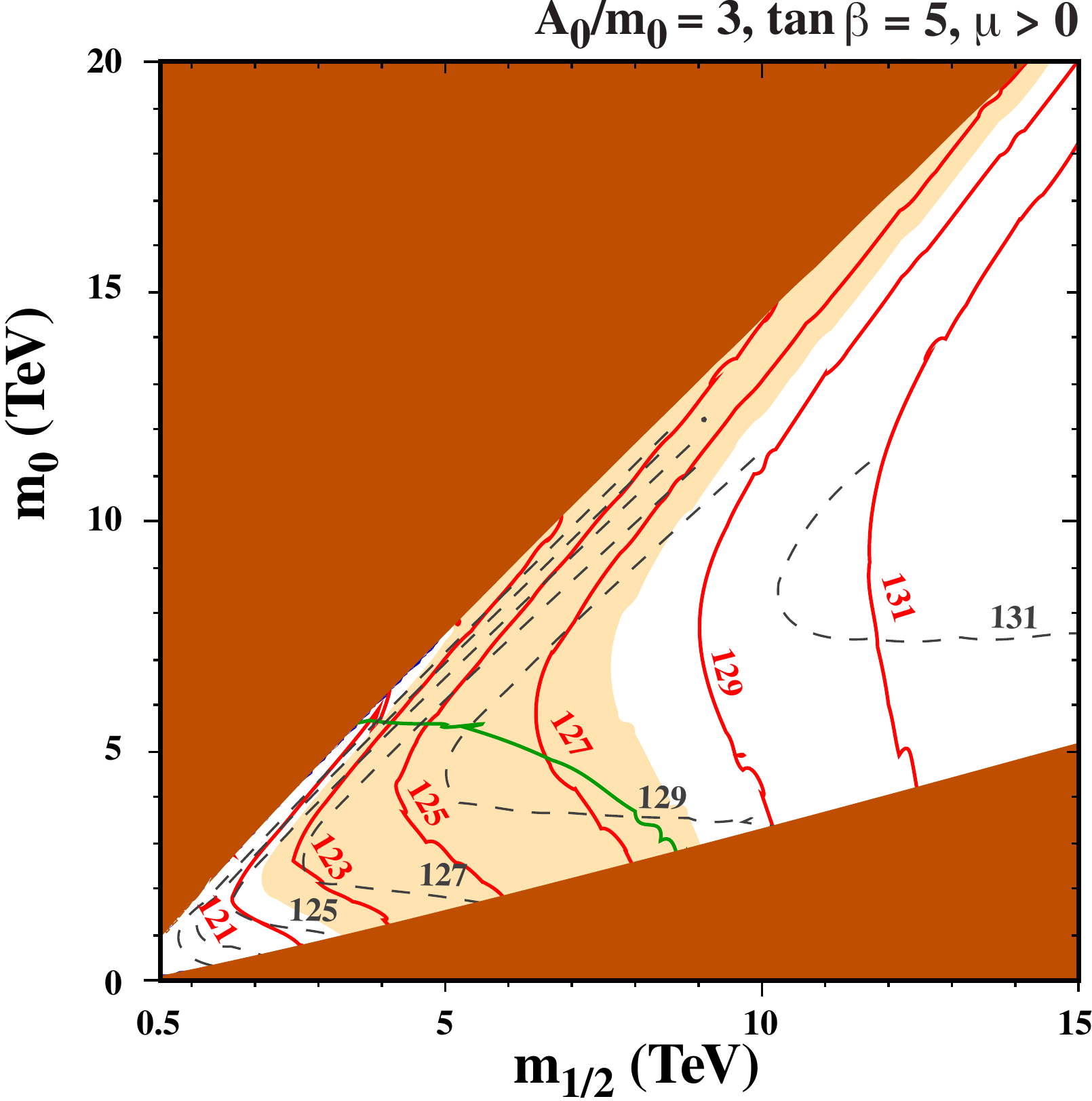}
\includegraphics[height=8cm]{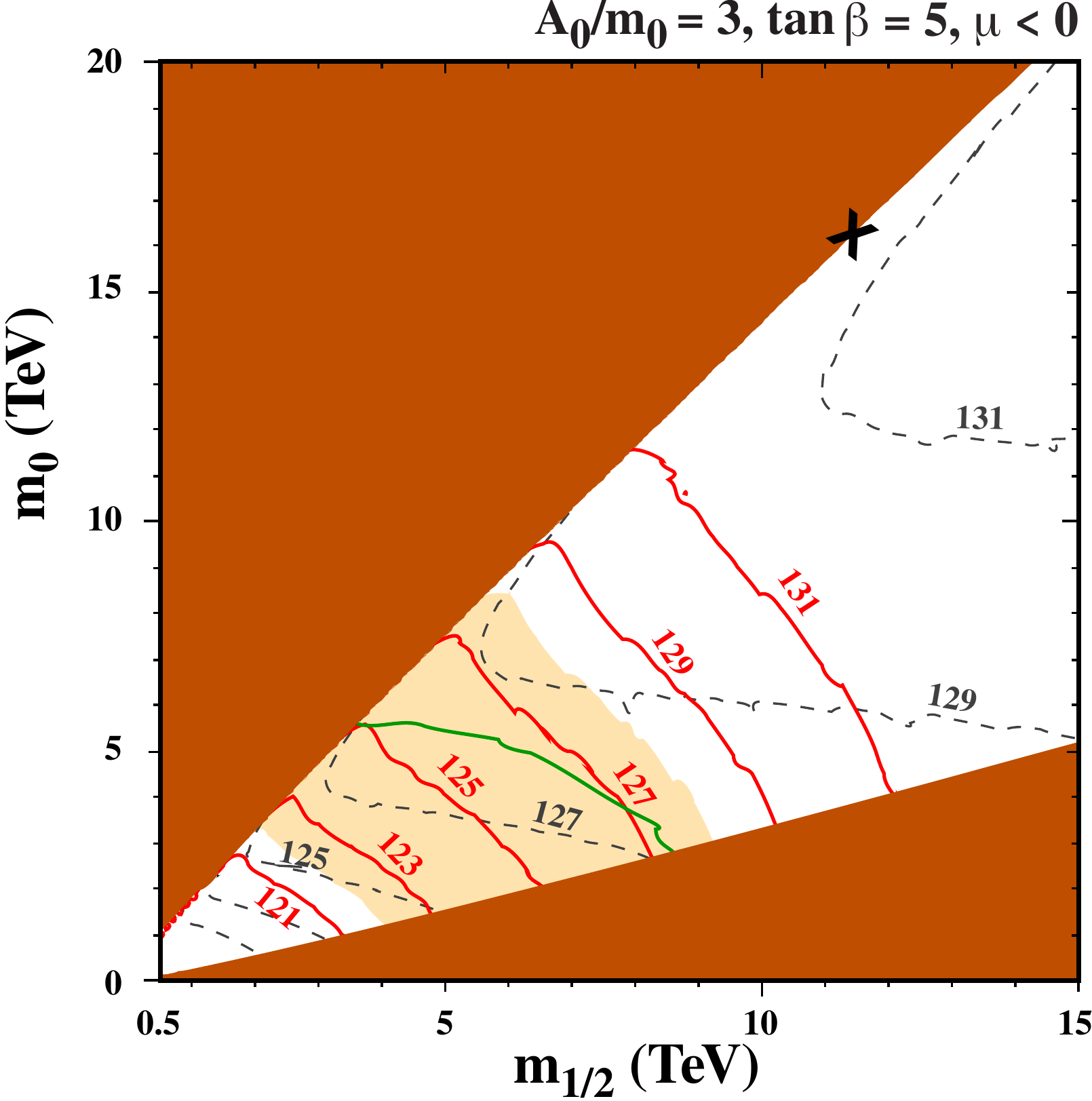}}
\centerline{
\includegraphics[height=8cm]{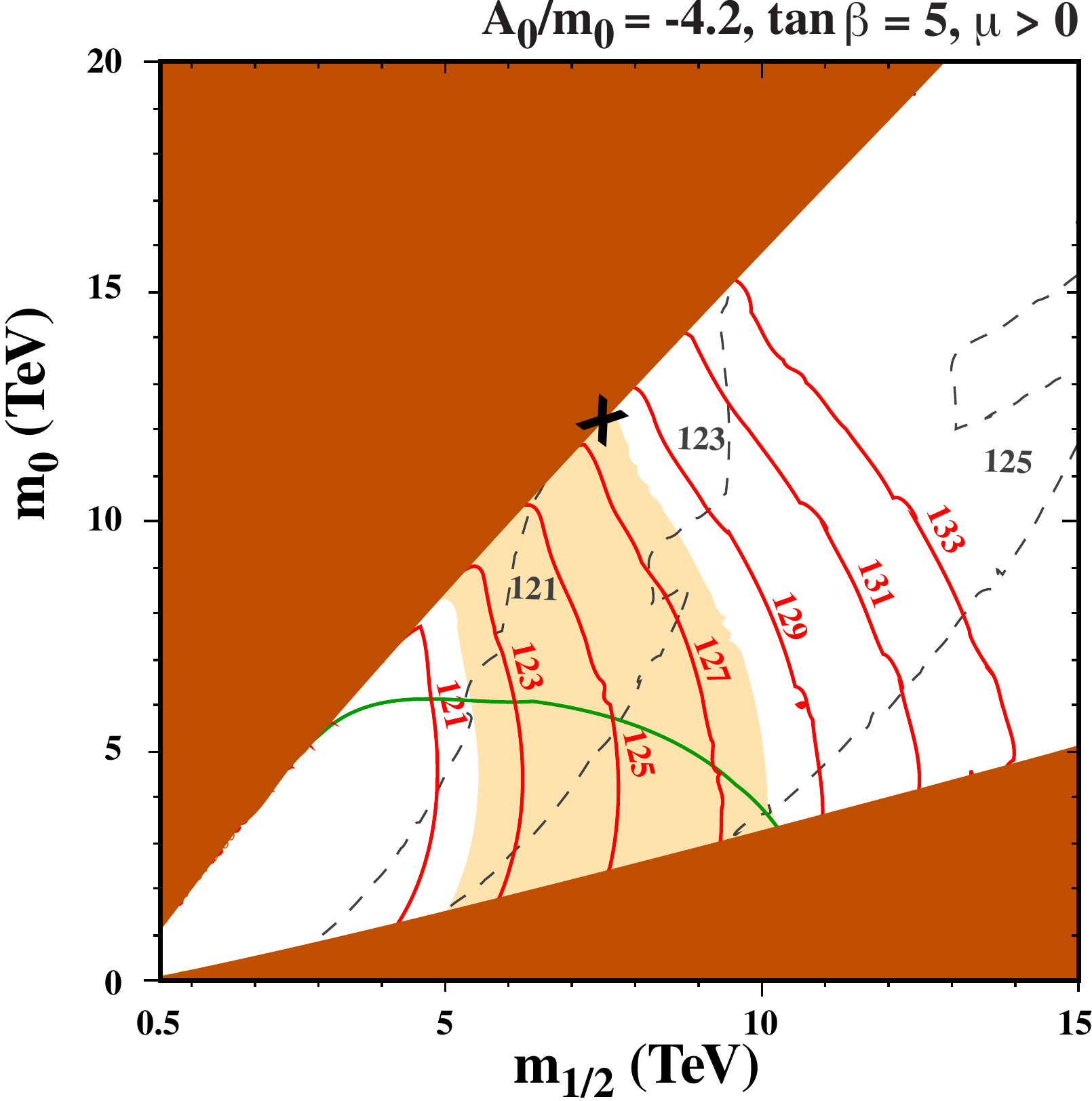}
\includegraphics[height=8cm]{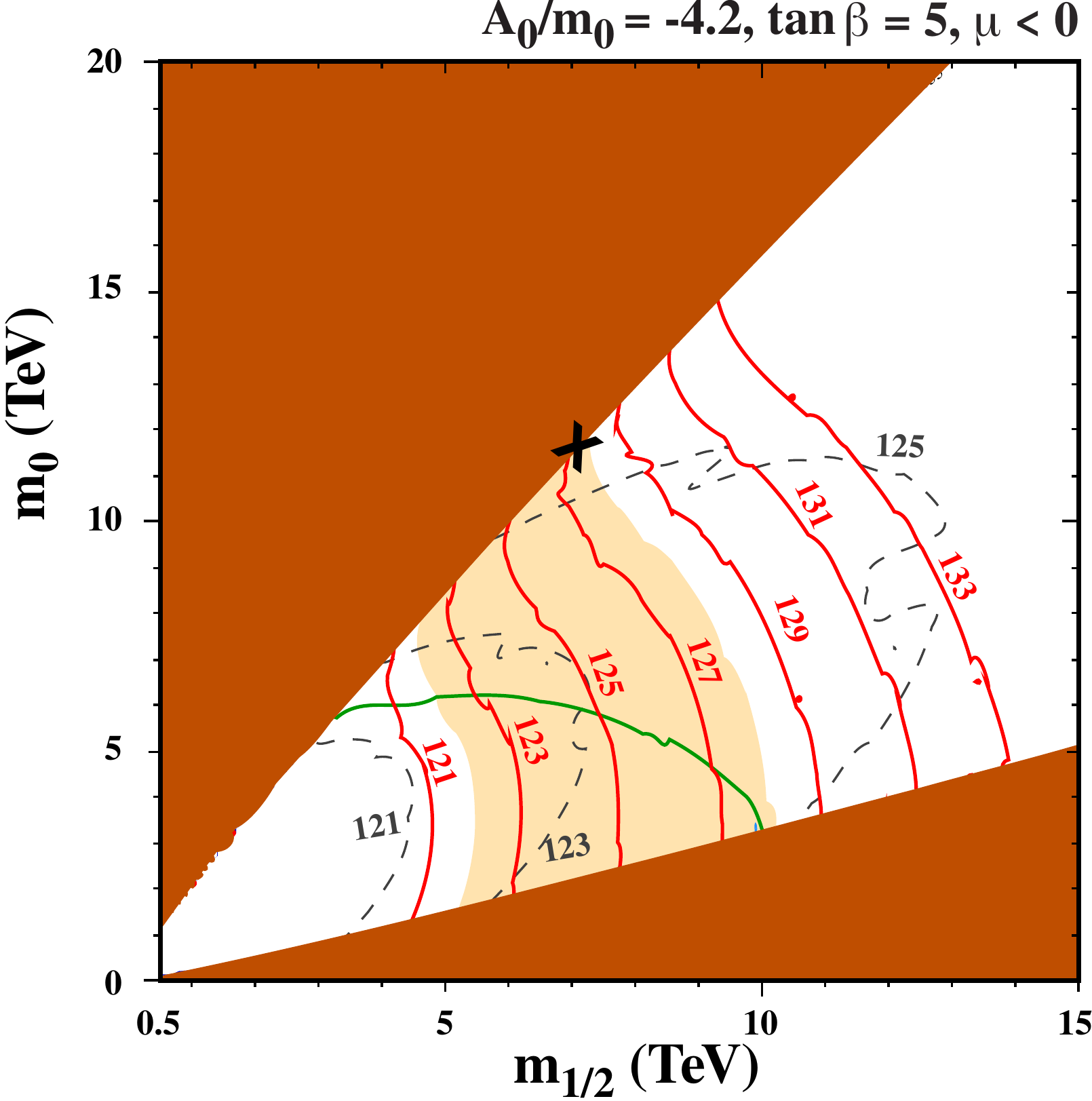}}
\caption{
\label{fig:m12vsm0tb5}
The \mbox{$(m_{1/2}, m_0)$}~planes in the~CMSSM for \mbox{$\tan \beta
  = 5$}, \mbox{$A_0 = 3\, m_0$} and~\mbox{$\mu > 0$} (upper left
panel) or \mbox{$\mu < 0$} (upper right panel), \mbox{$\tan \beta =
  5$}, \mbox{$A_0 = - 4.2\, m_0$} and \mbox{$\mu > 0$} (lower left
panel), and \mbox{$\tan \beta = 5$}, \mbox{$A_0 = -4.2\, m_0$} and
\mbox{$\mu < 0$} (lower right panel).  The brick-red shaded regions
are excluded because they feature a charged~LSP, and {the} panels
contain narrow dark blue strips close to these excluded regions
where~\mbox{$\Omega_\chi\, h^2 < 0.2$}, as well as contours of~$\Mh$
calculated using \FHnew\ (red solid lines) and \FHold\ (thin gray
dashed lines).  The light orange shaded region corresponds
to~\mbox{$\Mh \in [122,128] \gev$} using \FHnew.  The~{\bf X} marks
the position of the stop coannihilation endpoint.  The solid green
lines show the lower limit on the proton lifetime calculated in a
minimal supersymmetric SU(5)~GUT.}
\end{figure}

Across the \mbox{$(m_{1/2}, m_0)$}~planes we see very different
behaviors of the values of the Higgs mass calculated with {\tt
  FeynHiggs}~{\tt 2.14.1} and \FHold, particularly along the stop
coannihilation strip, where~$m_{\tilde t_1}$ and~$\mneu1$ may reach
several~TeV. In such a case, the values of~$\Mh$ given by \FHnew\ are
much more reliable than those obtained with \FHold.  In the absence of
a {detailed uncertainty estimate that depends on the considered
  region of the parameter space (the update of the uncertainty
  estimate of \FH\ taking into account the latest improvements in the
  Higgs-mass prediction is still a work in progress)}, here and later we
consider values of the input mass parameters {as acceptable} for
which \FHnew\ yields~\mbox{$\Mh = 125 \pm 3 \gev$}, \IE,~\mbox{$\Mh
  \in [122, 128] \gev$} {(where the additional experimental
  uncertainty is negligible in comparison).}  This range from
\FHnew\ is shaded light orange. We discuss this constraint in more
detail below, but it is already clear from Fig.~\ref{fig:m12vsm0tb5}
that \FHnew\ favors ranges of~$m_{\tilde t_1}$ and~$\mneu1$ that are
quite different from those that would have been indicated by \FHold.

For~\mbox{$A_0 = 3\,m_0$} and~\mbox{$\mu > 0$}, the Higgs mass
decreases rapidly as the stop LSP~boundary is approached.  In this
case, the Higgs mass calculated using \FHold\ is too small all along
the coannihilation strip.  Furthermore, we see that \FHold\ was not
able to produce a reliable result beyond~\mbox{$m_0 \gtrsim 13$}~TeV.
Since the endpoint of the coannihilation strip is at much
larger~$m_0$, \FHnew\ offers a significant improvement.  This version
of \FH\ yields values of the Higgs mass that are significantly larger
along the strip, rising as high as~\mbox{$\Mh = 128$}~GeV at the
endpoint which is not seen in this panel as it lies beyond the shown
range in~\mbox{$(m_{1/2},m_0)$}.  For~\mbox{$A_0 = 3\,m_0$}
and~\mbox{$\mu < 0$}, the Higgs mass is reduced in the newer version
of \FH\ for most of the strip, though it is larger for~\mbox{$m_{1/2}
  \gtrsim 6$}~TeV.  While both versions of \FH\ provide strip segments
with an acceptable Higgs mass, the location of the segment shifts
upwards in the new version.  In this case, the Higgs mass
is~\mbox{$\Mh = 135$}~GeV at the endpoint of the coannihilation strip,
so the Higgs mass itself provides a constraint~\mbox{$m_{1/2} \lesssim
  6$}~TeV, as seen more clearly in the profile plots discussed below.
The endpoint is marked by an~{\bf X} at~\mbox{$(m_{1/2},m_0) \sim
  (11.3,16.1) \tev$}. When~\mbox{$A_0 = -4.2\, m_0$} and~\mbox{$\mu >
  0$}, we clearly see a large difference between \FHold~and \FHnew.
In this case, the endpoint of the coannihilation strip is found at
lower~\mbox{$(m_{1/2}, m_0)$}.  With \FHold, we find~\mbox{$\Mh <
  122$}~GeV at the endpoint (as has also been found using {\tt
  FeynHiggs~2.11.3}~\cite{azar}), whereas with \FHnew, we
find~\mbox{$\Mh = 128$~GeV} at the endpoint.  When~\mbox{$A_0 = -4.2\,
  m_0$} and~\mbox{$\mu < 0$}, {the~\mbox{$\Mh = 127 \gev$}~contour
  from \FHold\ is beyond the frame,} whereas with \FHnew\ we find
\mbox{$\Mh = 128$}~GeV at the endpoint.

We also show in Fig.~\ref{fig:m12vsm0tb5} as green lines contours of
the lifetime for the proton decay~\mbox{$p \to K+ \nu$} of~\mbox{$6.6
  \times 10^{33}$}~yrs, the current lower limit for this decay
mode. These contours have been calculated in the minimal
SU(5)~GUT~model, neglecting possible effects due to new degrees of
freedom at the GUT~scale. Even though this calculation is probably
inapplicable in a realistic GUT~completion of the~CMSSM, it does
indicate that proton stability is unlikely to be a headache along the
stop coannihilation strip in the~CMSSM with~\mbox{$\tb = 5$} with
TeV~scale masses~\cite{EvNO,azar,eelnos}.  The position of this
contour is similar in all four panels as the proton lifetime is mostly
sensitive to~$\tan \beta$ rather than the signs of~$A_0$ or~$\mu$.

Fig.~\ref{fig:m12vsm0tb20} shows a similar set of plots
for~\mbox{$\tan \beta = 20$} and~\mbox{$A_0 = 2.75\, m_0$} (upper
panels) and for~\mbox{$\tan \beta = 20$} and~\mbox{$A_0 = - 3.5\,
  m_0$} (lower panels), with~\mbox{$\mu > 0$} (left panels)
and~\mbox{$\mu < 0$} (right panels). For specific values of~$m_{1/2}$
and~$m_0$, the calculated values of~$\Mh$ are generally larger
for~\mbox{$\tb = 20$} than for~\mbox{$\tb = 5$}, as was to be
expected.  We see again substantial differences between the values
of~$\Mh$ obtained from \FHnew\ (red solid lines) and from
\FHold\ (thin gray dashed lines), in particular along the stop
coannihilation strip.  Once again, we see that when~\mbox{$A_0 > 0$}
and~\mbox{$\mu > 0$}, the contours of~$\Mh$ run almost parallel to the
boundary of the LSP~region, implying that the values of~$\Mh$ along
the stop coannihilation strip are very sensitive to the input
parameters and the level of sophistication of the
$\Mh$~calculation. Since the coannihilation strip extends beyond the
range of the plot, both versions of \FH\ yield acceptable segments
along the strip, albeit with different mass ranges.  For~\mbox{$A_0 <
  0$}, the Higgs-mass contours no longer run parallel to the
coannihilation strip, and \FHnew\ predicts~\mbox{$\Mh = 130$}~GeV at
the endpoint, which is found at much lower~$m_{1/2}$ and~$m_0$ as
marked by the~{\bf X} in the figure. At this higher value of~$\tan
\beta$, there is not a large difference in the Higgs mass when the
sign of~$\mu$ is reversed, since the contribution to~$X_t$ depends
on~$\mu/\tan \beta$. Although the difference may appear small,
when~\mbox{$A_0 > 0$}, the Higgs mass is significantly larger along
the strip as one approaches the endpoint at large~$m_{1/2}$
and~$m_0$. We note that for~\mbox{$A,\mu < 0$}, at high~$m_{1/2}$ and
low~$m_0$ there is a lack of convergence of the~RGEs, due to a
divergent $b$-quark Yukawa coupling, shown by the gray shading.

\begin{figure}[hbtp!]
\centerline{
\includegraphics[height=8cm]{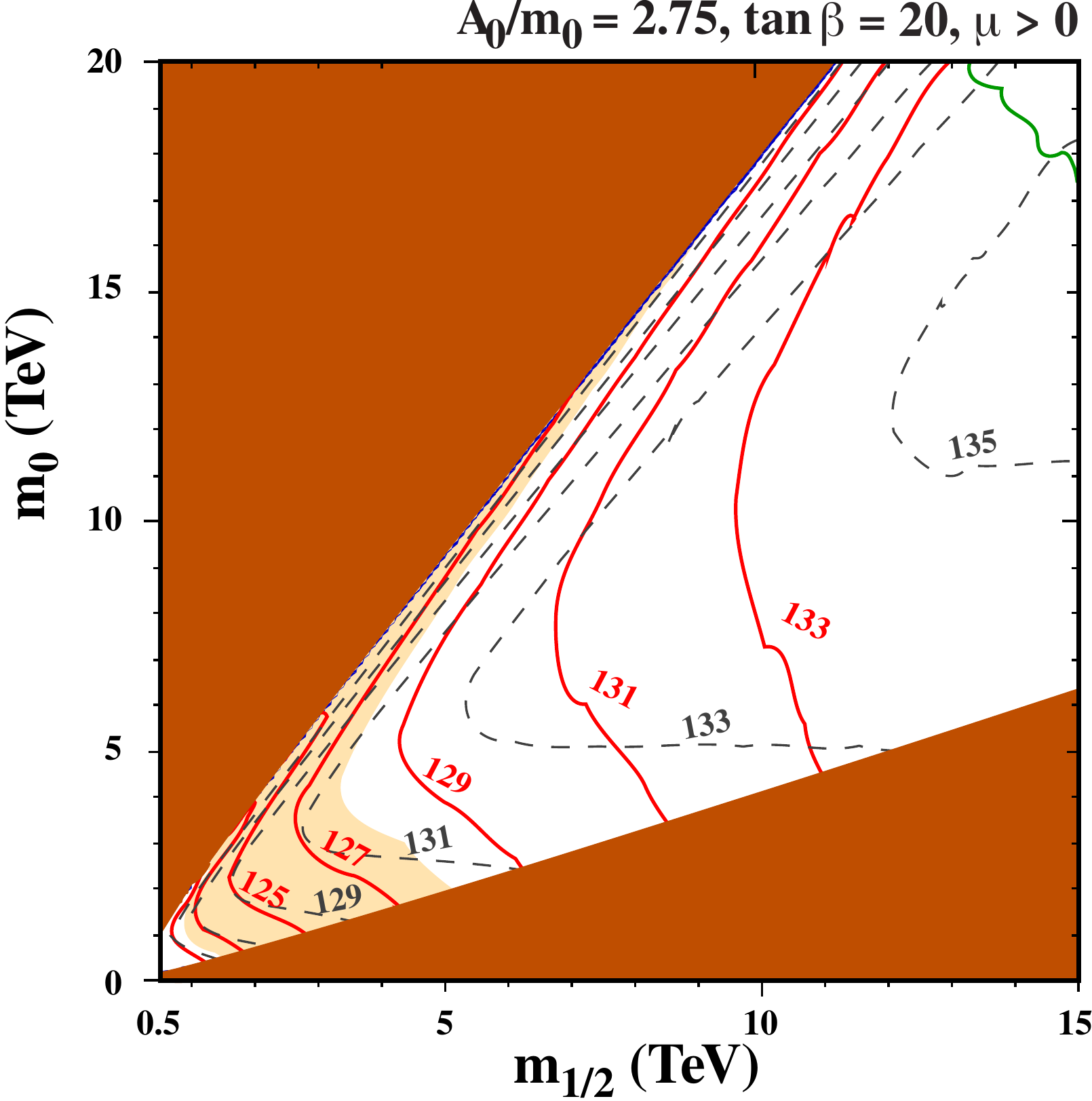}
\includegraphics[height=8cm]{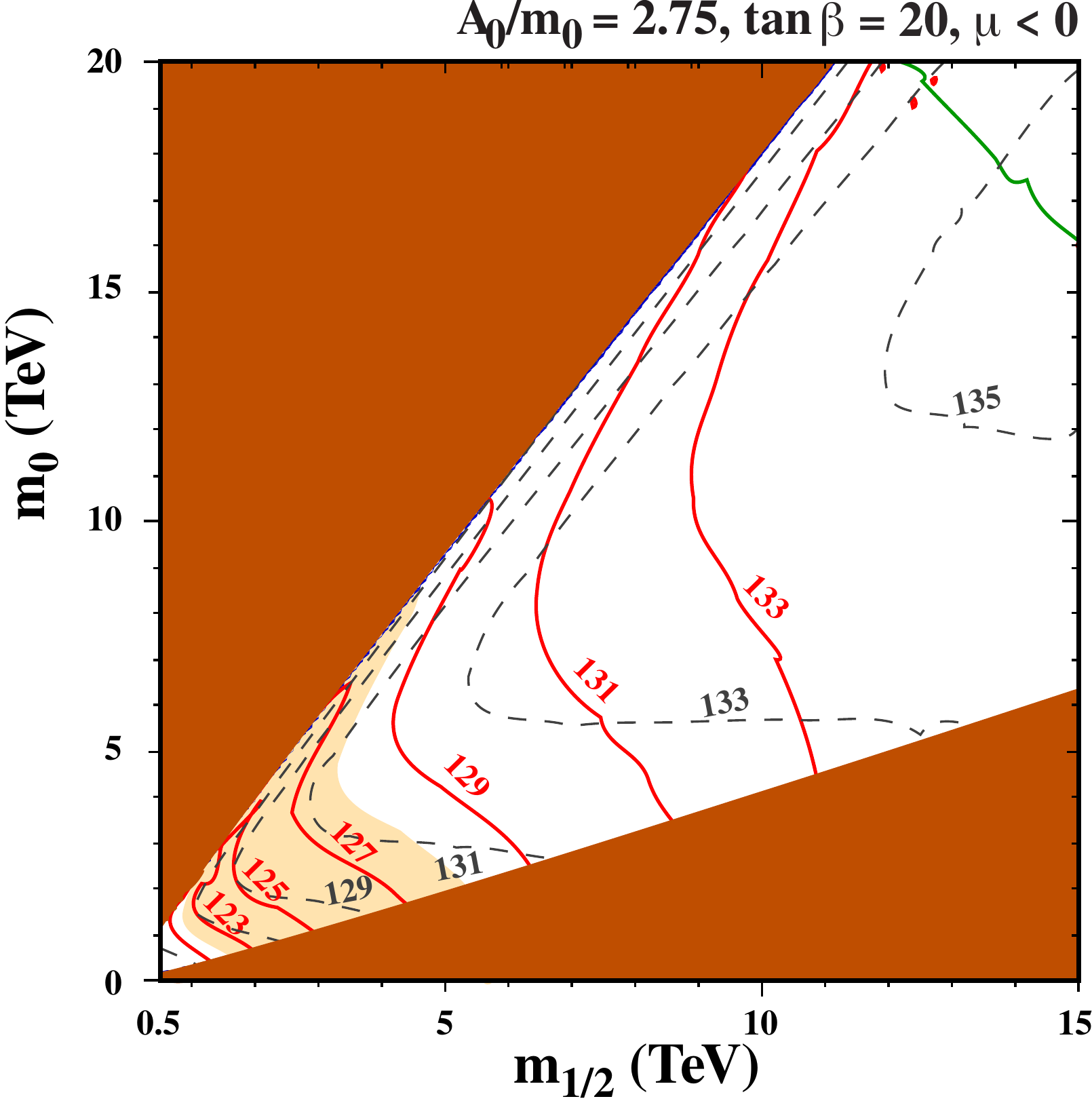}}
\centerline{
\includegraphics[height=8cm]{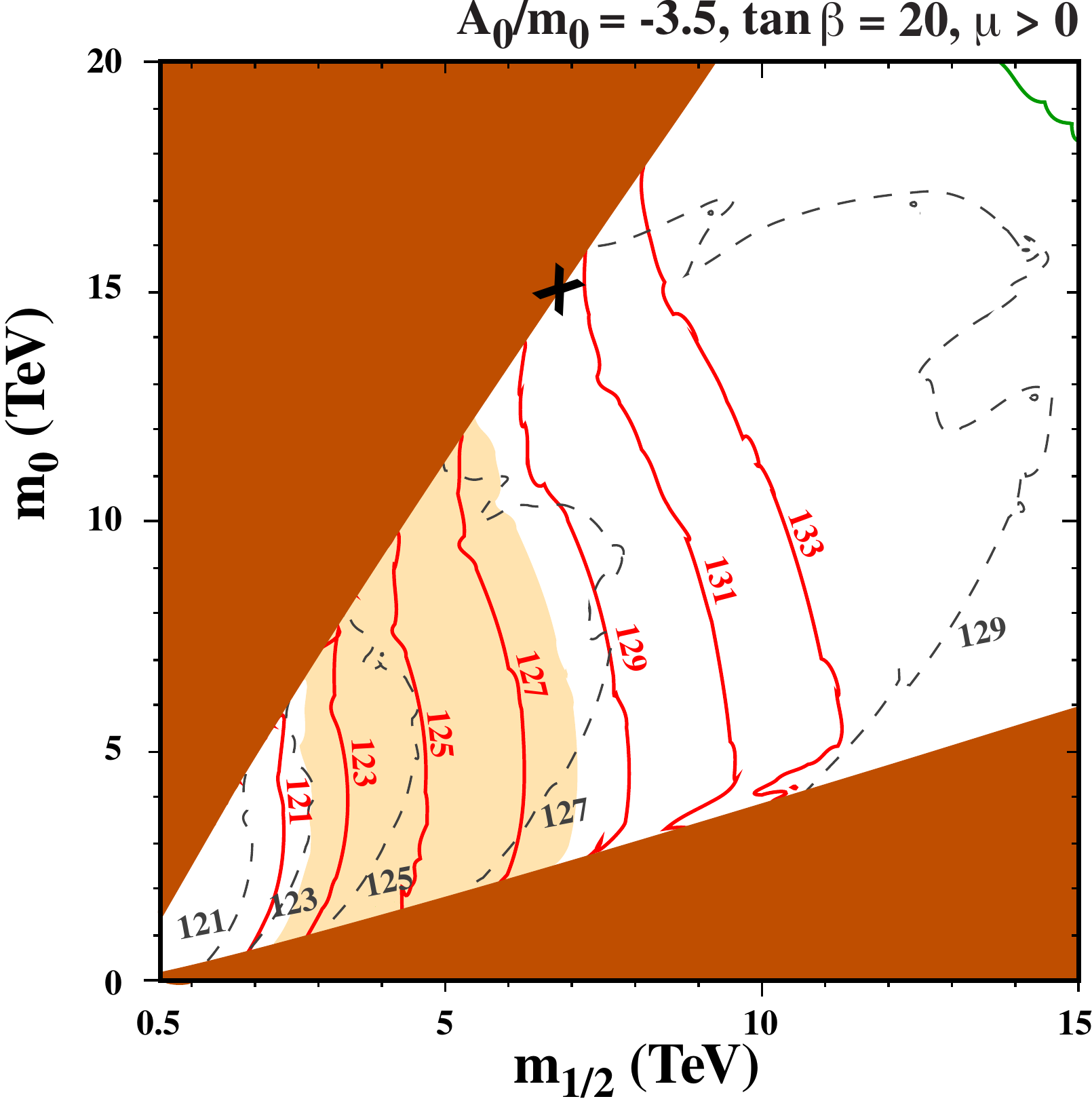}
\includegraphics[height=8cm]{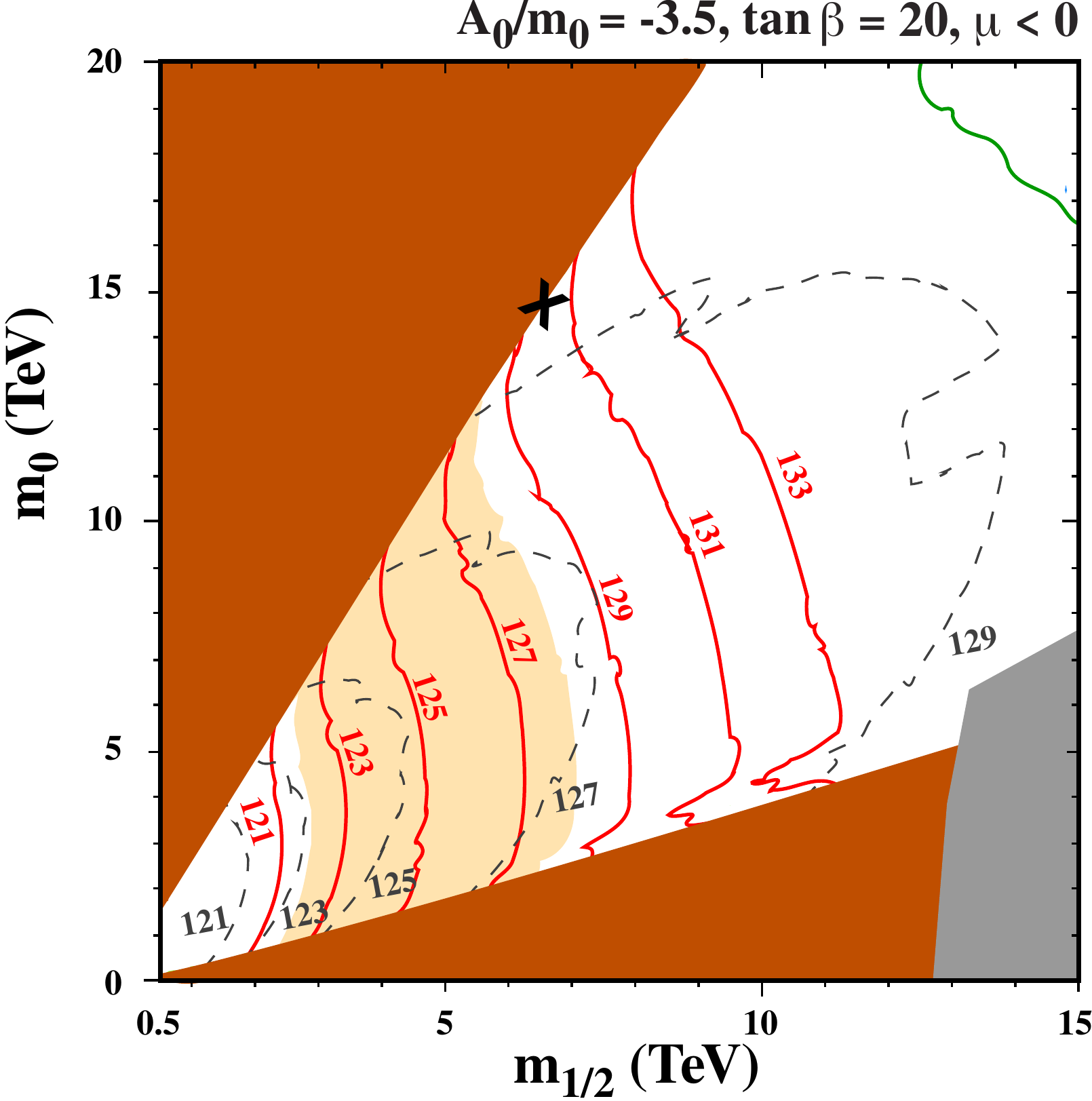}}
\caption{
\label{fig:m12vsm0tb20}
As in Fig.~\protect\ref{fig:m12vsm0tb5}, but for the cases \mbox{$\tan
  \beta = 20$}, \mbox{$A_0 = 2.75\, m_0$} and \mbox{$\mu > 0$} (upper
left panel), \mbox{$\tan \beta = 20$}, \mbox{$A_0 = 2.75\, m_0$} and
\mbox{$\mu < 0$} (upper right panel), \mbox{$\tan \beta = 20$},
\mbox{$A_0 = - 3.5\, m_0$} and \mbox{$\mu > 0$} (lower left panel) and
\mbox{$\tan \beta = 20$}, \mbox{$A_0 = - 3.5\, m_0$} and \mbox{$\mu <
  0$} (lower right panel).  Contours of~$\Mh$ calculated using
\FHnew\ are shown as red solid lines those using \FHold\ as thin gray
dashed lines.  The light orange shaded region corresponds
to~\mbox{$\Mh \in [122,128] \gev$} using \FHnew.  In each panel,
the~{\bf X} marks the position of the stop coannihilation endpoint.
The solid green lines show the lower limit on the proton lifetime
calculated in a minimal supersymmetric SU(5)~GUT.  For~\mbox{$A,\mu <
  0$}, the gray shading at high~$m_{1/2}$ denotes the lack of
convergence of the~RGEs due to a divergent $b$-quark Yukawa coupling.
}

\end{figure}

We note that the green contours where~\mbox{$\tau(p \to K+ \nu) = 6.6
  \times 10^{33}$}~yrs in the minimal SU(5) GUT model are at much
larger values of~$m_{1/2}$ and~$m_0$ for~\mbox{$\tb = 20$} than they
were for~\mbox{$\tb = 5$}, as was also to be expected. However, we
emphasize that the calculation of the proton lifetime is sensitive to
the details of the GUT~dynamics, and that proton stability may be an
issue but is not necessarily a problem for the~CMSSM with~\mbox{$\tb =
  20$}.\footnote{Corrections to the gauge couplings from
  Planck-suppressed operators can change significantly the estimate of
  the grand-unification scale and hence the proton lifetime.}

Details of the coannihilation strips and endpoints are seen more
clearly in Fig.~\ref{fig:stopstrips5}, which shows the profiles of the
stop coannihilation strips for~\mbox{$\tb = 5$} that were shown in
Fig.~\ref{fig:m12vsm0tb5}.  The values of~$m_{1/2}$ are indicated
along the lower horizontal axes, and the corresponding values
of~$\mneu1$ are shown along the upper horizontal axes. For each
value of $m_{1/2}$ we use {\tt SSARD} to calculate the value of $m_0$ that
yields the correct neutralino dark matter density, which we then use to
calculate the other quantities shown. The left
vertical axes show the scales for the mass difference~\mbox{$m_{\tilde
    t_1} - \mneu1$}, which is shown as the blue curve in each panel.
Here and in subsequent analogous figures, the right vertical axes are
the scales for the values of~$\Mh$, the {``allowed''} range~\mbox{$\Mh
  \in [122, 128] \gev$} is indicated by the horizontal light orange
shaded region.  The other lines show the values of~$\Mh$ calculated
using \FHnew\ (solid red) and \FHold\ (dashed black). Since we assign a
theoretical uncertainty of~$\pm 3 \gev$ {to} the \FHnew\ calculation
of~$\Mh$, as indicated by the 
light orange shaded band, the portions of the horizontal axes
corresponding to the \FHnew\ calculation of~\mbox{$\Mh \in [122, 128]
  \gev$} should be regarded as consistent with experiment.

\begin{figure}[btp!]
\centerline{
\includegraphics[height=6cm]{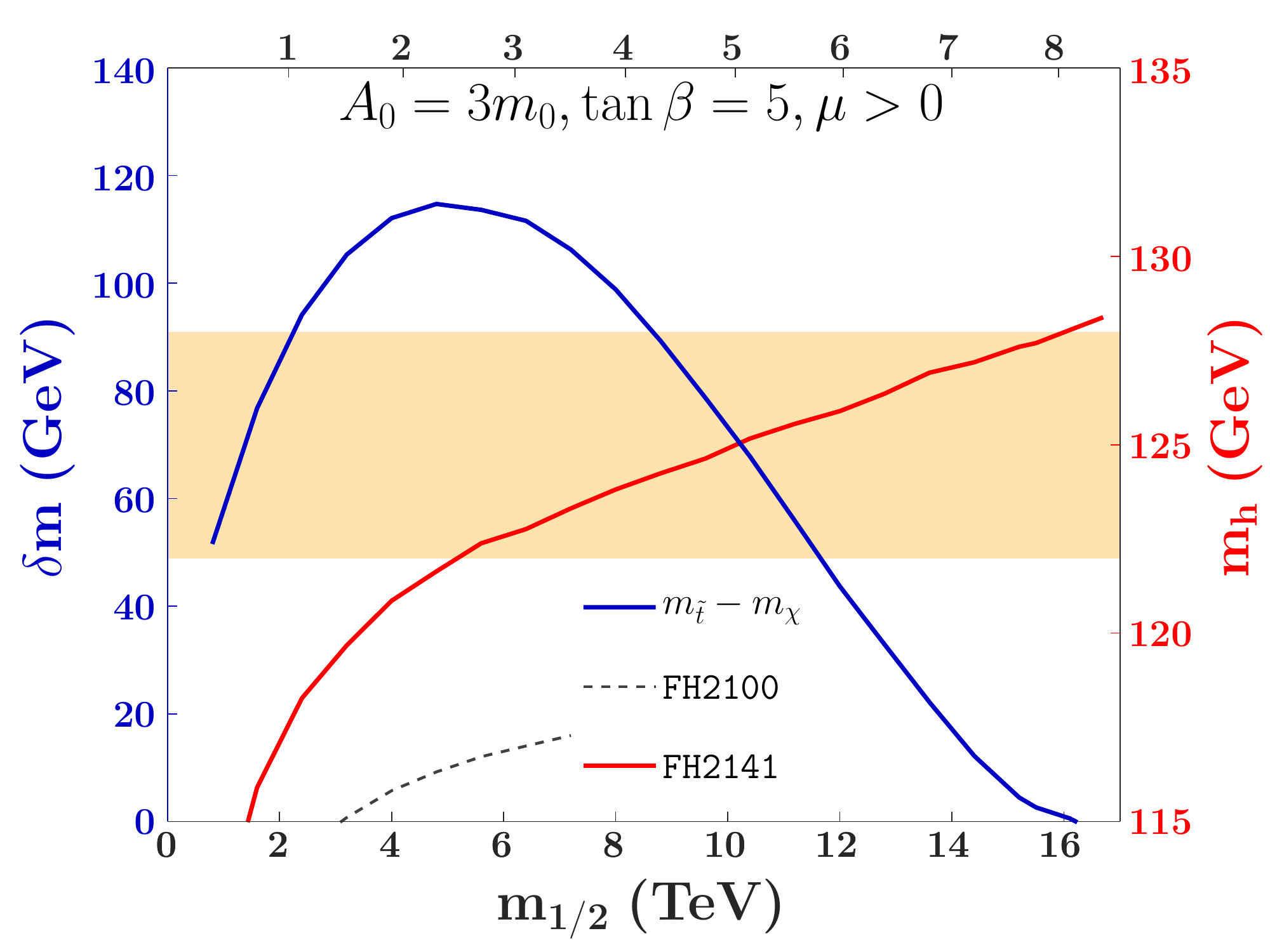}
\includegraphics[height=6cm]{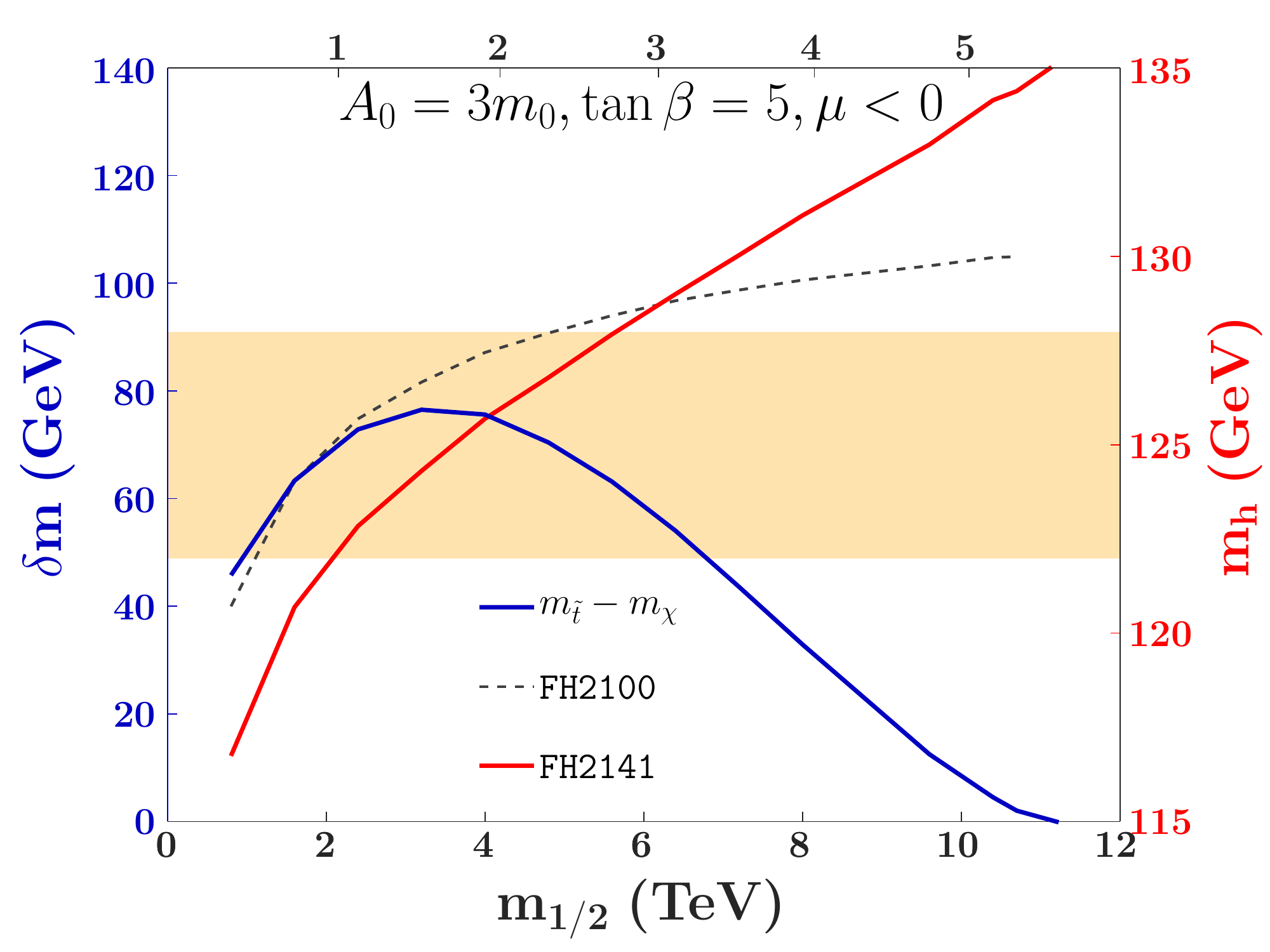}}
\centerline{
\includegraphics[height=6cm]{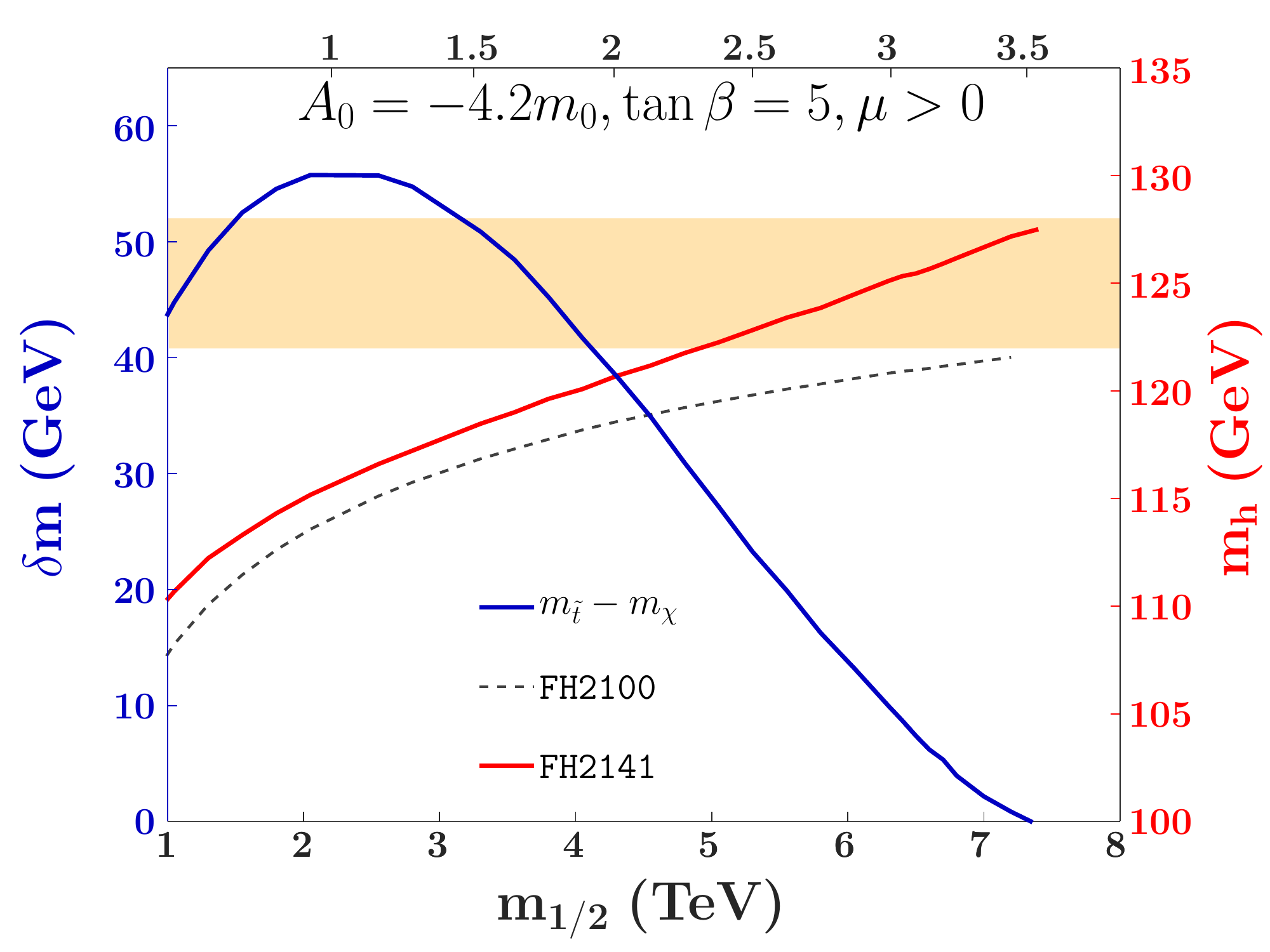}
\includegraphics[height=6cm]{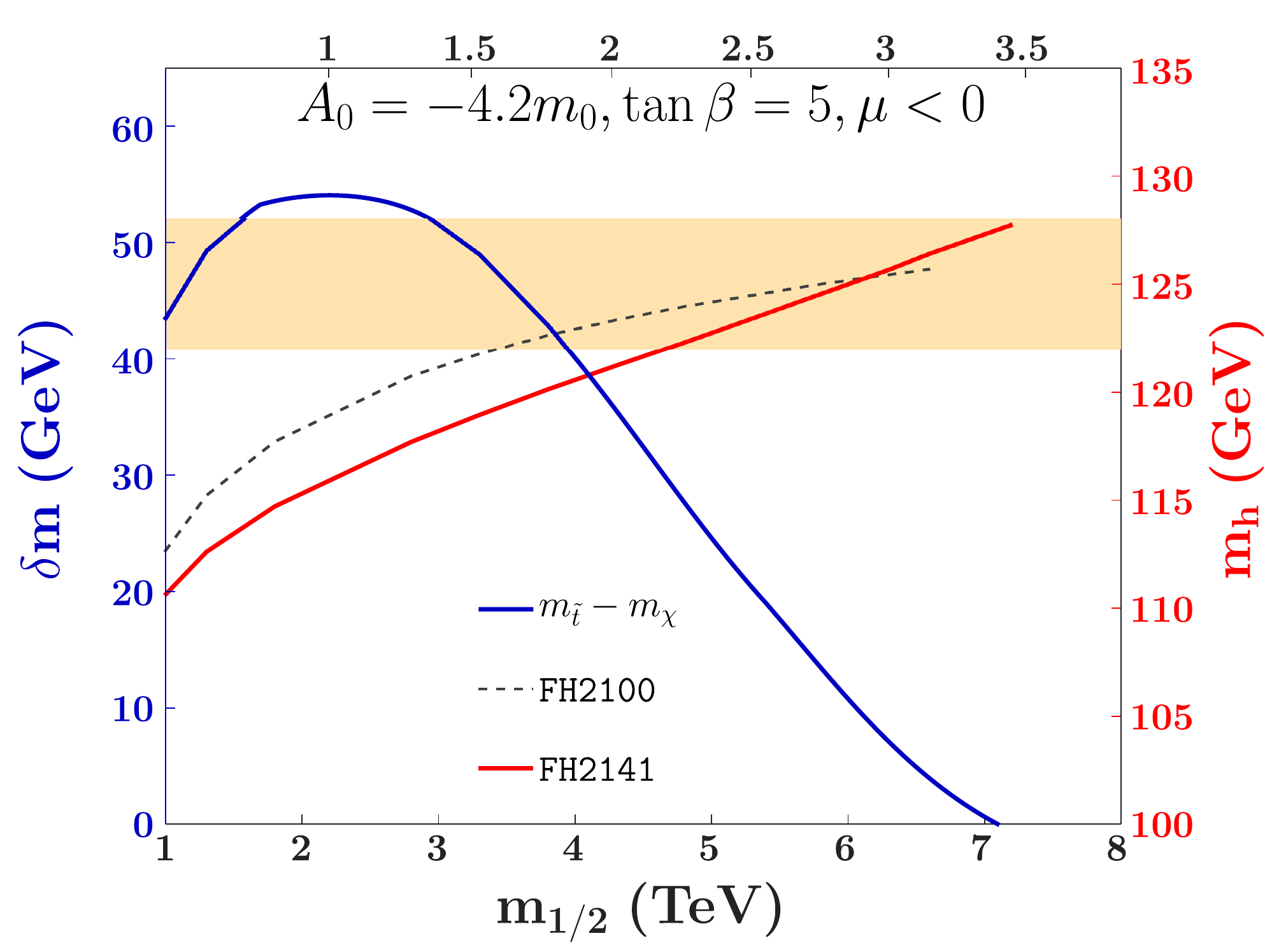}}
\caption{
  \label{fig:stopstrips5}
  The profiles of the CMSSM stop coannihilation strips for \mbox{$\tan
    \beta = 5$}, \mbox{$A_0 = 3\, m_0$} and \mbox{$\mu > 0$} (upper
  left), \mbox{$\tan \beta = 5$}, \mbox{$A_0 = 3\, m_0$} and
  \mbox{$\mu < 0$} (upper right), \mbox{$\tan \beta = 5$}, \mbox{$A_0
    = - 4.2\, m_0$} and \mbox{$\mu > 0$} (lower left), and \mbox{$\tan
    \beta = 5$}, \mbox{$A_0 = - 4.2\, m_0$} and \mbox{$\mu < 0$}
  (lower right).  The lower horizontal axes {show}~$m_{1/2}$, and
  the upper horizontal axes show the corresponding values of~$\mneu1$
  in~TeV.  The blue curves show the mass difference~\mbox{$m_{\tilde
      t_1} - \mneu1$}, to be read from the left vertical axes.  The
  horizontal light orange shaded band between~\mbox{$\Mh = 122, 128
    \gev$} corresponds to {predictions for~$\Mh$ that may be
    regarded as consistent with experiment.}  The other lines show the
  values of~$\Mh$ calculated using \FHnew\ (red) and \FHold\ (dashed
  black), to be read from the right vertical axes.  }
\end{figure}

The upper limits of the stop coannihilation strips shown in
Fig.~\ref{fig:stopstrips5} range from~\mbox{$m_{1/2} \simeq 16 \tev$}
(\mbox{$\mneu1 \simeq 8 \tev$}) for~\mbox{$\tb = 5$}, \mbox{$A_0 = 3\,
  m_0$} and~\mbox{$\mu > 0$} (upper left panel) down to~\mbox{$m_{1/2}
  \simeq 7.1 \tev$} (\mbox{$\mneu1 \simeq 3.4 \tev$}) for~\mbox{$\tb =
  5$}, \mbox{$A_0 = -4.2\, m_0$} and~\mbox{$\mu < 0$} (lower right
panel). In the case of \mbox{$\tb = 5$}, \mbox{$A_0 = 3\, m_0$},
\mbox{$\mu > 0$} (upper left panel of Fig.~\ref{fig:stopstrips5}),
\FHnew\ yields acceptable values of~$\Mh$ for~\mbox{$\mneu1 \gtrsim
  2.5 \tev$} to the end of the strip. On the other hand,
\FHold\ yielded values of~$\Mh$ that are unacceptably low
for~\mbox{$\mneu1 < 4 \tev$}, and unstable values of~$\Mh$
for~\mbox{$\mneu1 > 4 \tev$}. For the other sign of~$\mu$ (upper right
panel of Fig.~\ref{fig:stopstrips5}), both versions of {\tt FeynHiggs}
yield larger values of~$\Mh$, with \FHnew\ now yielding acceptable
values {for~\mbox{$\mneu1 \in [{1.1}, 2.7] \tev$}}, whereas
\FHold\ would have yielded acceptable values {for~\mbox{$\mneu1
    \in [{0.7}, 2.3] \tev$}}.  The differences between the two
versions of {\tt FeynHiggs} are also significant for~\mbox{$A_0 = -
  4.2\, m_0$} (lower panels of Fig.~\ref{fig:stopstrips5}), with
\FHnew\ yielding acceptable values of~$\Mh$ for~\mbox{$\mneu1 \gtrsim
  2.4 \tev$}.  In contrast, \FHold\ predicted a Higgs mass which was
below~$122$~GeV over the entire strip.  When the sign of~$\mu$ is
reversed for this value of~$A_0$, \mbox{$\mneu1 \gtrsim 2.4 \tev$} is
viable with the new version of \FH.

Fig.~\ref{fig:stopstrips20} displays an analogous set of profiles of
stop coannihilation strips for~\mbox{$\tb = 20$}, with~\mbox{$A_0 =
  2.75\, m_0$} in the upper panels, \mbox{$A_0 = - 3.5\, m_0$} in the
lower panels, \mbox{$\mu > 0$} in the left panels and~\mbox{$\mu < 0$}
in the right panels. The upper limits on~$m_{1/2}$
{in the stop coannihilation strip imposed by~$\Mh$ range}
between~$\simord 9 \tev$ for the case~\mbox{$\tb = 20$}, \mbox{$A_0 =
  2.75\, m_0$}, \mbox{$\mu > 0$} and~$\simord 5.5 \tev$ for the
case~\mbox{$\tb = 20$}, \mbox{$A_0 = - 3.5\, m_0$}, \mbox{$\mu <
  0$}. As in the case of~\mbox{$\tb = 5$}, the differences between
\FHnew\ and \FHold\ are larger for~\mbox{$A_0 > 0$} than
for~\mbox{$A_0 < 0$}. {Values} of~$\mneu1$ allowed by the
\FHnew\ calculation of~$\Mh$ range from~$\simord 0.8$ to~$\simord 4.5
\tev$ when~\mbox{$A_0 = 2.75\, m_0$} and~\mbox{$\mu > 0$}, $\simord
0.6$ to~$\simord 2.2 \tev$ when~\mbox{$A_0 = 2.75\, m_0$}
and~\mbox{$\mu < 0$}, and~$\simord 1.3$ to~$\simord 2.6 \tev$
when~\mbox{$A_0 =-3.5\, m_0$} for both signs of~$\mu$.  The
calculation of~$\Mh$ using \FHold\ would have favored different
ranges of~$\mneu1$ in general, \EG, allowing~\mbox{$\mneu1 \gtrsim 1.3
  \tev$} for~\mbox{$A_0 = 2.75\, m_0$} and~\mbox{$\mu > 0$}.  We also
note that, at the larger value of~$\tan \beta$ in this figure, the
sign of~$\mu$ plays a smaller role than in Fig.~\ref{fig:stopstrips5}
with~\mbox{$\tan \beta = 5$}.

As seen in Figs.~\ref{fig:stopstrips5} and~\ref{fig:stopstrips20}, in
general \FHnew\ yields values of~$\Mh$ that increase more rapidly
with~$m_{1/2}$ than the values calculated with \FHold\ along the stop
coannihilation strips we have studied. As a consequence, the
\FHnew\ values of~$\Mh$ lie within the ``allowed'' range for~$\Mh$ in
a smaller interval of~$m_{1/2}$ in some cases. Furthermore, they tend
to be larger than the \FHold\ values at large~$m_{1/2}$. These
differences change substantially the ranges of~$m_{1/2}$ that are
consistent with the experimental measurement of~$\Mh$.

\begin{figure}[bt!]
\centerline{
\includegraphics[height=6cm]{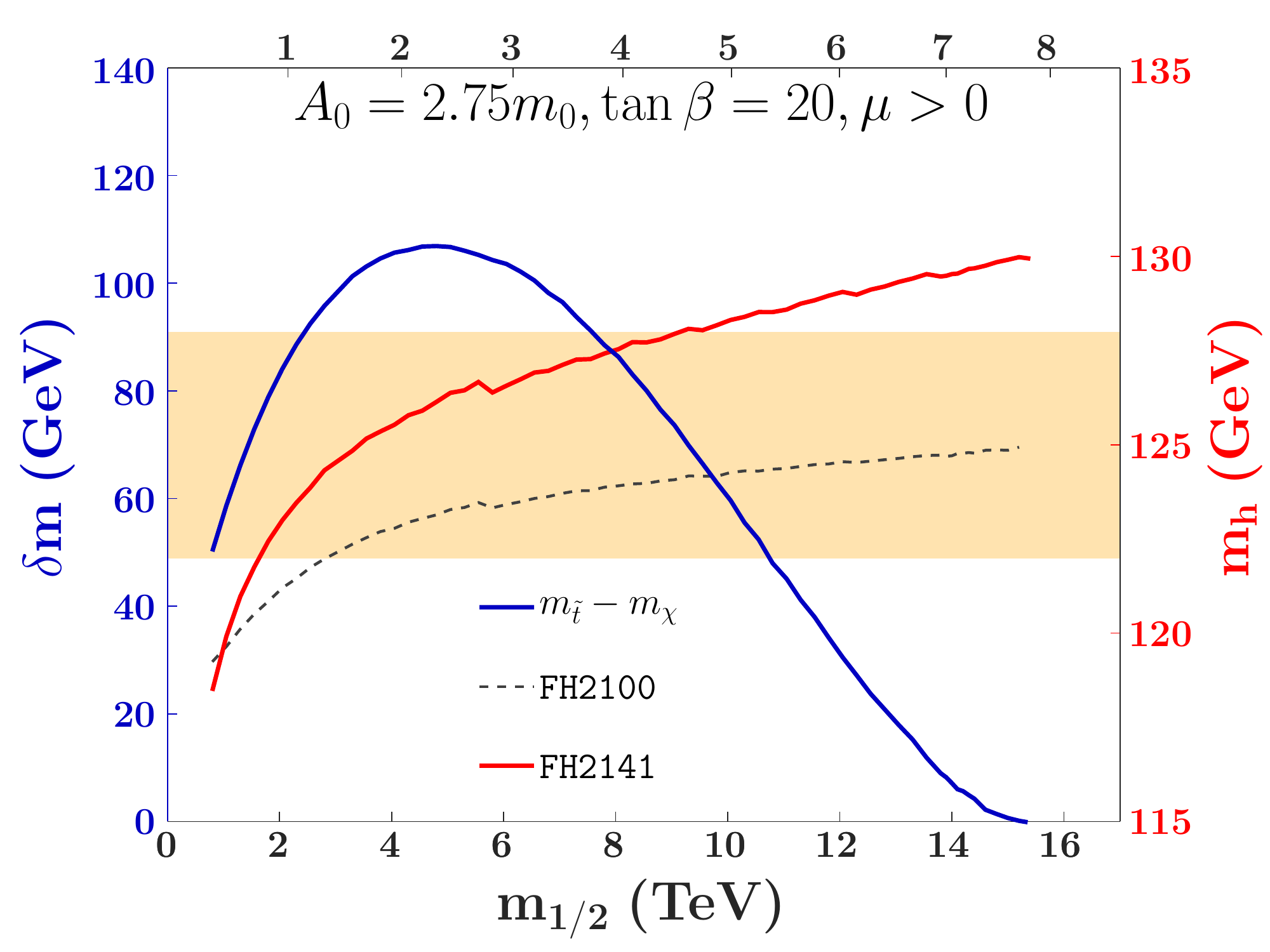}
\includegraphics[height=6cm]{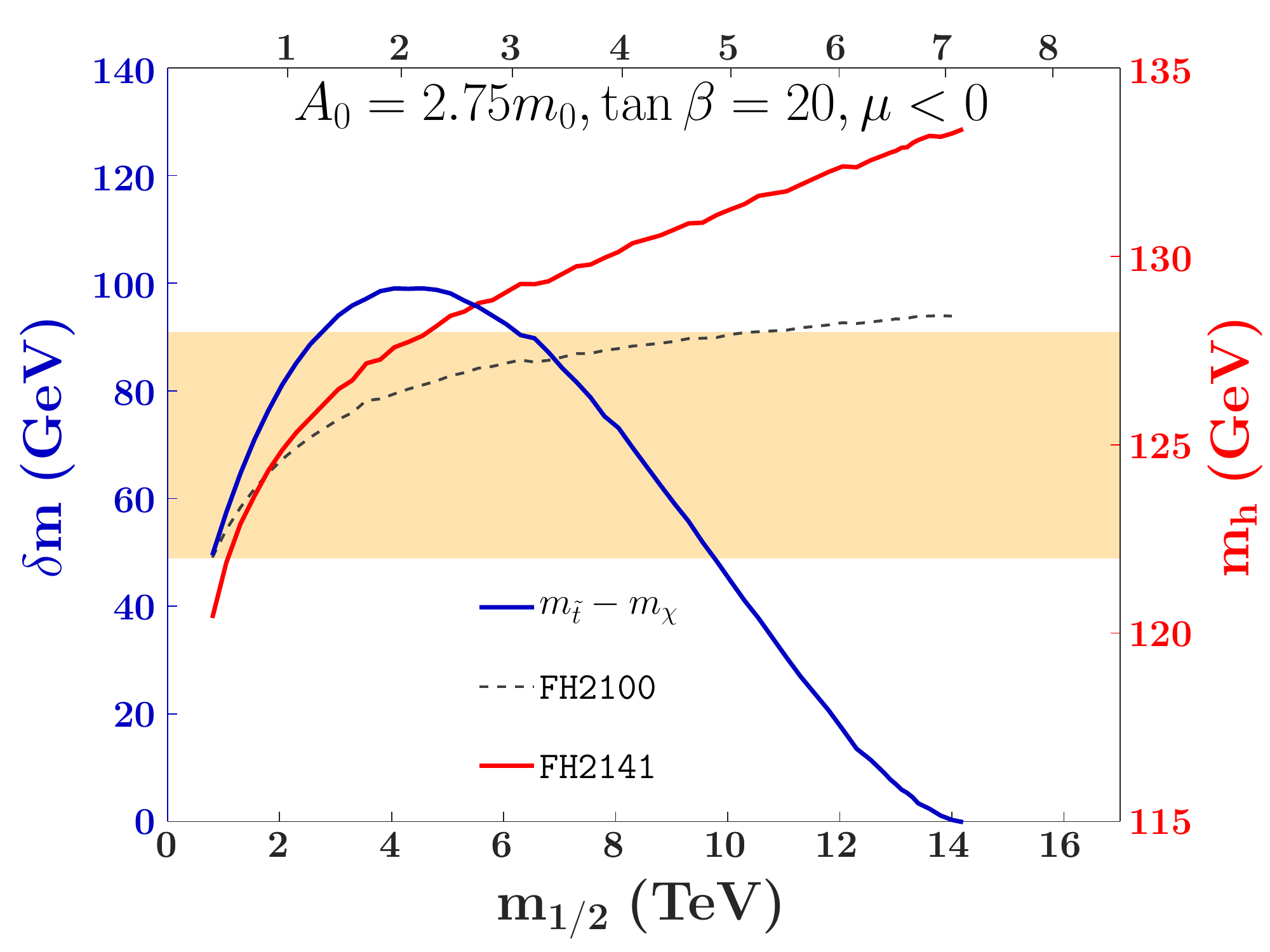}
}
\centerline{
\includegraphics[height=6cm]{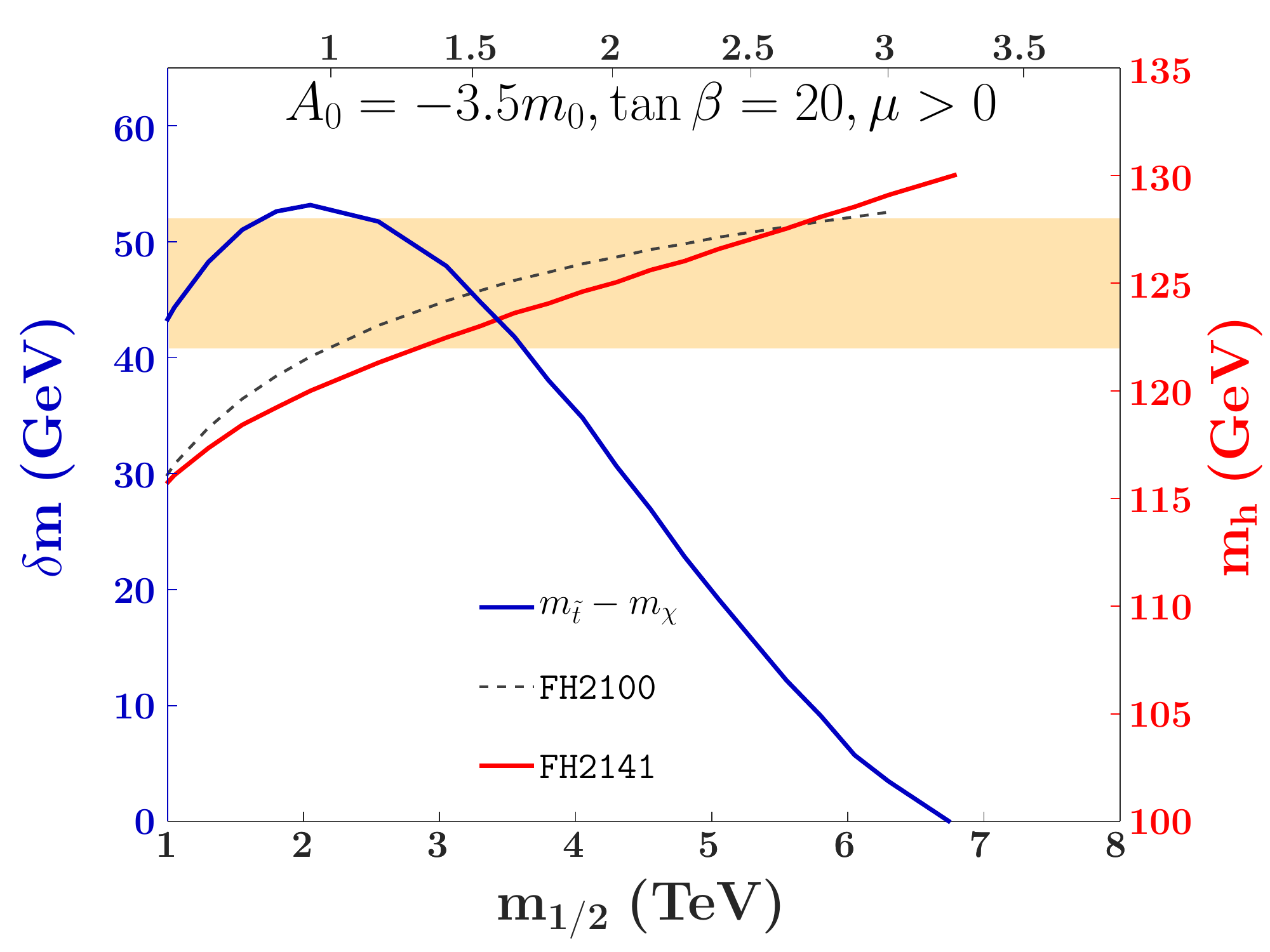}
\includegraphics[height=6cm]{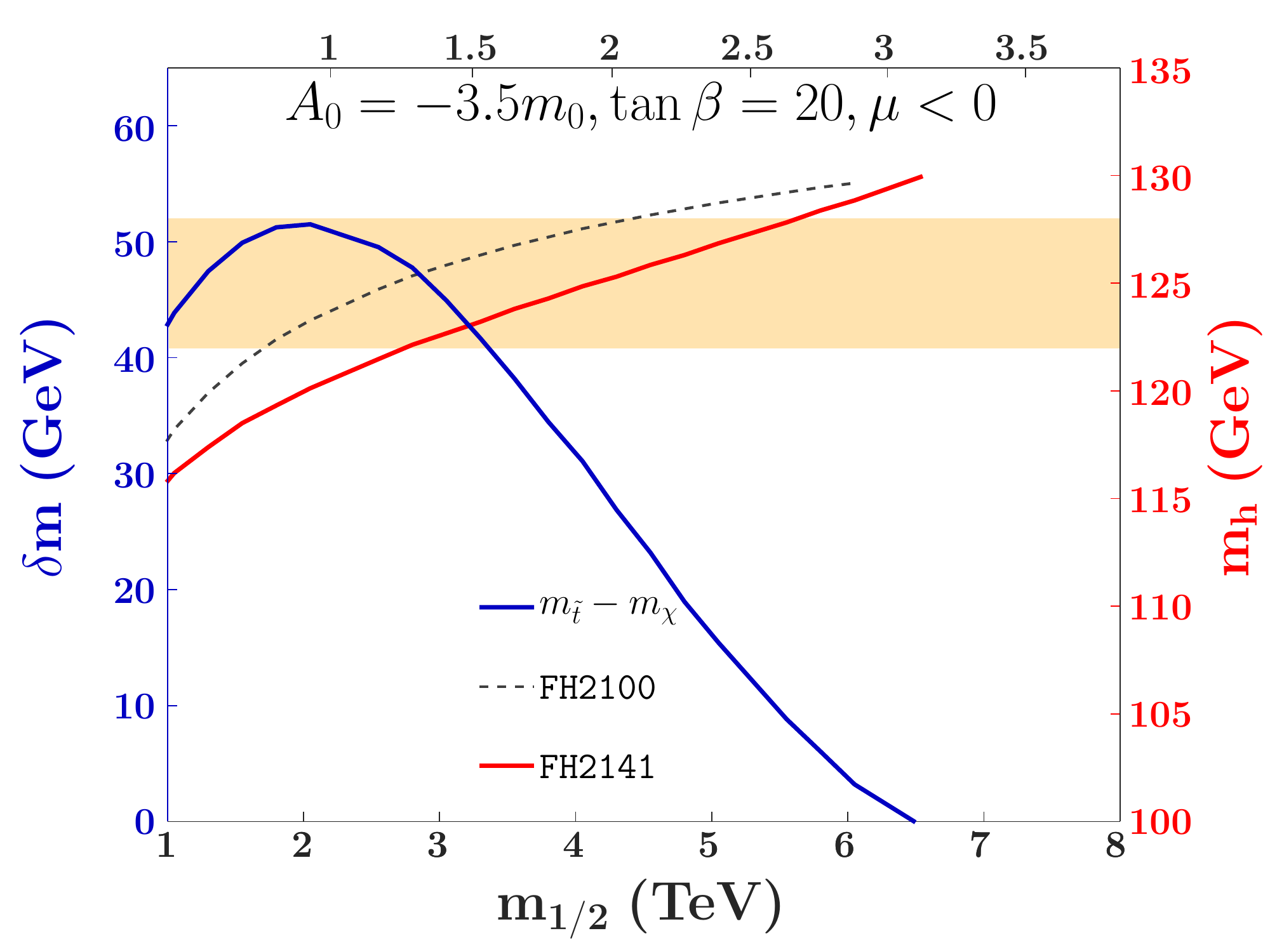}}
\caption{
  \label{fig:stopstrips20}
  The profiles of the CMSSM stop coannihilation strips for \mbox{$\tan
    \beta = 20$}, \mbox{$A_0 = 2.75\, m_0$} and \mbox{$\mu > 0$}
  (upper left), \mbox{$\tan \beta = 20$}, \mbox{$A_0 = 2.75\, m_0$}
  and \mbox{$\mu < 0$} (upper right), \mbox{$\tan \beta = 20$},
  \mbox{$A_0 = - 3.5\, m_0$} and \mbox{$\mu > 0$} (lower left), and
  \mbox{$\tan \beta = 20$}, \mbox{$A_0 = - 3.5\, m_0$} and \mbox{$\mu
    < 0$} (lower right).  The lower horizontal axes show~$m_{1/2}$,
  and the upper horizontal axes show the corresponding values
  of~$\mneu1$ in~TeV.  The blue curves show the mass
  difference~\mbox{$m_{\tilde t_1} - \mneu1$}, to be read from the
  left vertical axes.  The horizontal light orange shaded band
  between~\mbox{$\Mh = 122, 128 \gev$} corresponds to {predictions
    for~$\Mh$ that may be regarded as consistent with experiment.}
  The other lines show the values of~$\Mh$ calculated using
  \FHnew\ (red) and \FHold\ (dashed black), to be read from the right
  vertical axes.  }
\end{figure}

\subsubsection{Focus-Point Strips in the CMSSM\label{sec:focuspoint}}

We now turn to an alternative mechanism in the CMSSM that can yield an
acceptable cold dark matter density even for large values of (some)
input parameters. This is the focus-point region, where the neutralino
LSP acquires a significant Higgsino component that enhances
(co)annihilation rates, thereby bringing the relic density down into
the allowed range.

Examples of focus-point strips are visible in the \mbox{$(m_{1/2},
  m_0)$}~planes shown in Fig.~\ref{fig:focus-point}.  The regions
shaded pink in these plots are where the electroweak symmetry-breaking
conditions cannot be satisfied, and the dark blue strips running along
the boundaries of these regions (now clearly visible) are the
focus-point strips. To make these strips more visible, we used the
range~\mbox{$0.06 < \Omega_\chi\, h^2 < 0.2$}.  As before, the brick
red shaded regions are where the~LSP is charged. In addition to the
stau-LSP~regions in the lower right parts of the planes, we see in the
upper panels for~\mbox{$\tan \beta = 10$}, \mbox{$A_0 = 0$} and the
two signs of~$\mu$ additional brick red strips where the~LSP is a
chargino. In the lower right panel for~\mbox{$\tan \beta = 30$},
\mbox{$A_0 = 0$} and~\mbox{$\mu < 0$} there is a gray shaded region at
large~$m_{1/2}$ where the~RGEs for the Yukawa coupling of the
$b$~quark break down. This region expands as~$\tan \beta$ is increased
when~\mbox{$\mu < 0$}.

\begin{figure}[hbtp!]
\centerline{
\includegraphics[height=8cm]{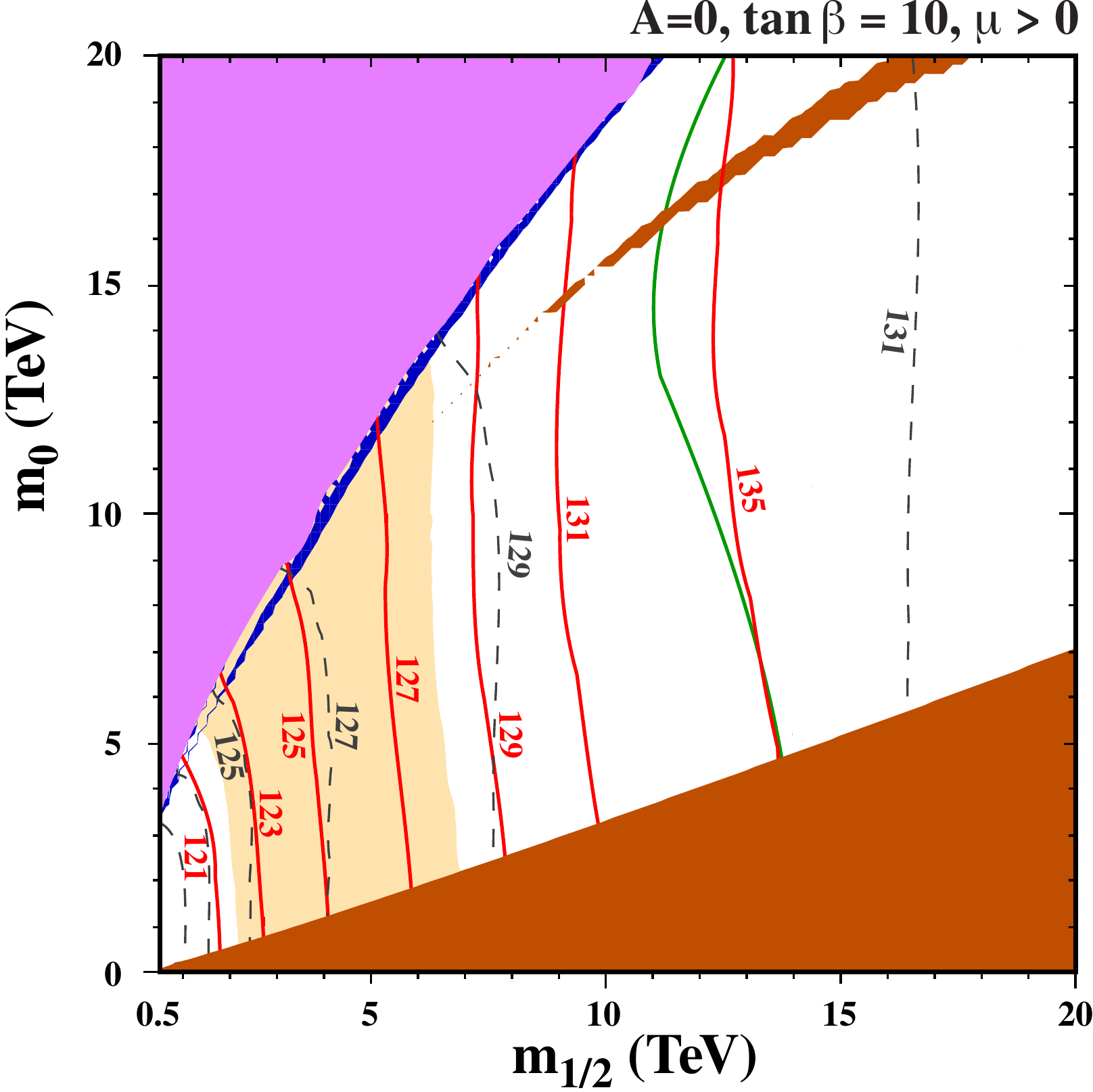}
\includegraphics[height=8cm]{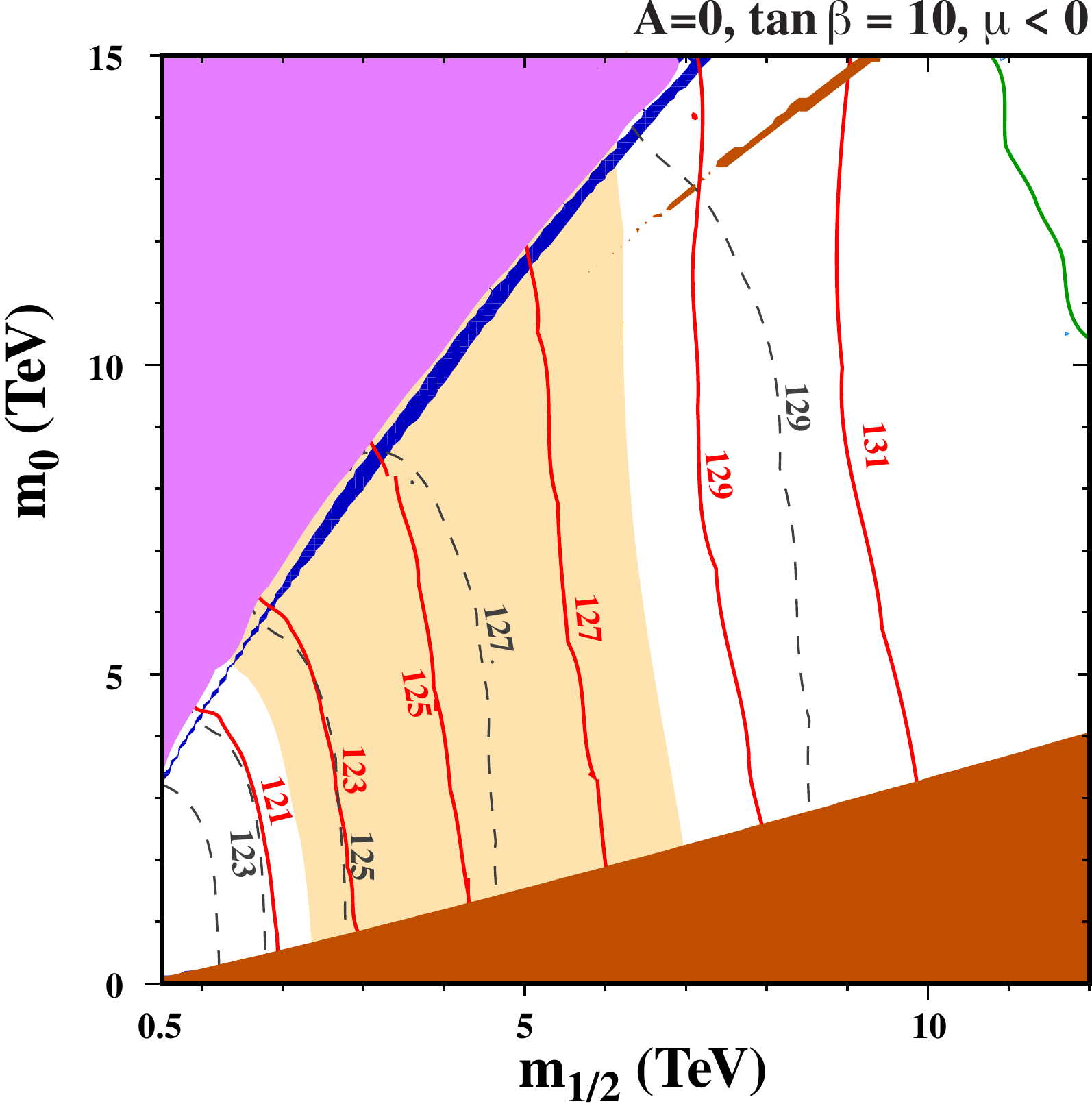}}
\centerline{
\includegraphics[height=8cm]{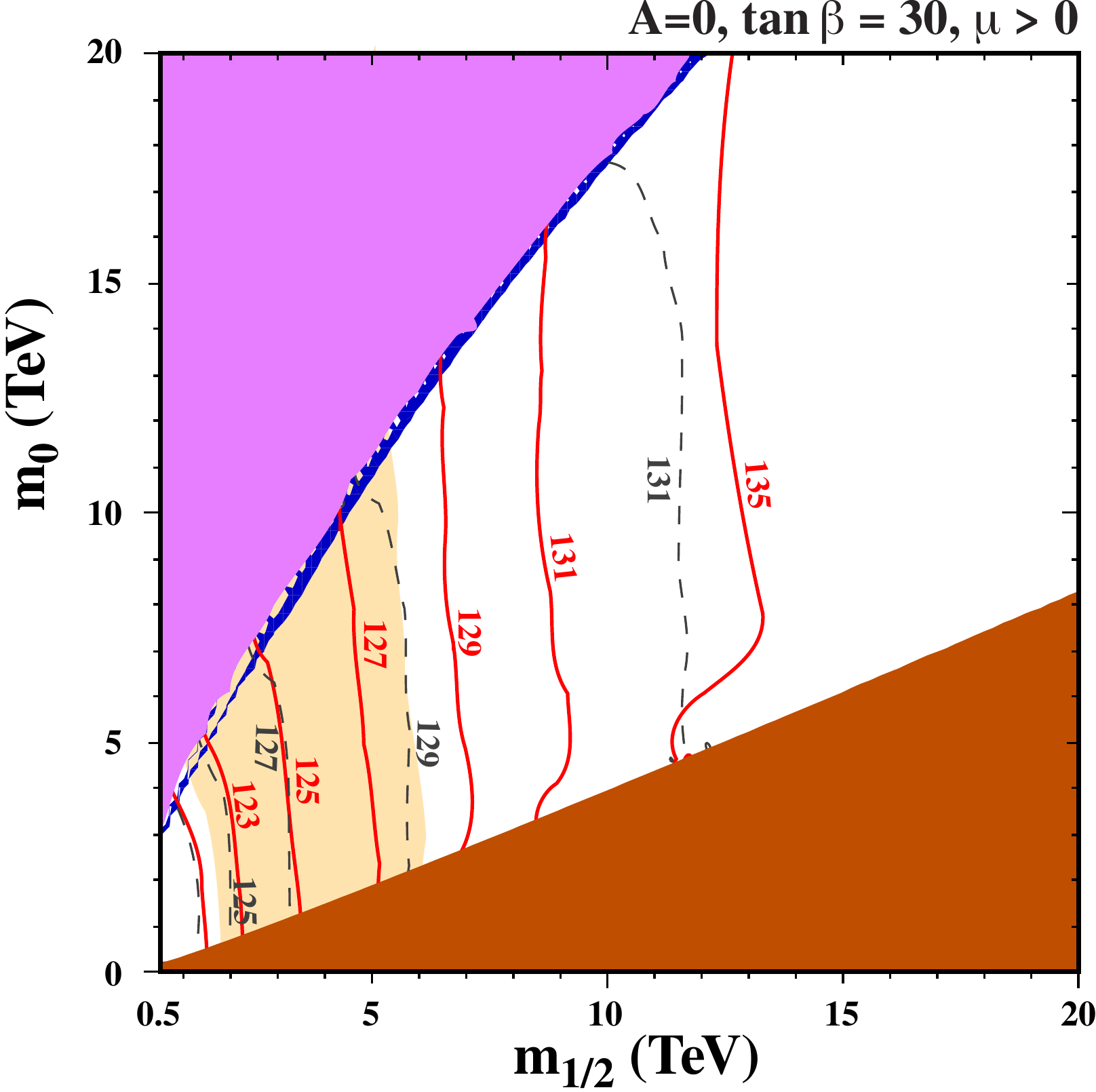}
\includegraphics[height=8cm]{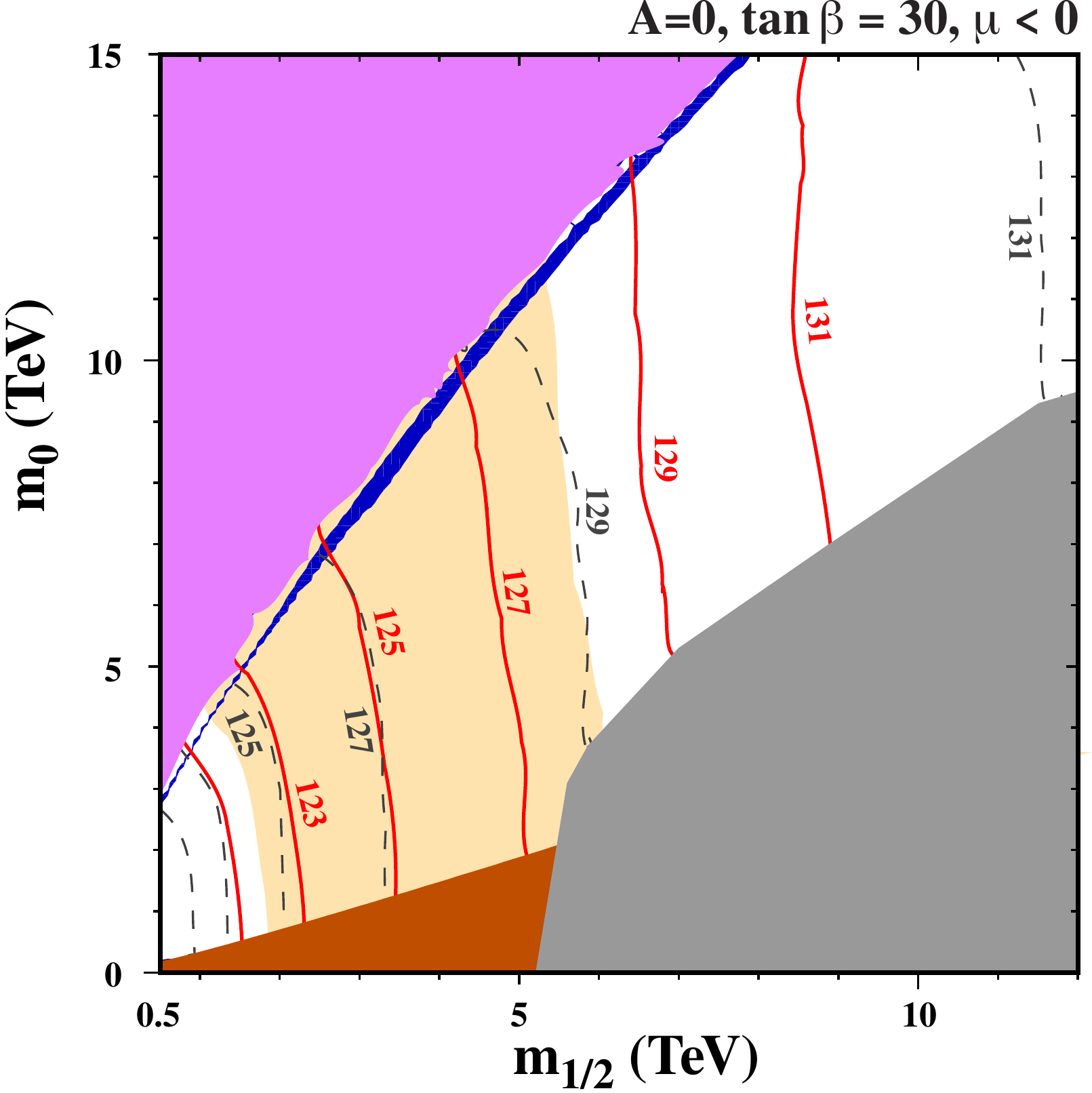}}
\caption{
  \label{fig:focus-point}
  As in Fig.~\protect\ref{fig:m12vsm0tb5}, but for the cases
  \mbox{$\tan \beta = 10$}, \mbox{$A_0 = 0$} and \mbox{$\mu > 0$}
  (upper left panel), \mbox{$\tan \beta = 10$}, \mbox{$A_0 = 0$} and
  \mbox{$\mu < 0$} (upper right panel), \mbox{$\tan \beta = 30$},
  \mbox{$A_0 = 0$} and \mbox{$\mu > 0$} (lower left panel), and
  \mbox{$\tan \beta = 30$}, \mbox{$A_0 = 0$} and \mbox{$\mu < 0$}
  (lower right panel).  The electroweak symmetry-breaking conditions
  cannot be satisfied in the regions shaded pink in these plots.
  Contours of~$\Mh$ calculated using \FHnew\ are shown as red solid
  lines, those using \FHold\ as gray dashed lines.  The light orange
  shaded region corresponds to~\mbox{$\Mh \in [122,128] \gev$} using
  \FHnew.  The blue strips show the region with~\mbox{$0.06 <
    \Omega_\chi\, h^2 < 0.2$}.  The solid green lines show the lower
  limit on the proton lifetime calculated in a minimal supersymmetric
  SU(5)~GUT.  For~\mbox{$\tan \beta = 30$}, these lie beyond the range
  of the plot.  For large~$\tan \beta$ and~\mbox{$\mu < 0$}, the gray
  shading at high~$m_{1/2}$ denotes the lack of convergence of
  the~RGEs due to a divergent $b$-quark Yukawa coupling.  }
\end{figure}

Fig.~\ref{fig:focus-point} displays examples of focus-point strips
that extend to~$m_{1/2}$ and~\mbox{$m_0 \gtrsim 10 \tev$}.  Although
the values of~\mbox{$\mste1 - \mneu1$} can become very large along
this strip, the relic density is instead controlled by the value
of~$\mu$, which tends towards zero as the pink region is
approached. For small~$\mu$, the~LSP becomes Higgsino-like, and the
relic density is determined by Higgsino annihilations and
coannihilations with the second Higgsino and chargino, which are
nearly degenerate in mass with the~LSP. While the extent of the strips
is very large, as one can see in each of the panels, it is limited by
the Higgs mass which differs in the two versions of \FH.  The profiles
of these strips are shown in Fig.~\ref{fig:FPstrips} for~\mbox{$\tan
  \beta = 10$} upper panels) and~\mbox{$\tan \beta = 30$} (lower
panels), for~\mbox{$\mu > 0$} (left panels) and for~\mbox{$\mu < 0$}
(right panels), with~\mbox{$A_0 = 0$} in all cases. There is little
difference between the calculations of~$\Mh$ using \FHnew\ (red lines)
and \FHold\ (black dashed lines) for the different signs of~$\mu$,
with \FHnew\ yielding lower~$\Mh$ for~\mbox{$m_{1/2} \lesssim
  6$}~or~$7 \tev$.  As expected, the calculations generally produce
higher Higgs masses for~\mbox{$\tb = 30$} than for~\mbox{$\tb = 10$},
particularly at small~$m_{1/2}$. In both cases the values of~$\Mh$
obtained with \FHnew\ are compatible with experiment for ranges
of~\mbox{$1.2$~to~$1.5\tev \lesssim m_{1/2} \lesssim 5.4$~to~$6.4
  \tev$}, whereas the larger values of~$\Mh$ obtained with
\FHold\ would have been problematic for~\mbox{$m_{1/2} \gtrsim
  3$}~or~$4 \tev$.

Fig. \ref{fig:focus-point} also shows the minimal SU(5) proton decay
limits (as green contours) for~\mbox{$\tan \beta = 10$}.
For~\mbox{$\tan \beta = 30$}, the contour would lie beyond the scope
of the plot.  As a consequence, the proton decay limit is in conflict
with the upper limit derived from the Higgs mass.  However, we stress
again that this should be viewed as a constraint on the~GUT rather
than a problem for the low-energy supersymmetric model.

\begin{figure}[bt!]
\centerline{ \includegraphics[height=6cm]{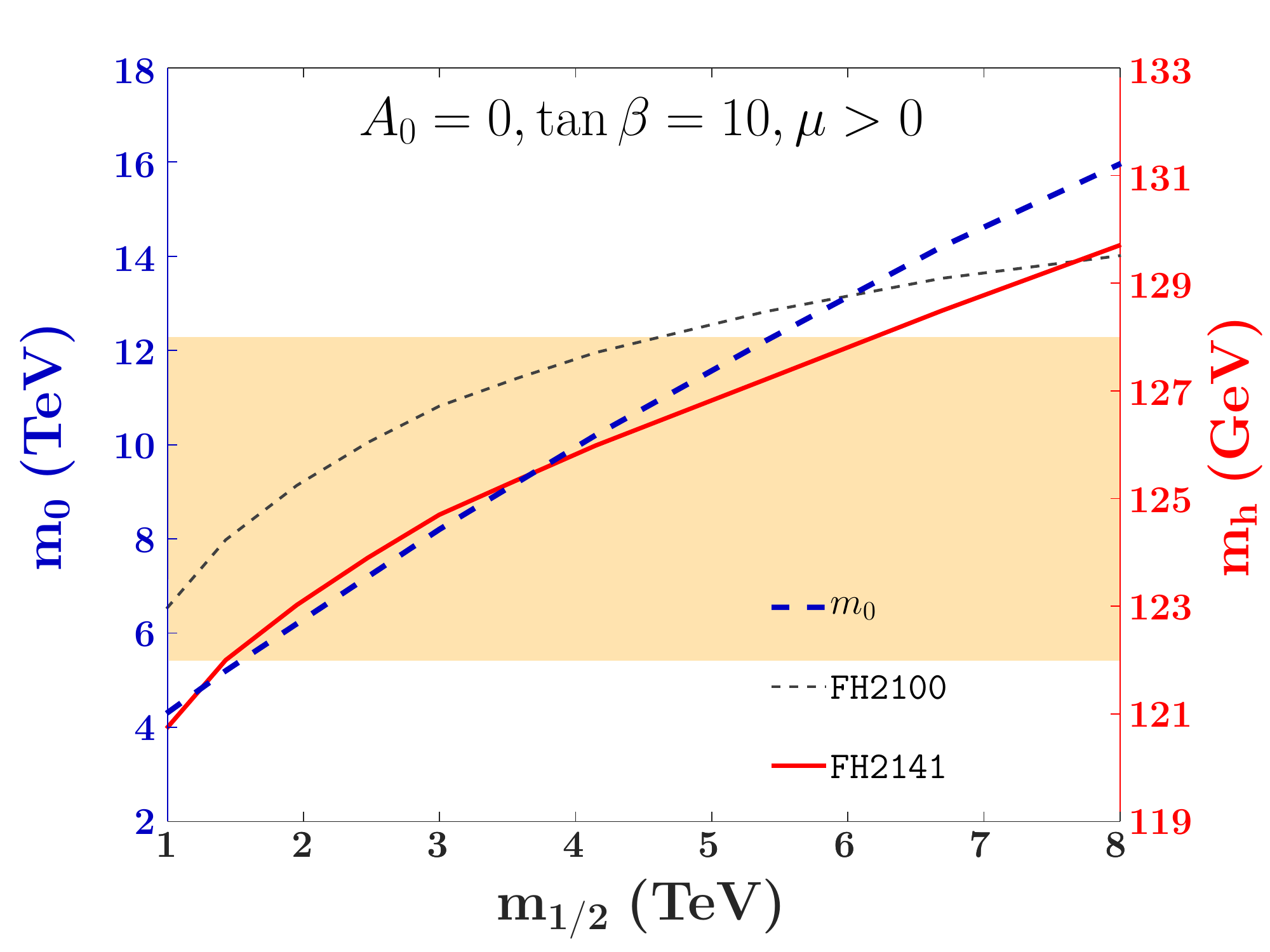}
\includegraphics[height=6cm]{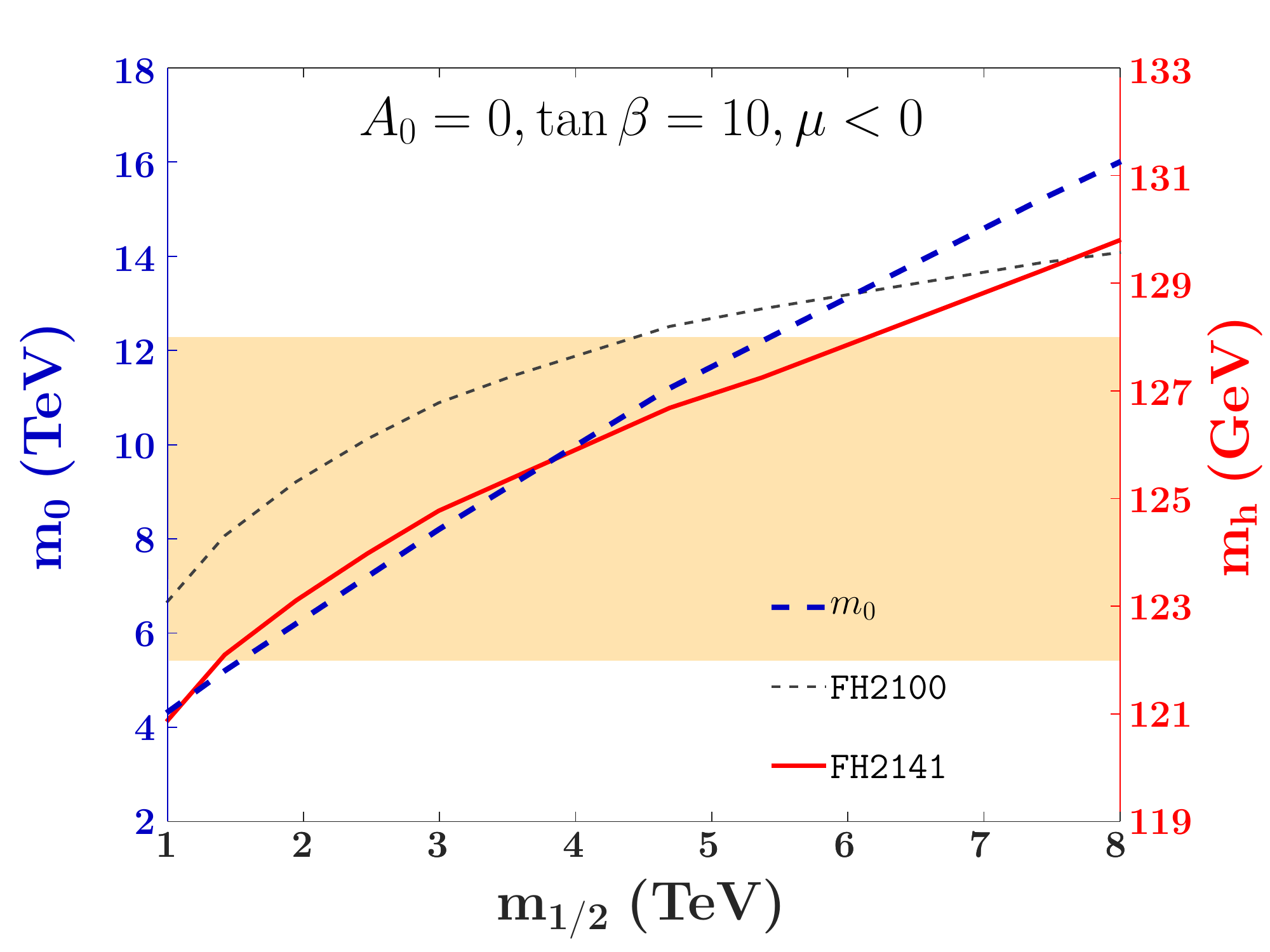}}
\centerline{
\includegraphics[height=6cm]{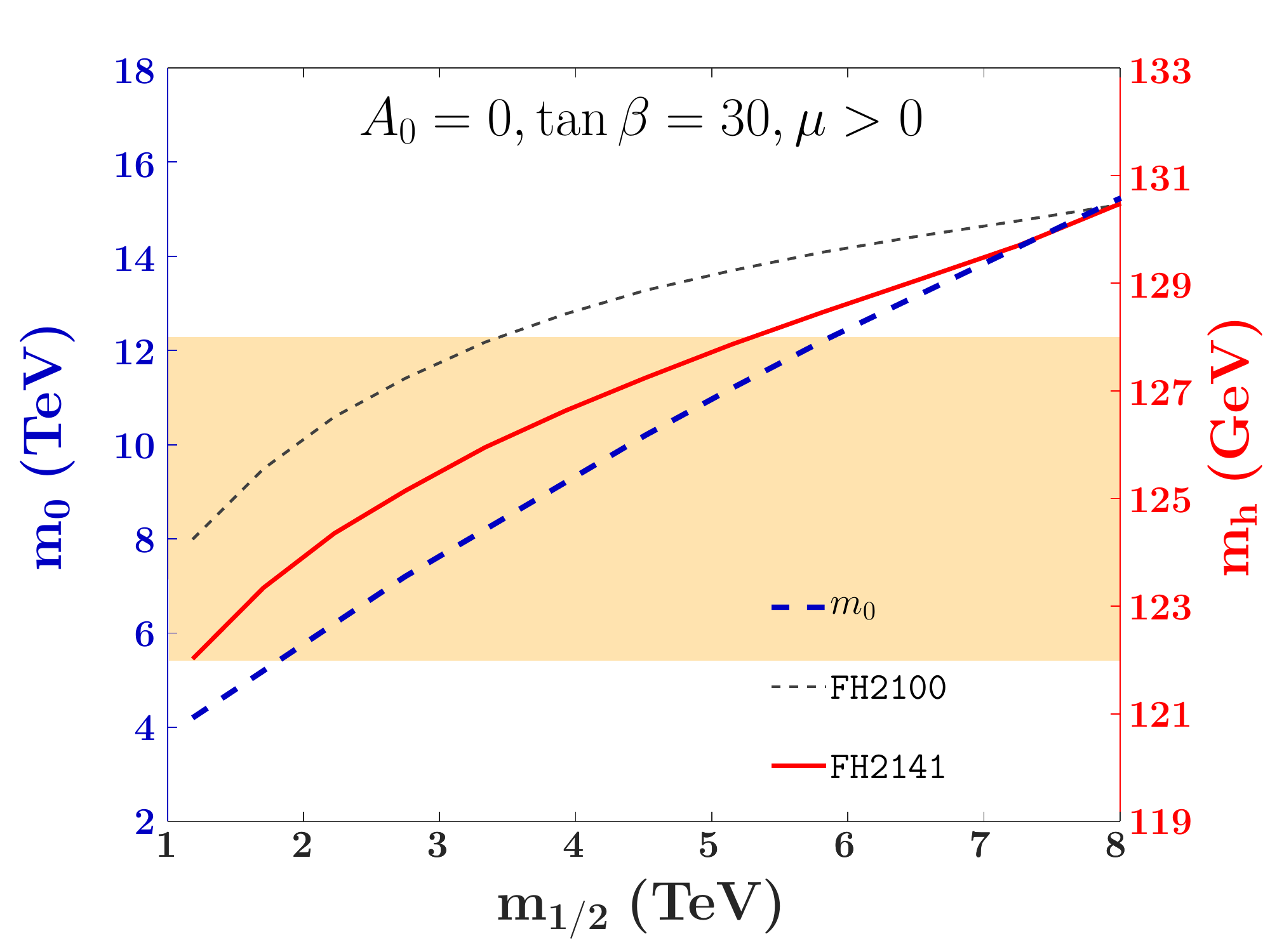}
\includegraphics[height=6cm]{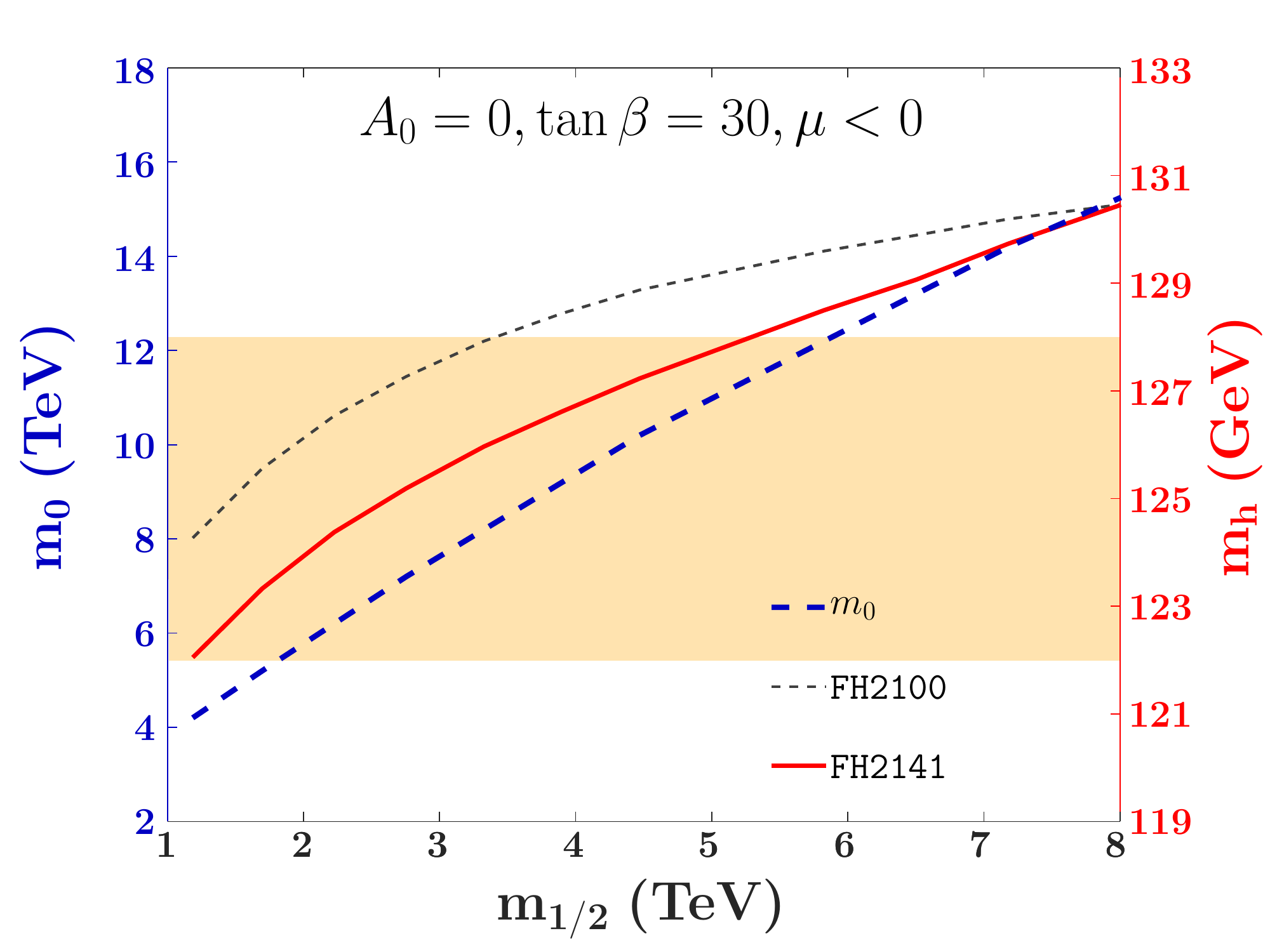}
}
\caption{
  \label{fig:FPstrips}
The profiles of the focus-point strips for \mbox{$A_0 = 0$} and
\mbox{$\mu > 0$} (left panels), \mbox{$\mu < 0$} (right panels) and
\mbox{$\tb = 10$} (upper panels) and \mbox{$\tb = 30$} (lower panels).
The lower horizontal axes show~$m_{1/2}$, the blue dashed curves show
the value of~$m_0$, to be read from the left vertical axes.  The
horizontal light orange shaded band between~\mbox{$\Mh = 122, 128
  \gev$} corresponds to {predictions for~$\Mh$ that may be
  regarded as consistent with experiment.}  The other lines show the
values of~$\Mh$ calculated using \FHnew\ (red) and \FHold\ (dashed
black), to be read from the right vertical axes.}
\end{figure}


\subsection{Sub-GUT Models\label{sec:subGUT}}

We now extend the previous discussion to a `sub-GUT' class of SUSY
models, in which the soft SUSY-breaking parameters are universal at
some input scale~\Min\ below the GUT~scale~\MGUT\ but above the
electroweak scale~\cite{sub-GUT,ELOS,eelnos}.  Models in this class
may arise if the soft SUSY-breaking parameters in the visible sector
are induced by gluino condensation or some dynamical mechanism that
becomes effective below the GUT scale. Examples of sub-GUT models
include those with mirage mediation~\cite{mirage} of soft SUSY
breaking, and certain scenarios for moduli stabilization~\cite{KKLT}.

The reduced RG~running below~\Min, relative to that below~\MGUT\ in
the~CMSSM and related models, leads in general to SUSY~spectra that
are more compressed~\cite{sub-GUT}. These lead, in particular, to
increased possibilities for coannihilation processes. The reduced
RG~running also suggests a stronger lower limit on~$\mneu1$, because
of a smaller hierarchy to the gluino mass, and there are also smaller
hierarchies between the squark and slepton masses.  For a discussion
of the implications for LHC~searches for sparticles in sub-GUT models,
see~\cite{MCsubGUT}.

The five-dimensional parameter space of the sub-GUT~MSSM that we
consider here includes, besides~$\Min$ and~$\tb$, the three soft
supersymmetry-breaking parameters~$m_{1/2}$, $m_0$ and~$A_0$ that are
familiar from the~CMSSM, but which are now assumed to be universal at
the sub-GUT input mass scale~\mbox{$\Min < \MGUT$}.

Fig.~\ref{fig:sub-GUT} illustrates some of the possibilities that
appear in this five-dimensional space. The panels in the top and
middle rows are all for~\mbox{$\tan \beta = 20$}, \mbox{$A_0 = 2.75\,
  m_0$} and~\mbox{$\mu > 0$}, with different choices of~\mbox{$\Min =
  10^7$}~GeV (top left), $10^8$~GeV (top right), $10^9$~GeV (middle
left), and~$10^{10}$~GeV (middle right).  Similar parameter planes
were considered in \citere{eeloz} using {\tt FeynHiggs~2.13.0}.
As~\Min\ increases, we see that a double-lobed brick red region at
low~$m_{1/2}$ and~$m_0$ expands to larger mass values.  The upper left
lobe is a stop-LSP~region, and the lower right lobe is a
stau-coannihilation region.  With some imagination one can anticipate
that for larger~\Min\ the plane would evolve towards the~CMSSM case
shown in the upper left panel of Fig.~\ref{fig:m12vsm0tb20}, which has
the same values of~\mbox{$\tan \beta = 20$} and~\mbox{$A_0 = 2.75\,
  m_0$}, but~\mbox{$\Min \to \MGUT$}.  As in that plane, there are
dark blue stop coannihilation strips that border the upper left lobes
in the top and middle left panels of
Fig.~\ref{fig:sub-GUT}.\footnote{There are in principle also stau
  coannihilation strips bordering the lower right lobes.}  Once again,
to improve the visibility of the relic density strips, we show the
values of~\mbox{$0.06 < \Omega_\chi\,{h^2} < 0.2$} in the blue
shaded region with the exception of the two panels with~\mbox{$A_0 =
  2.75\, m_0$} and~\mbox{$\Min = 10^9 \gev$} where the
range~\mbox{$0.1151 < \Omega_\chi\,{h^2} < 0.1235$} is used.  This
is possible as the neutralino-stop mass difference varies very slowly
with increasing~$m_0$, allowing for a visible coannihilation strip.

\begin{figure}[hbtp!]
\vspace{-1cm}
\centerline{
\includegraphics[height=6cm]{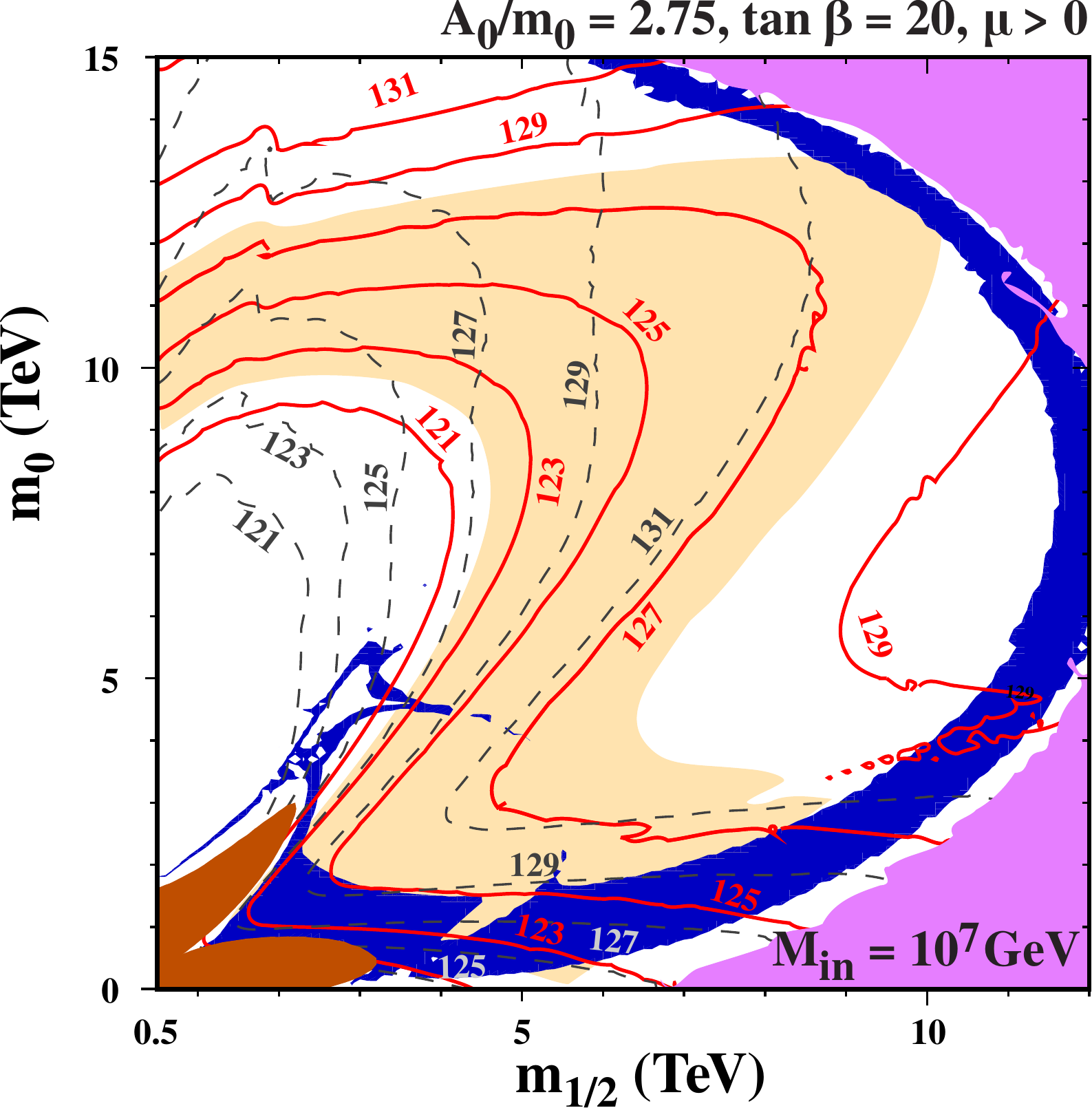}
\includegraphics[height=6cm]{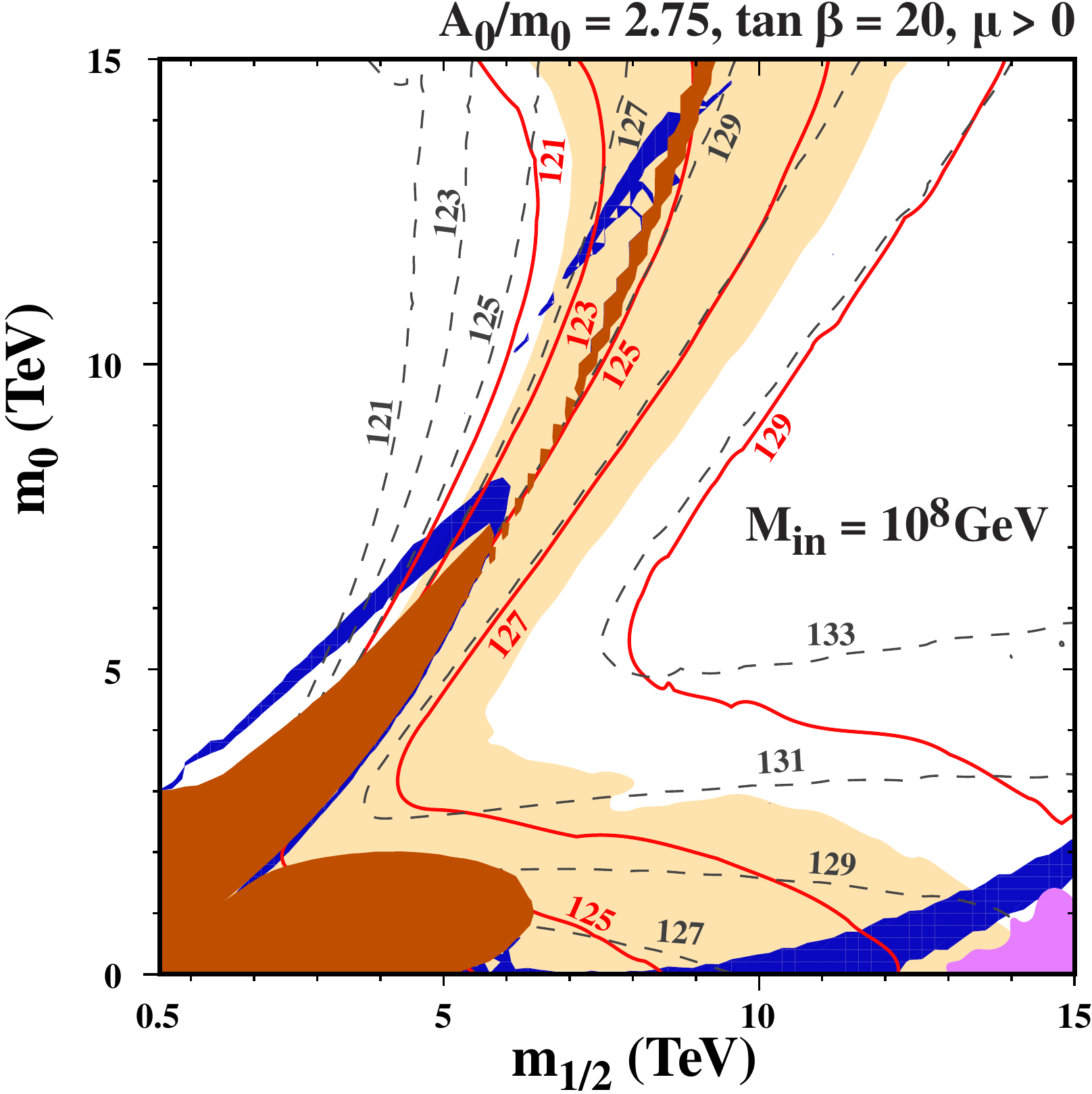}}
\vspace{0.3cm}
\centerline{
\includegraphics[height=6cm]{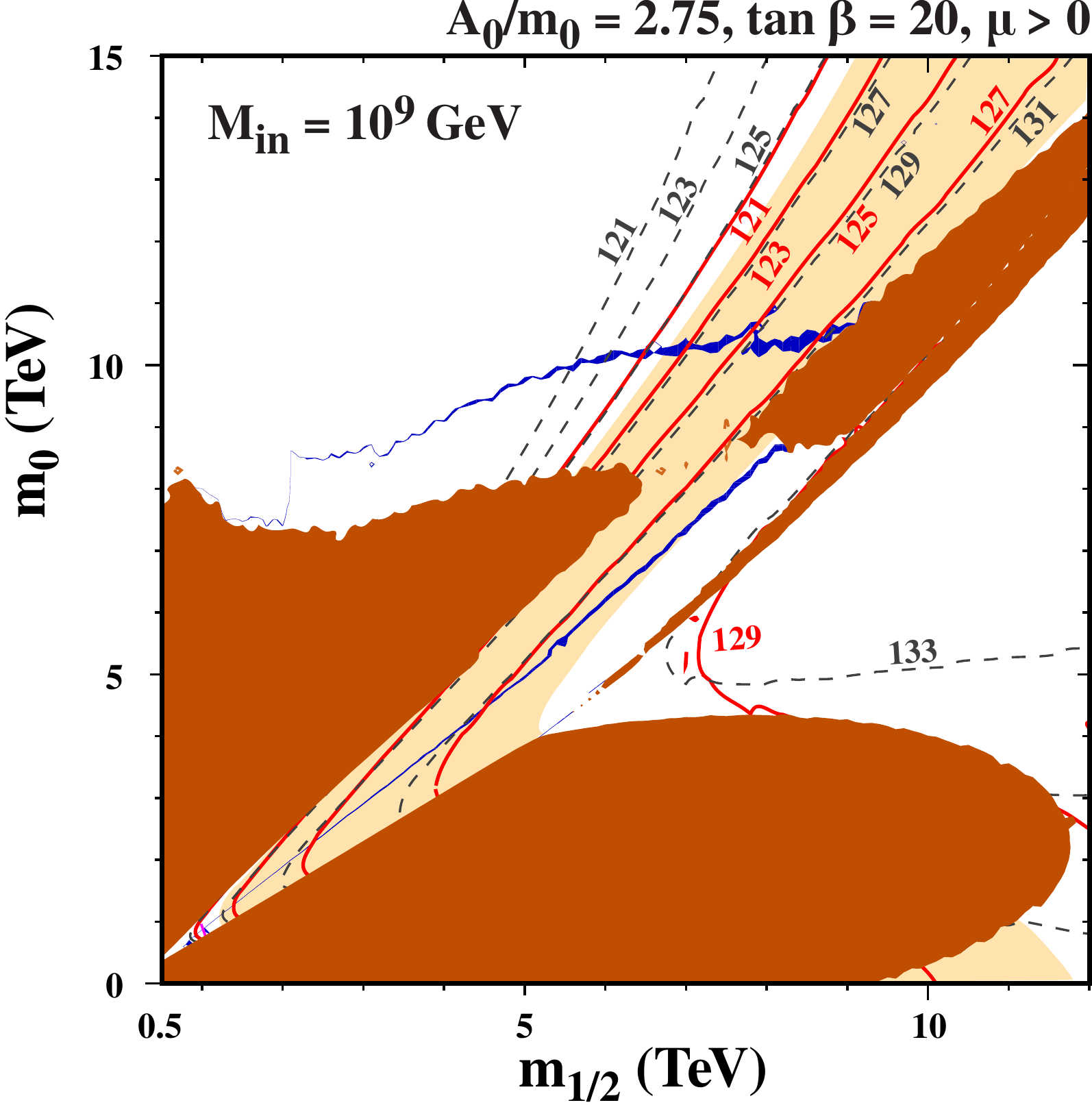}
\includegraphics[height=6cm]{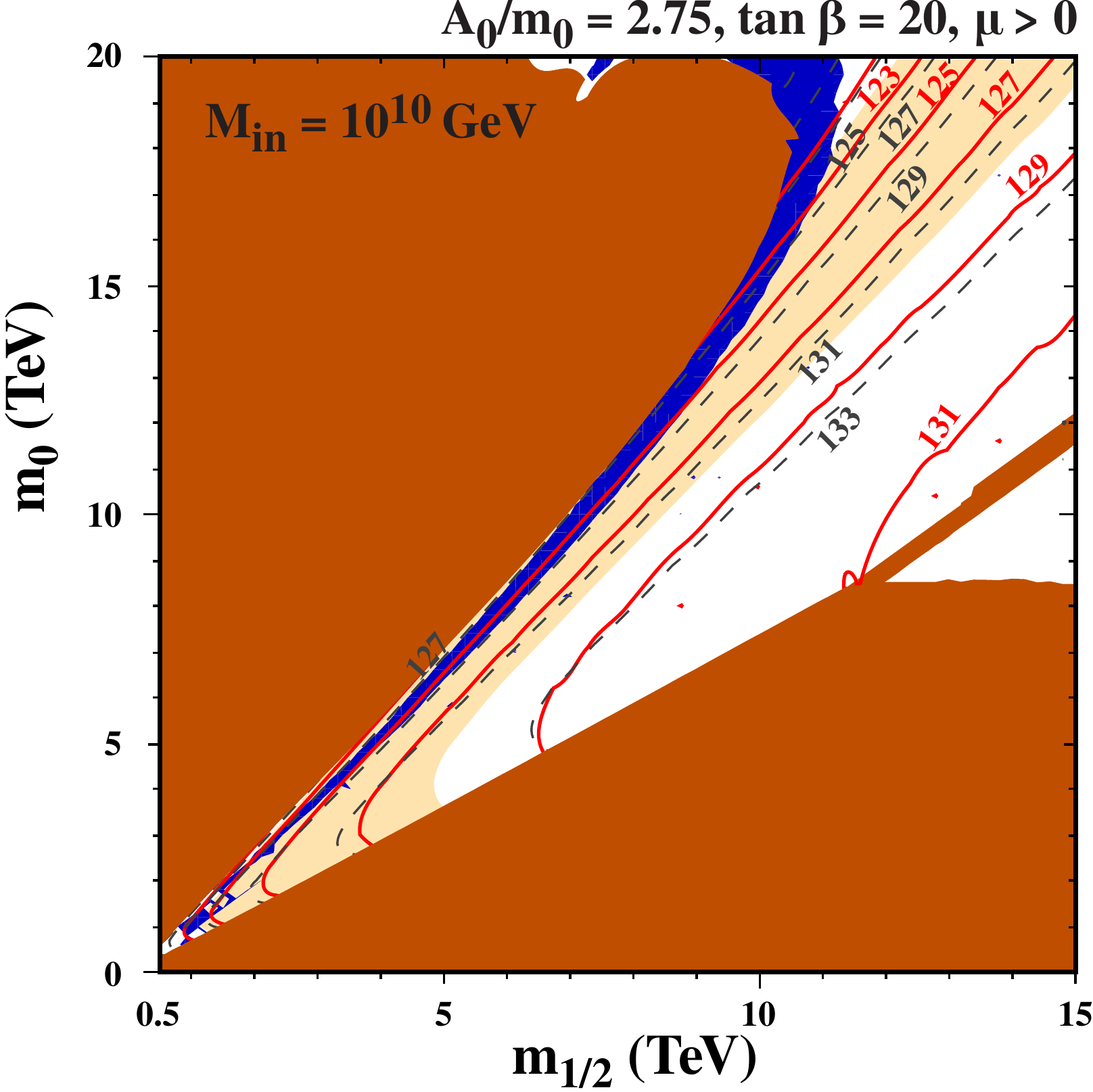}}
\vspace{0.3cm}
\centerline{
\includegraphics[height=6cm]{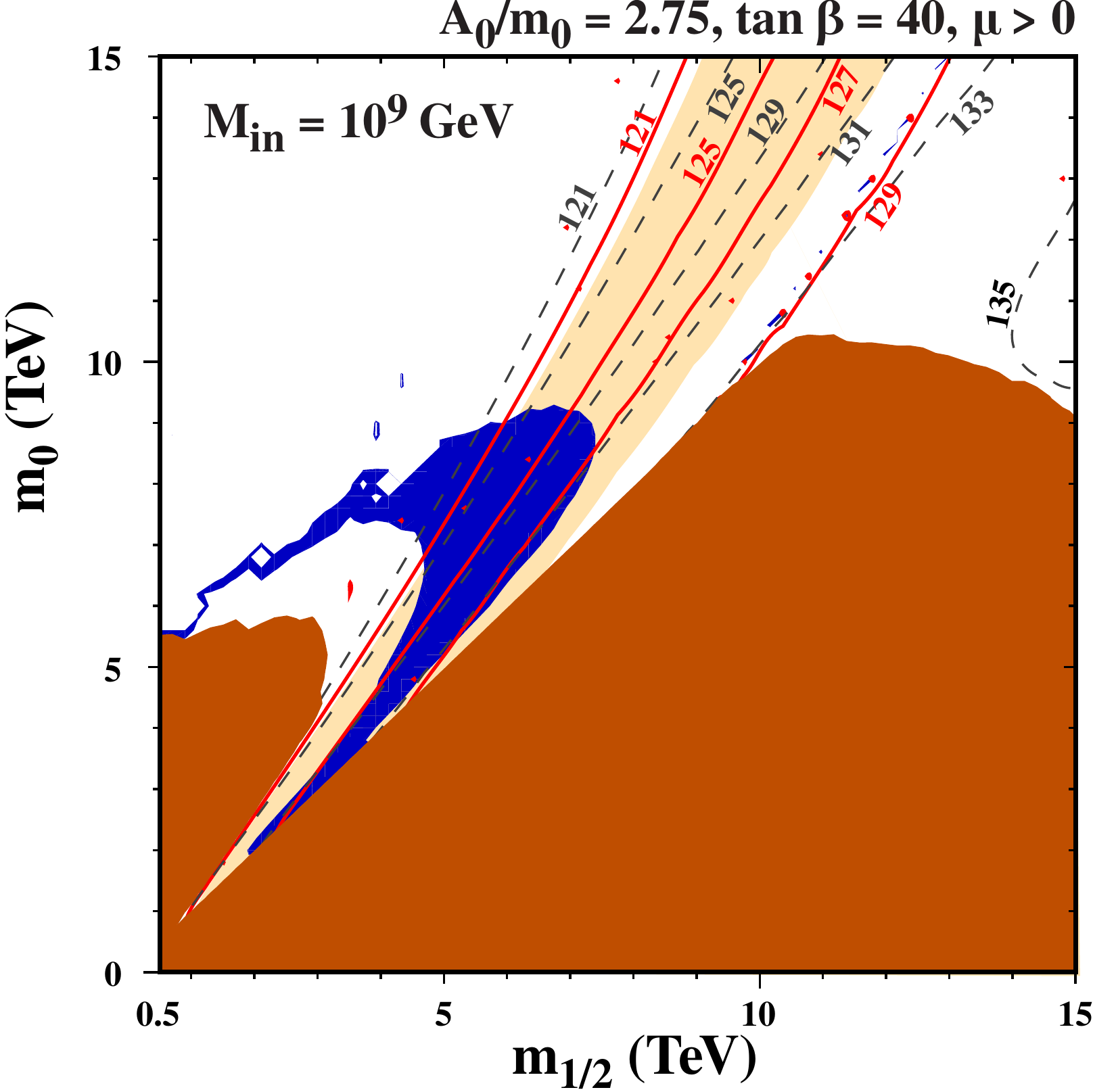}
\includegraphics[height=6cm]{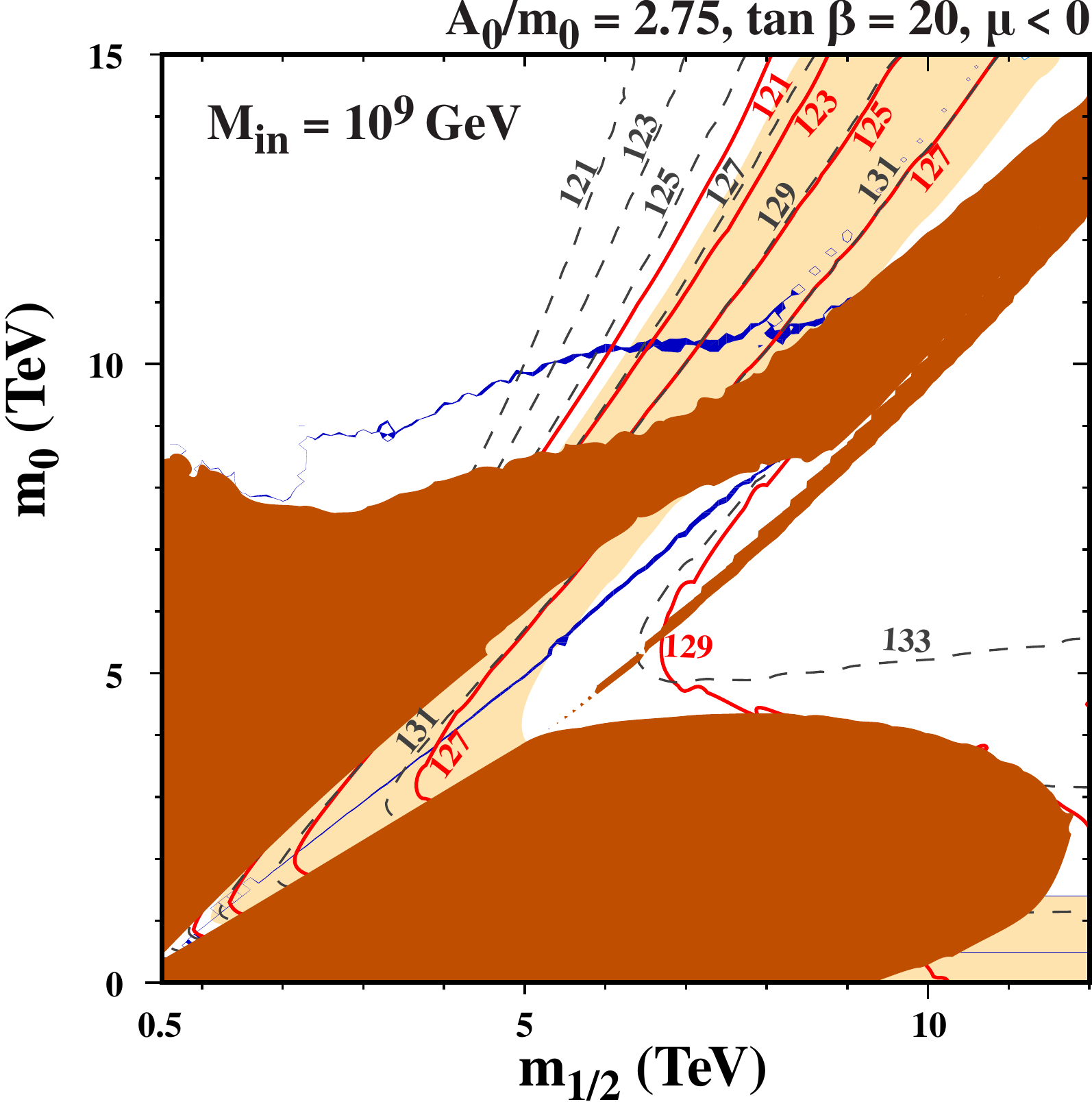}
}
\caption{
  \label{fig:sub-GUT}
  As in Fig.~\protect\ref{fig:m12vsm0tb5}, but for sub-GUT scenarios
  with \mbox{$\tan \beta = 20$}, \mbox{$A_0 = 2.75\, m_0$},
  \mbox{$\Min = 10^7$}~GeV (top left), \mbox{$\Min = 10^8$}~GeV (top
  right), \mbox{$\Min = 10^9$}~GeV (middle left), and \mbox{$\Min =
    10^{10}$}~GeV (middle right), also for \mbox{$\tan \beta = 40$},
  \mbox{$A_0 = 2.75\, m_0$}, \mbox{$\Min = 10^9$}~GeV (bottom left),
  all with \mbox{$\mu > 0$}, and \mbox{$\tan \beta = 20$}, \mbox{$A_0
    = 2.75\, m_0$}, \mbox{$\Min = 10^9$}~GeV, \mbox{$\mu < 0$} (bottom
  right).  The electroweak symmetry-breaking conditions cannot be
  satisfied in the regions shaded pink in these plots.  Contours
  of~$\Mh$ calculated using \FHnew\ are shown as red solid lines,
  those using \FHold\ as gray dashed lines.  The light orange shaded
  region corresponds to~\mbox{$\Mh \in [122,128] \gev$} using \FHnew.
  The blue strips show the region with~\mbox{$0.06 < \Omega_\chi\, h^2
    < 0.2$} except when \mbox{$A_0 = 2.75\, m_0$} and \mbox{$\Min =
    10^9 \gev$} where~\mbox{$0.1151 < \Omega_\chi\, h^2 < 0.1235$} is
  used.  }
\end{figure}

Some other features are worth noting. In the top left panel of
Fig.~\ref{fig:sub-GUT}, for~\mbox{$\Min = 10^7$}~GeV, there are a pair
of pink regions at large~$m_{1/2}$ where the electroweak vacuum
conditions cannot be satisfied, which shrinks away at
larger~\Min. Bordering these pink regions there is a crescent-shaped
focus-point band.  We also note in the top left panel for~\mbox{$\Min
  = 10^7$}~GeV an irregular blue ring-shaped region extending above
the stop strip, and in the top right panel for~\mbox{$\Min =
  10^8$}~GeV there is a blue strip that crosses a brick red chargino
LSP strip when~\mbox{$m_0 \sim 14 \tev$}.  As discussed in
\citere{eeloz}, the masses of the three lightest neutralinos are quite
similar in these regions of parameter space, and multiple
coannihilations interplay, enhanced through the heavy Higgs funnel.
The chargino LSP region expands when~\mbox{$\Min = 10^9 \gev$} (middle
left) and merges with the stau-LSP~region when~\mbox{$\Min = 10^{10}
  \gev$} (middle right).

Contours of~$\Mh$ with values determined by \FHnew\ are shown by red
solid curves and determined by \FHold\ with dashed gray curves.  In
much of the parameter space, the \FHnew\ values are lower by
about~$4$~GeV than the values produced by \FHold.  This difference
shifts the viable regions of the parameter space, as is seen more
clearly in Fig.~\ref{fig:sub-GUTprofiles}, which shows the profiles
along the dark matter strips in these sub-GUT scenarios.  In the top
left panel for~\mbox{$\tan \beta = 20$}, \mbox{$A_0 = 2.75\, m_0$},
\mbox{$\Min = 10^7$}~GeV and~\mbox{$\mu > 0$} we distinguish two
groupings of lines, one extending up to~\mbox{$m_{1/2} \sim 3$}~TeV,
and the other from~\mbox{$m_{1/2} \sim 3$}~TeV to~\mbox{$m_{1/2} \sim
  12$}~TeV corresponding, respectively, to the near-vertical
`peninsula' at low~$m_{1/2}$ in Fig.~\ref{fig:sub-GUT} that extends up
to~\mbox{$m_{0} \sim 5$}~TeV and to the arc that lies close to the
boundary of the region where electroweak symmetry breaking is
possible.
In the low-$m_{1/2}$~grouping, corresponding to the dark matter strip
lying above the stop-LSP~region, the blue dashed lines show
that~\mbox{$\mneu1 \lesssim 2.7 \tev$} along these strips, and the red
lines show that \FHnew\ generally yields values of~$\Mh$ that are
smaller than the experimental value, whereas the dashed black lines
show that \FHold\ would have yielded acceptable values of~$\Mh$ on the
lower-$m_0$~side of the `peninsula' and part of the higher-$m_0$~side.
In the high-$m_{1/2}$~grouping, the blue dashed lines show
that~\mbox{$\mneu1 \sim 1 \tev$}, which is characteristic of Higgsino
dark matter. The red solid line indicates that \FHnew\ yields
acceptable values of~$\Mh$ along most of the lower-$m_0$~part of the
arc up to~\mbox{$m_{1/2} \gtrsim 10 \tev$}, whereas it yields values
of~$\Mh$ that are too high along the upper part of the arc. In
contrast, \FHold\ would have yielded acceptable values of~$\Mh$ only
for~\mbox{$m_{1/2} \lesssim 7 \tev$} along the lower arc.

\begin{figure}[hbtp!]
\vspace{-0.5cm}
\centerline{
\includegraphics[height=6cm]{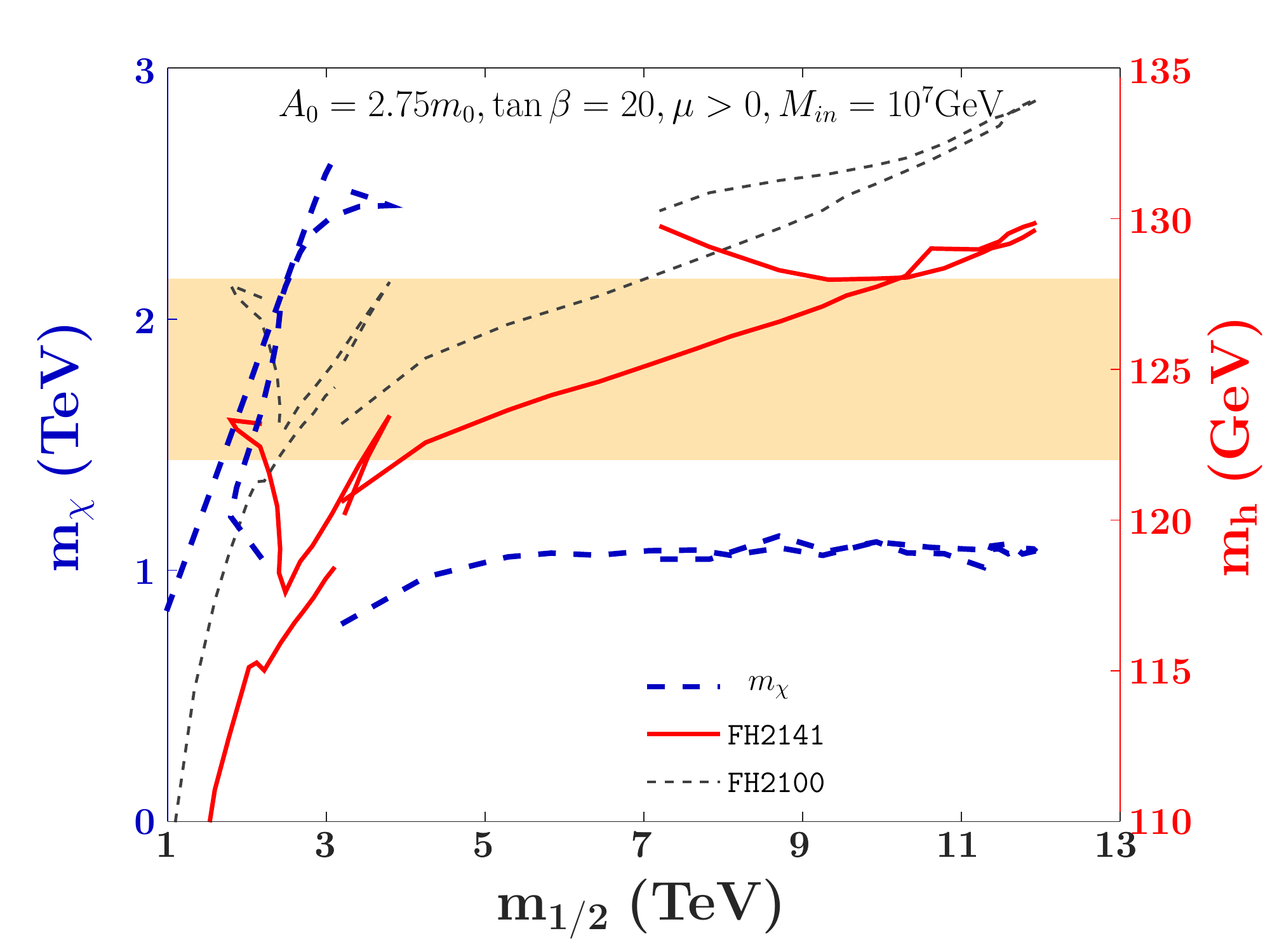}
\includegraphics[height=6cm]{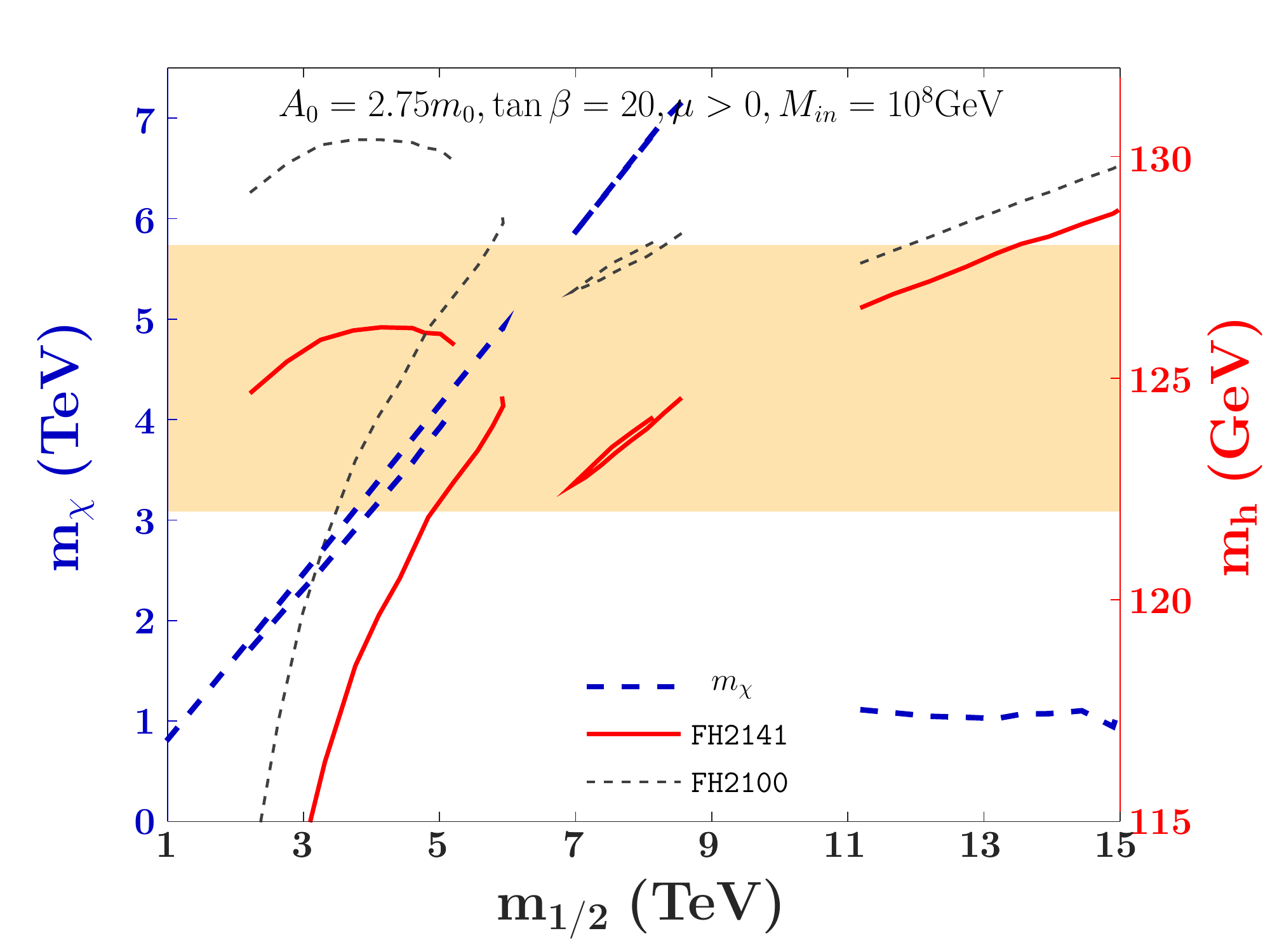}}
\centerline{
\includegraphics[height=6cm]{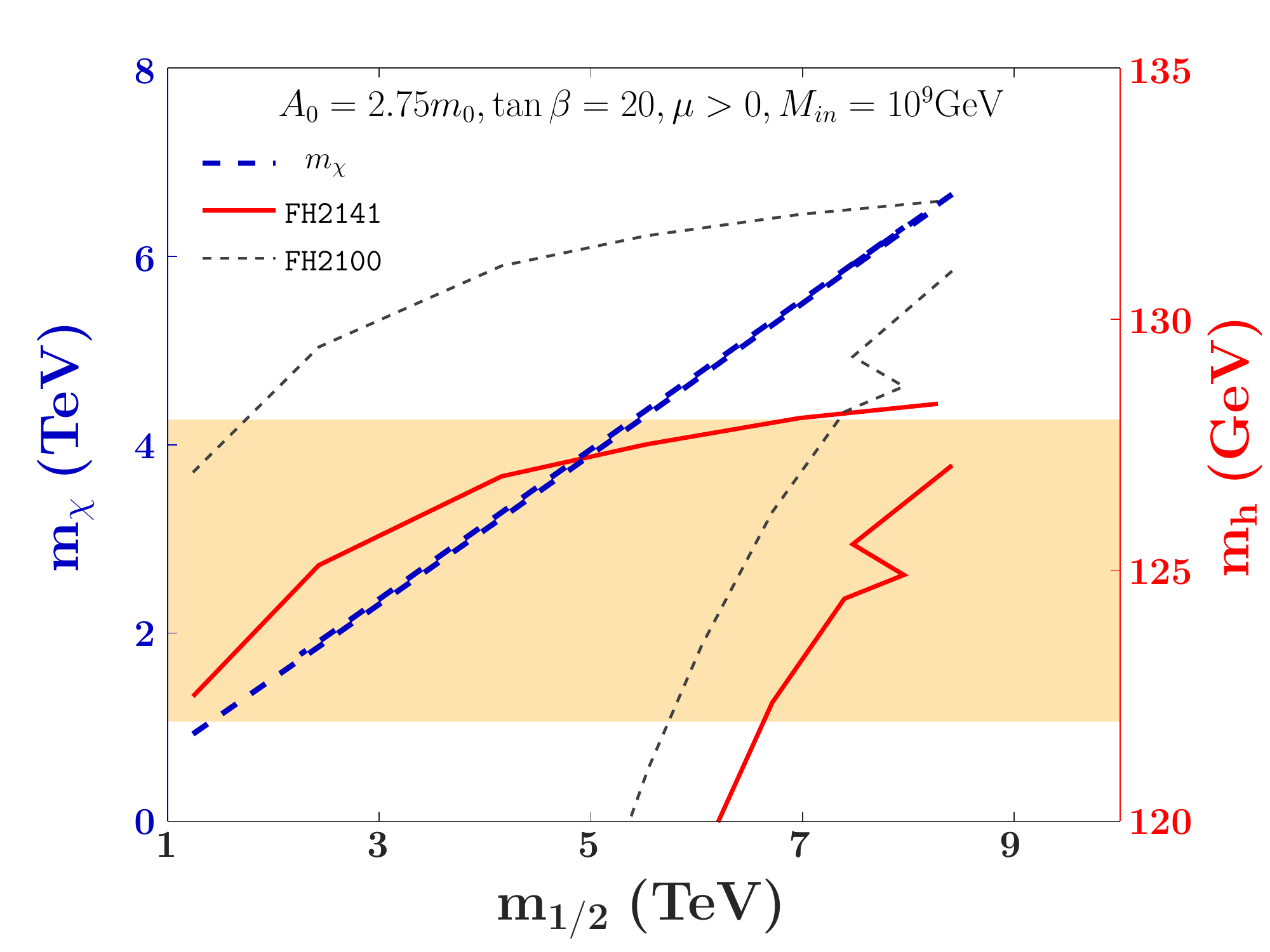}
\includegraphics[height=6cm]{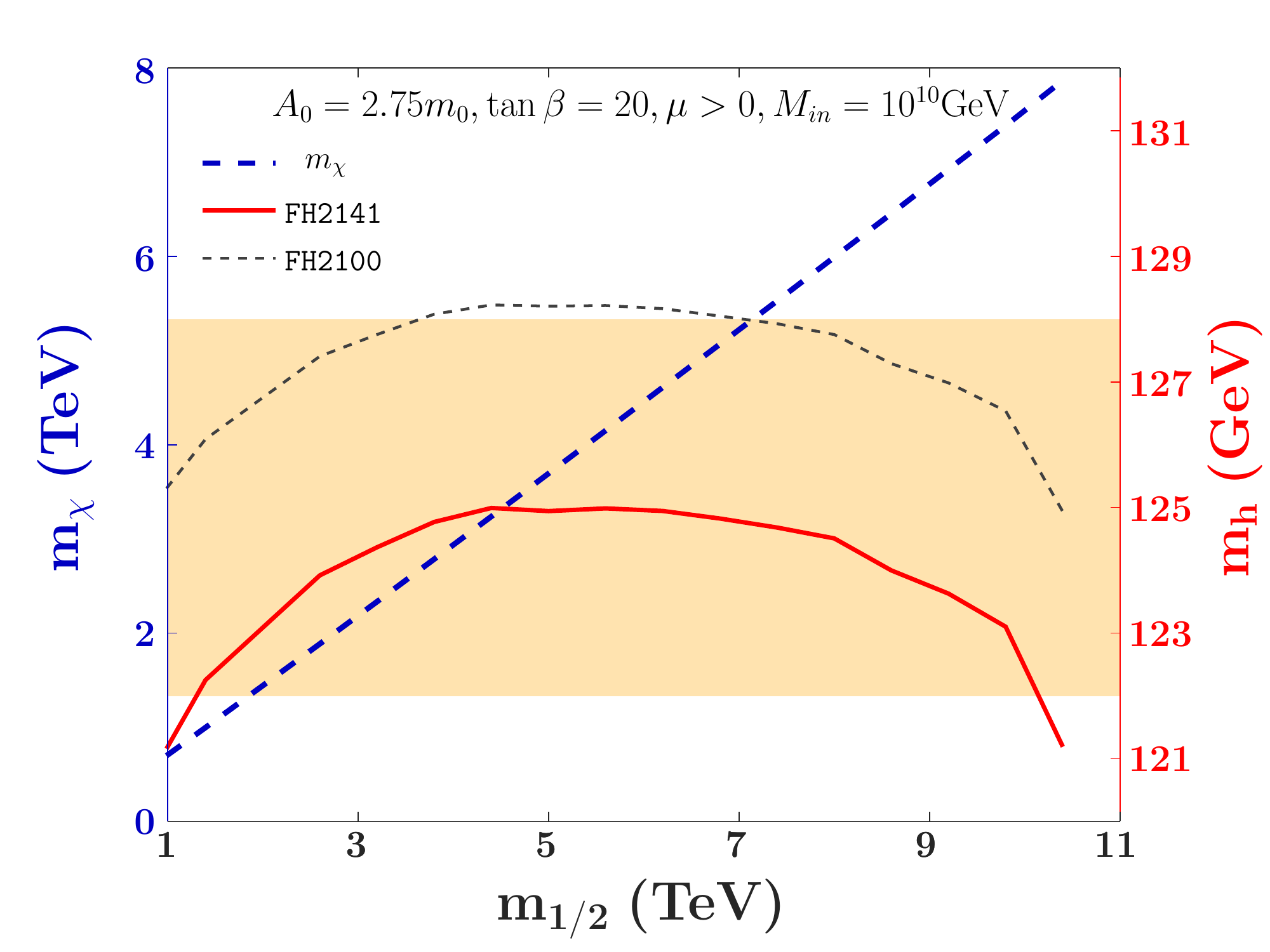}}
\centerline{
\includegraphics[height=6cm]{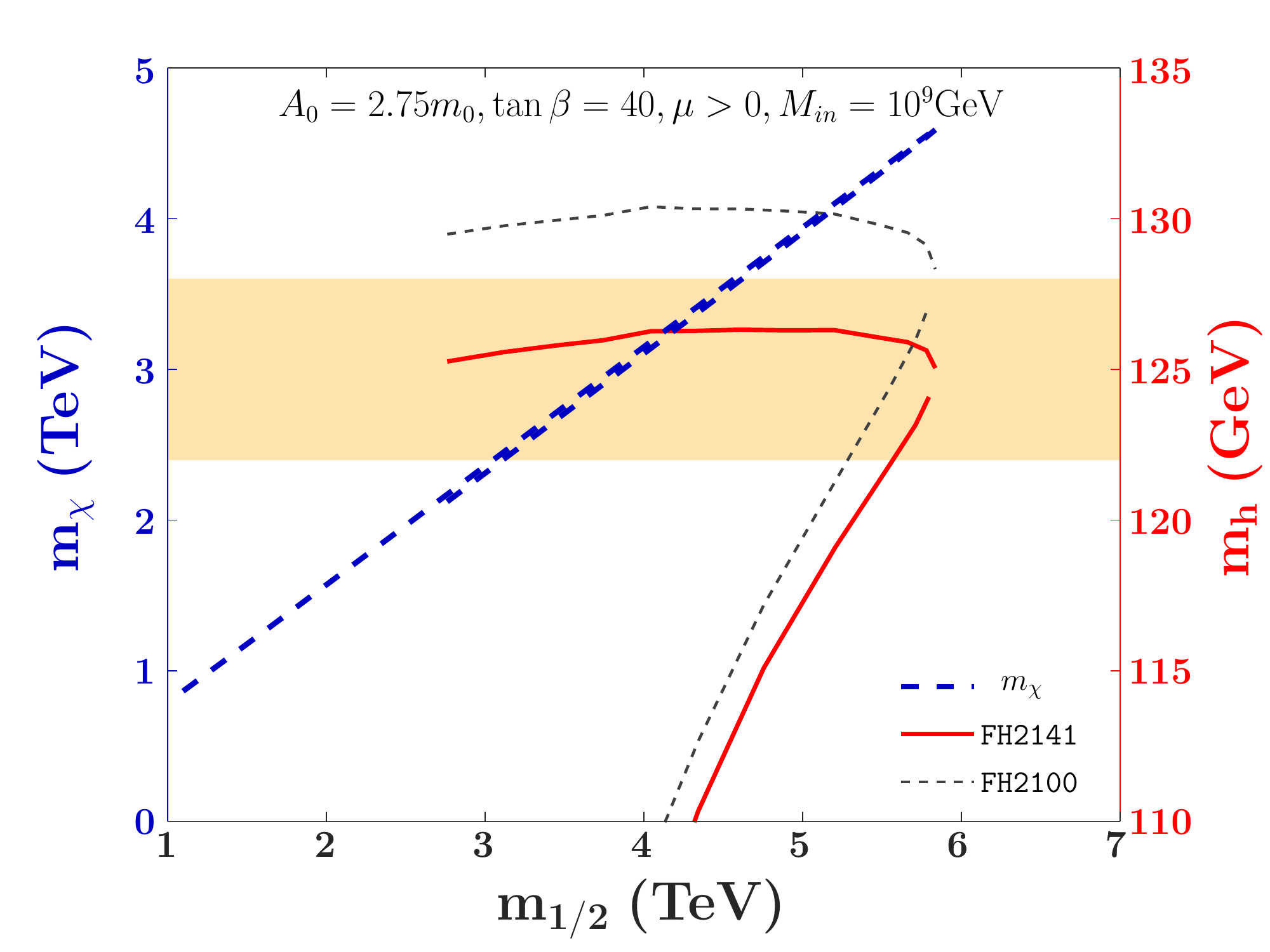}
\includegraphics[height=6cm]{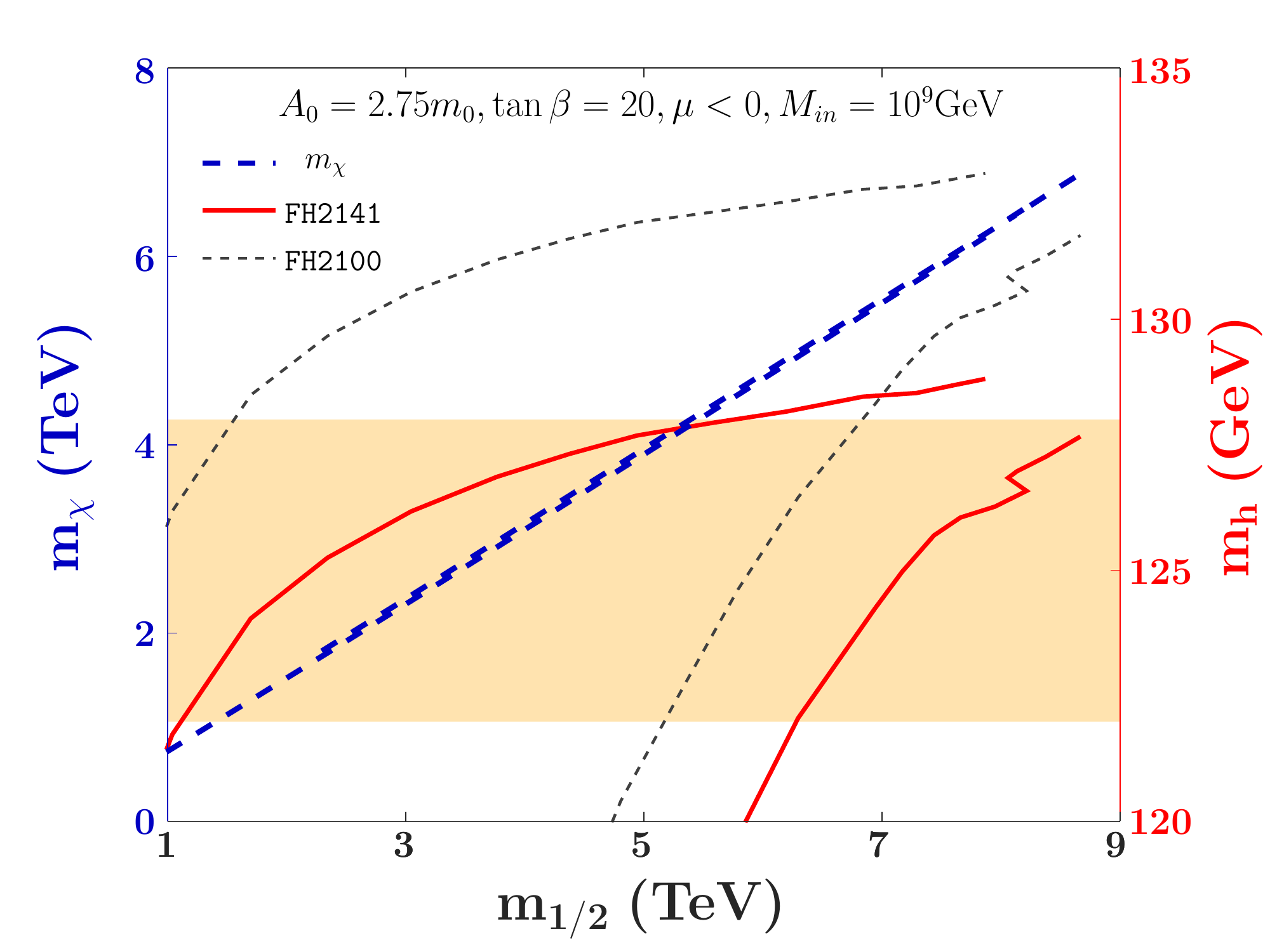}}
\caption{
  \label{fig:sub-GUTprofiles}
  The profiles of the sub-GUT dark matter strips for \mbox{$A_0 =
    2.75\, m_0$} and, going from top left to bottom right,
  \mbox{$\{\Min (\mathrm{GeV}), \tb, {\rm sgn}\mu\} =$} \mbox{$\{10^7,
    20, +\}$}, \mbox{$\{10^8, 20, +\}$}, \mbox{$\{10^9, 20, +\}$},
  \mbox{$\{10^{10}, 20, +\}$}, \mbox{$\{10^9, 40, +\}$},
  \mbox{$\{10^9, 20, -\}$}.
The lower horizontal axes show~$m_{1/2}$,
the blue dashed curves show~$\mneu1$ (left vertical axes).
The ``allowed'' range for~$\Mh$ is indicated by the horizontal light
orange shaded band for~\mbox{$\Mh \in [122, 128] \gev$}.  Calculated
values for~$\Mh$: \FHnew\ (red) and \FHold\ (dashed black), to be read
from the right vertical axes.  }
\end{figure}

As one can see by comparing the results in the upper four panels of
Fig.~\ref{fig:sub-GUTprofiles}, the strips and values of~$\Mh$ are
very sensitive to~$\Min$. For~\mbox{$\Min = 10^8 \gev$}, the high,
middle and low~$m_{1/2}$~regions are clearly separate.  Comparing with
Fig.~\ref{fig:sub-GUT}, we can associate the curves at
low~\mbox{$m_{1/2} \lesssim 6$}~TeV with the relic density strip
corresponding to stop coannihilation. The higher flatter curves and
the lower steeper curves represent the strips below and above the
stop-LSP~region respectively. The change from \FHold\ to
\FHnew\ lowers~$\Mh$ by a few~$\gev$ and brings the higher strip into
good agreement with experiment, while pushing the lower strip to lower
values of~$\Mh$.  We see that~\mbox{$\mneu1 \lesssim 5 \tev$} along
the upper strips, but~\mbox{$ 2 \tev \lesssim \mneu1 \lesssim 4 \tev$}
along the lower strip. For the heavy-Higgs-funnel region
at~\mbox{$m_{1/2}\sim 7$--$9\tev$}, \mbox{$\mneu1 \sim 6$--$7
  \tev$}. While the results of both versions of \FH\ are consistent
with the experimental value in this region, \FHnew\ decreases~$\Mh$
by~$3.5\gev$ compared to \FHold.  The Higgsino strip at
high~\mbox{$m_{1/2}\gtrsim 11\tev$} is only slightly affected by the
change, but \FHnew\ (unlike \FHold) yields acceptable~$\Mh$ along much
of the high-$m_{1/2}$~strip.

At~\mbox{$\Min = 10^9 \gev$}, there is again a substantial change
in~$\Mh$. The upper, flatter profiles correspond to the strip that
threads between the stop- and stau-LSP~regions and, in contrast with
the result from \FHold, now lies at an acceptable value of~$\Mh$ for a
wide range in~$m_{1/2}$. In contrast, the steeper profiles correspond
to the nearly horizontal stop-coannihilation strip in
Fig.~\ref{fig:sub-GUT}, and are now only viable at~\mbox{$m_{1/2}
  \gtrsim 6.5 \tev$}. We see that~\mbox{$\mneu1 \lesssim 6.6 \tev$}
along both the dark matter strips.  At still larger~\mbox{$\Min =
  10^{10} \gev$}, the lower strip in the previous plot has now morphed
into a stop-coannihilation strip reminiscent of those in the~CMSSM,
which runs nearly parallel with the Higgs-mass contours.  In this case
we find that~\mbox{$\mneu1 \lesssim 8 \tev$}.  Both \FHnew\ and
\FHold\ give acceptable values of~$\Mh$ along this strip but, for most
of this strip,~$\Mh$ is significantly lower and greatly improved in
\FHnew.

The two bottom panels of Fig.~\ref{fig:sub-GUT} illustrate other
features of the sub-GUT parameter space.  The bottom left panel
has~\mbox{$A_0 = 2.75\, m_0$}, \mbox{$\Min = 10^9$}~GeV and~\mbox{$\mu
  > 0$} as before, but~\mbox{$\tan \beta = 40$}.  Comparing with the
middle left panel for~\mbox{$\tan \beta = 20$}, we see that the
stop-LSP~lobe has contracted whereas the stau-LSP~lobe has expanded,
the stop-coannihilation band has broadened, and the
chargino-LSP~region has disappeared.  The Higgs-mass profiles in
Fig.~\ref{fig:sub-GUTprofiles} in this case correspond to the two
sides of the stop-coannihilation region in Fig.~\ref{fig:sub-GUT}, the
upper, flatter profiles corresponding to the lower strip running
parallel to the Higgs-mass contours and the steeper profiles to the
upper stop-coannihilation strip. While~\Mh\ is not very sensitive to
the version of \FH\ for the former strip, \FHnew\ improves the latter
by lowering~$\Mh$ by~\mbox{$3$--$4$}~GeV.  Finally, the bottom right
panel has~\mbox{$\tb = 20$}, \mbox{$A_0 = 2.75\, m_0$}, \mbox{$\Min =
  10^9$}~GeV and~\mbox{$\mu < 0$}. It is relatively similar to the
middle left panel, which has the opposite sign of~$\mu$ but identical
values of the other parameters.  The main difference is the appearance
of a `causeway' between the chargino-LSP~`island' and the
stop-LSP~lobe.  We see that the results for the Higgs-mass profiles
are also very similar, indicating that the sign of~$\mu$ is less
important than the values of the other sub-GUT parameters.

As seen in Fig.~\ref{fig:sub-GUTprofiles}, in general \FHnew\ yields
lower values of~$\Mh$ compared to \FHold\ along both the upper and
lower sub-GUT~strips we have studied. In the cases of the
lower-$m_0$~strips (solid lines) this reduction improves consistency
with the experimental value of~$\Mh$ over a wider range of~$m_{1/2}$.
The picture is more mixed for the higher-$m_0$~strips, where the
preferred ranges of~$m_{1/2}$ change, but are not necessarily more
extensive when \FHnew\ is used.


\subsection{Minimal AMSB Models\label{sec:amsb}}

As a contrast to the previous~CMSSM and sub-GUT models, now we analyze
the minimal scenario for anomaly-mediated SUSY breaking
(the~mAMSB)~\cite{anom,mAMSB}.  This has a very different spectrum,
and a different composition of the~LSP, giving sensitivity to
different aspects of the calculation of~$\Mh$. The~mAMSB has
{three} relevant continuous parameters, with the overall scale of
SUSY breaking being set by the gravitino mass,~\mgrav.  In pure~AMSB
the soft SUSY-breaking scalar masses~$m_0$, like the gaugino masses,
are proportional to~$\mgrav$ before renormalization.  However, in this
case renormalization leads to negative squared masses for
sleptons. Thus, the pure~AMSB is unrealistic, and some additional
contributions to the scalar masses~$m_0$ are postulated. It is
simplest to assume that these are universal, as in
minimal~AMSB~(mAMSB) models.  In the~mAMSB~model the soft trilinear
SUSY-breaking mass terms,~$A_i$, are determined by anomalies, like the
gaugino masses, and hence are also proportional to~\mgrav, resulting
in the following three free continuous parameters:~\mgrav, $m_0$ and
the ratio of Higgs vevs,~$\tb$.  The~$\mu$~term and the soft Higgs
bilinear SUSY-breaking term,~$B$, are determined phenomenologically
via the electroweak vacuum conditions, as in the~CMSSM and related
models, and may have either sign.

Since the gaugino masses~$M_{1, 2, 3}$ are induced by anomalous loop
effects, they are suppressed relative to the gravitino mass,~$\mgrav$,
which is quite heavy in this scenario:~\mbox{$\mgrav \gtrsim 20
  \tev$}. The gaugino masses have the following ratios
at~NLO:~\mbox{$|M_1| : |M_2| : |M_3| \approx 2.8:1:7.1$}.  We note
that the wino-like states are lighter than the~bino, which is
therefore not a candidate to be the~LSP. The~LSP may be either a
Higgsino-like or a wino-like neutralino~$\neu1$, and is almost
degenerate with a chargino partner,~$\cha1$, in both cases.  If
the~LSP is a wino-like~$\neu1$ and it is the dominant source of the
dark matter density, its mass has been shown to be~$\simeqord 3
\tev$~\cite{winomass1,winomass} once Sommerfeld-enhancement
effects~\cite{Sommerfeld1931} are taken into account. On the other
hand, if a Higgsino-like~$\neu1$ provides the
CDM~density,~\mbox{$\mneu1 \sim 1.1 \tev$}. In the~mAMSB~model, the
Higgsino-like~LSP still has a non-negligible wino~component, and is
therefore heavier than a pure Higgsino, with~\mbox{$m_\chi \gtrsim 1.5
  \tev$}.

The following are characteristic features of the~mAMSB~model: the
{superpartners of the} left- and right-{handed leptons} are
nearly degenerate in mass,~\mbox{$\msl{R} \approx \msl{L}$}, 
as are the lightest chargino and
neutralino,~\mbox{$\mcha1 \approx \mneu1$}. The ratio between the
slepton and gaugino masses depends on the input parameters, but the
squark masses are typically very heavy, since they receive a
contribution~\mbox{${\propto}\,g_3^4\, m^2_{3/2}$}
where~\mbox{$g_3^2/(4\,\pi) = \als$}.  The relatively small
loop-induced values of the trilinears~$A_i$ and the measured Higgs
mass also favor relatively high stop masses.

\begin{figure}[bt!]
\centerline{
\includegraphics[height=8cm]{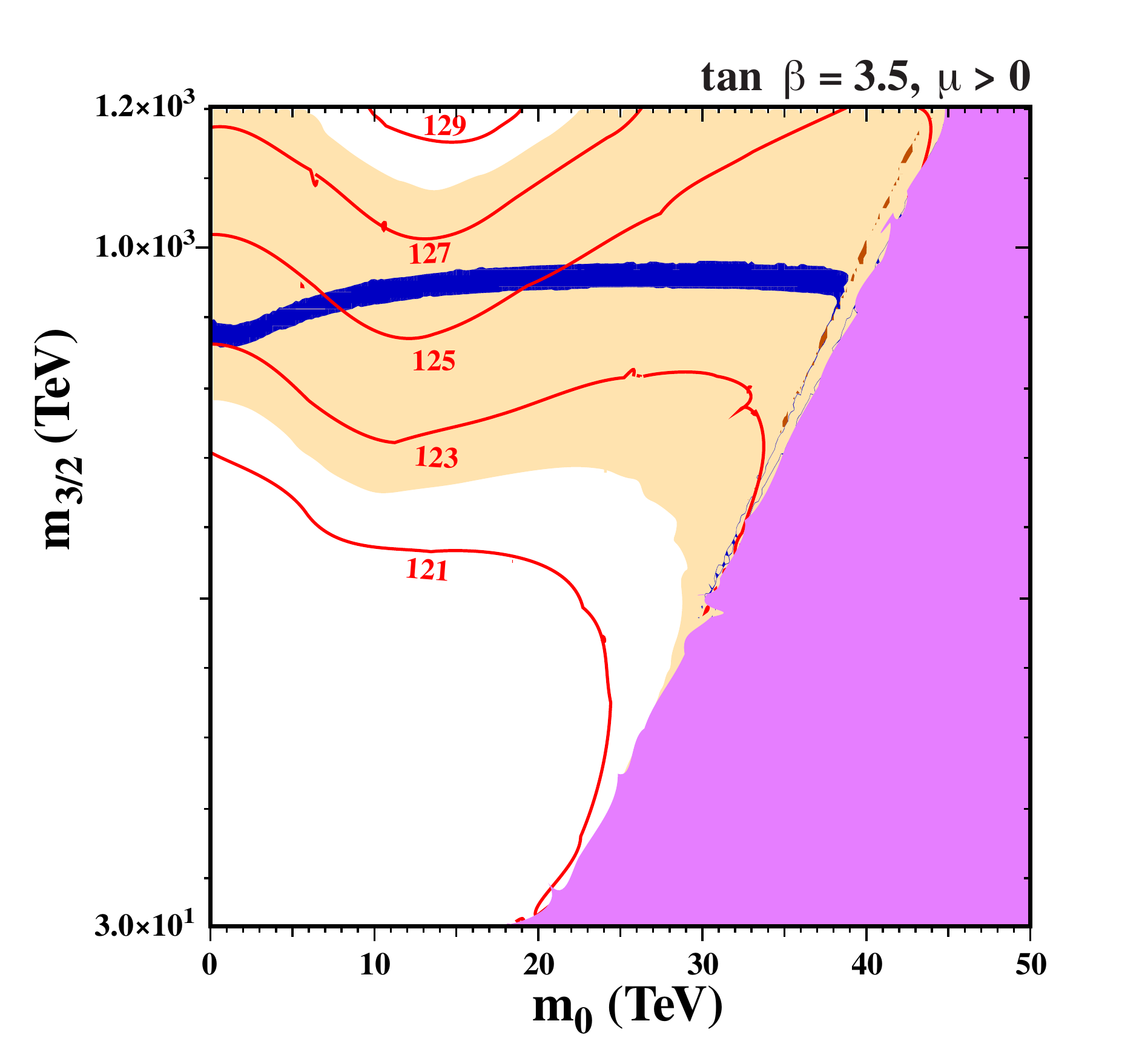}
\includegraphics[height=8cm]{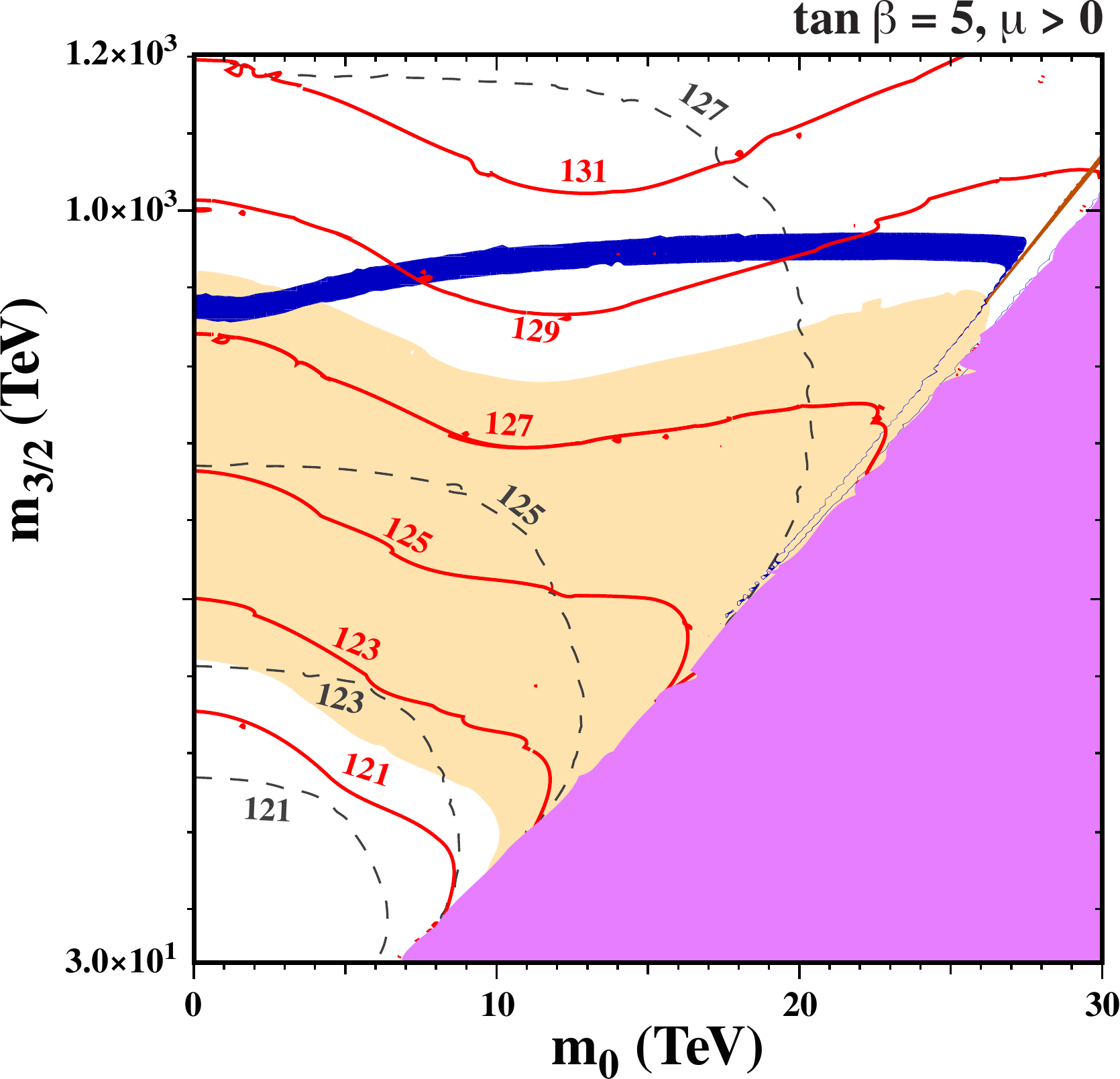}}
\vspace{0.5cm}
\centerline{
\includegraphics[height=8cm]{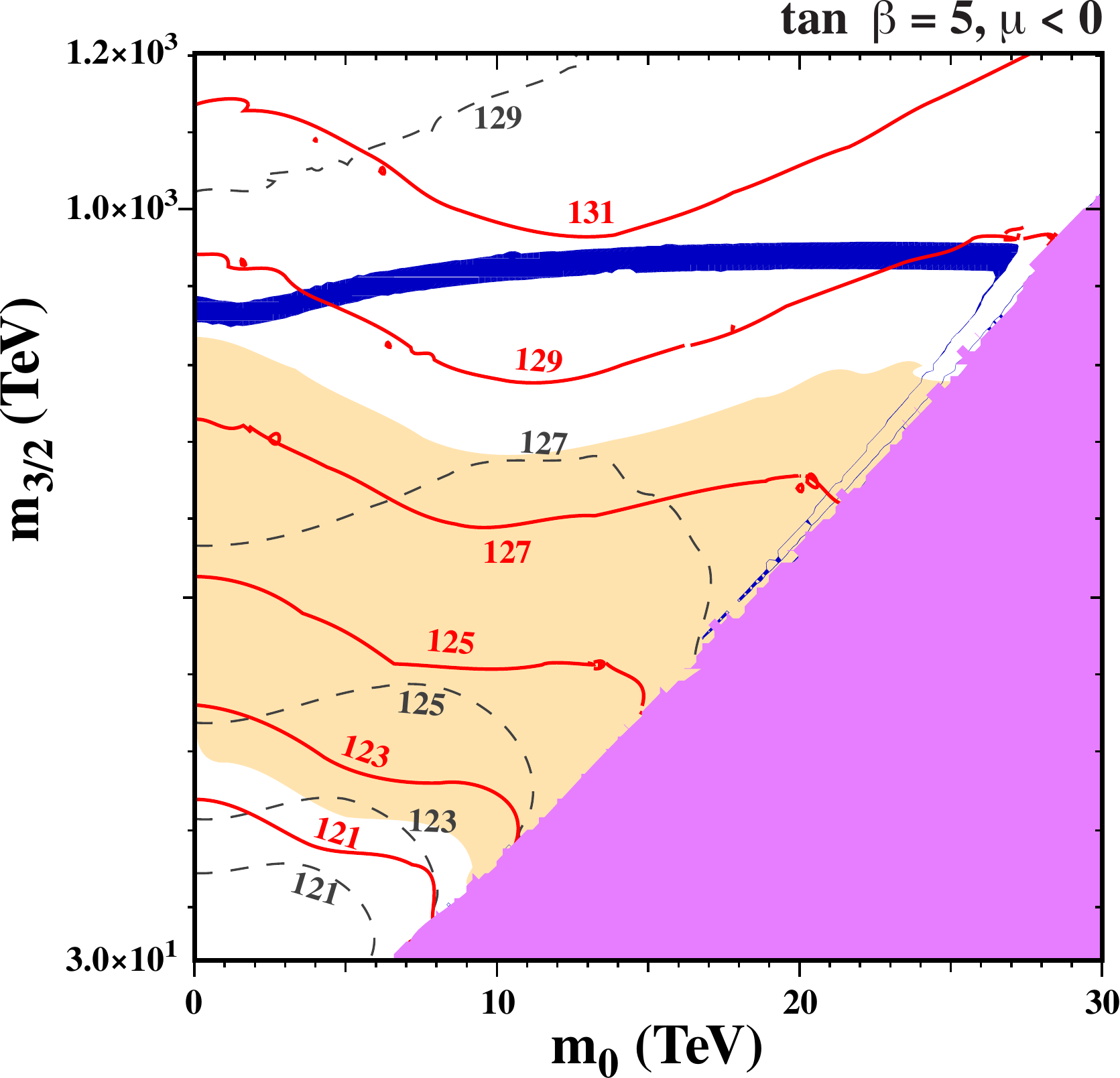}
\hspace{0.5cm}
\includegraphics[height=8cm]{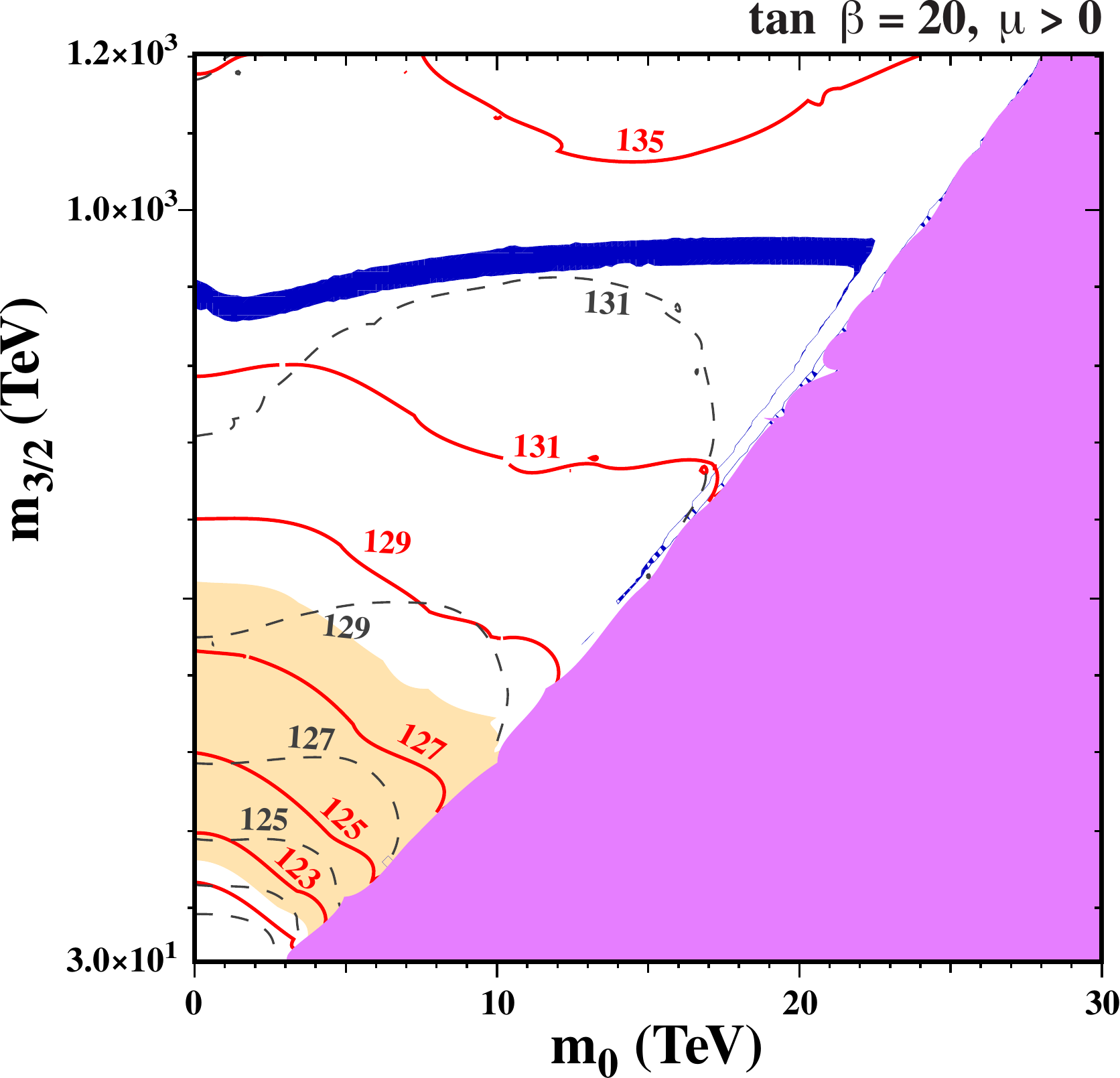}
}
\caption{
  \label{fig:amsb}
  The \mbox{$(m_0, m_{3/2})$}~planes in the mAMSB model for
  \mbox{$\tan \beta = 3.5$}, \mbox{$\mu > 0$} (upper left panel), for
  \mbox{$\tan \beta = 5$}, \mbox{$\mu > 0$} (upper right panel), for
  \mbox{$\tan \beta = 5$}, \mbox{$\mu < 0$} (lower left panel), and
  for \mbox{$\tan \beta = 20$}, \mbox{$\mu > 0$} (lower right panel).
  The electroweak symmetry-breaking conditions cannot be satisfied in
  the regions shaded pink in these plots.  Contours of~$\Mh$
  calculated using \FHnew\ are shown as red solid lines, those using
  \FHold\ as gray dashed lines.  The light orange shaded region
  corresponds to~\mbox{$\Mh \in [122,128] \gev$} found using \FHnew.
  The blue strips show the region with~\mbox{$0.06 < \Omega_\chi\, h^2
    < 0.2$}.  }
\end{figure}

We display in Fig.~\ref{fig:amsb} four \mbox{$(m_0, m_{3/2})$}~planes
in the~mAMSB~model.  They all have a pink shaded region at large~$m_0$
and relatively small~$m_{3/2}$ where the electroweak vacuum conditions
cannot be satisfied. Each panel also features a prominent
near-horizontal band with acceptable dark matter density where the~LSP
is mainly a~{wino} with mass~$\simeqord 3 \tev$. They also feature
narrower and less obvious strips close to the electroweak vacuum
boundary where the~LSP has a larger Higgsino fraction and a smaller
mass.  In this figure we use the range~\mbox{$0.1151 <
  \Omega_{{\chi}}\, h^2 < 0.1235$}.
As one can see, there is a strong preference for low~$\tan \beta$ for
the wino-dark-matter~strip.  At~\mbox{$\tan \beta > 5$}, most of the
wino~strip has Higgs masses in excess of~$128$~GeV. While portions of
the Higgsino~strip are acceptable at higher~$\tan \beta$,
at~\mbox{$\tan \beta = 20$} the pair of Higgsino~strips is also at
large~$\Mh$.

The profiles of the mAMSB-dark-matter~strips displayed in
Fig.~\ref{fig:amsb} are shown in Fig.~\ref{fig:mAMSBprofiles}.  In
each panel, the horizontal axis is~$m_0$, the left vertical axis
is~$\mneu1$, and the right vertical axis is~$\Mh$.  We can again
easily distinguish between the wino and Higgsino-like~strips. The
wino~strip spans a wide range in~$m_0$ as seen in Fig.~\ref{fig:amsb},
where the Higgsino-like~strip resides only at large~$m_0$.  In the
wino-like~strip, the neutralino mass is shown by the blue dashed
curves and~\mbox{$\mneu1 \simeq 3 \tev$} at large~$m_0$, falling
to~$\simord 2.7 \tev$ at low~$m_0$, whereas along the lower
strip~$\mneu1$ falls from~$\simeqord 3 \tev$ to~$\simord 1.5 \tev$
as~$m_0$ decreases towards the tip of the strip at~\mbox{$m_0 \sim
  15$} to~$30 \tev$.

\begin{figure}[bt!]
\centerline{
\includegraphics[height=6cm]{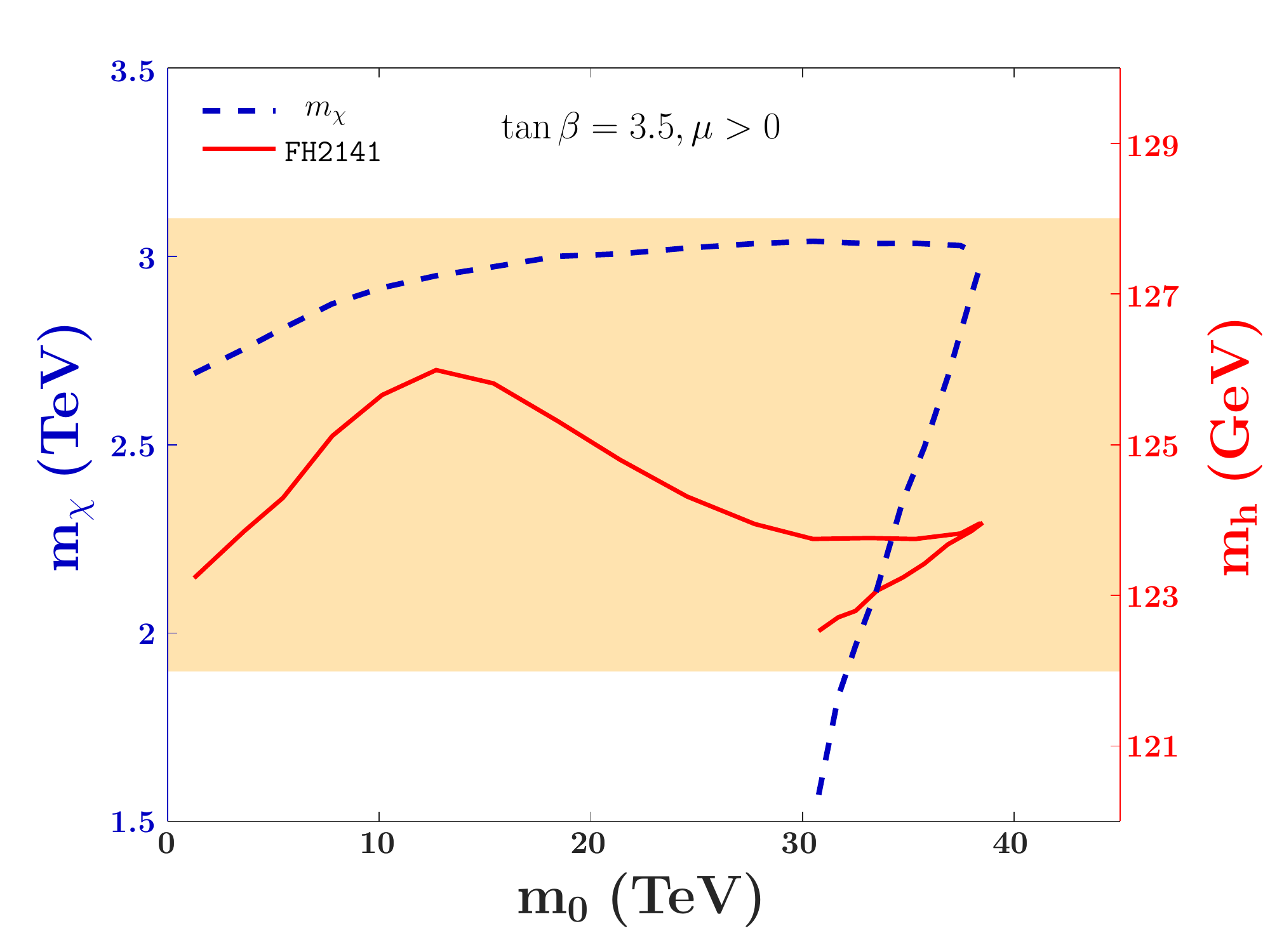}
\includegraphics[height=6cm]{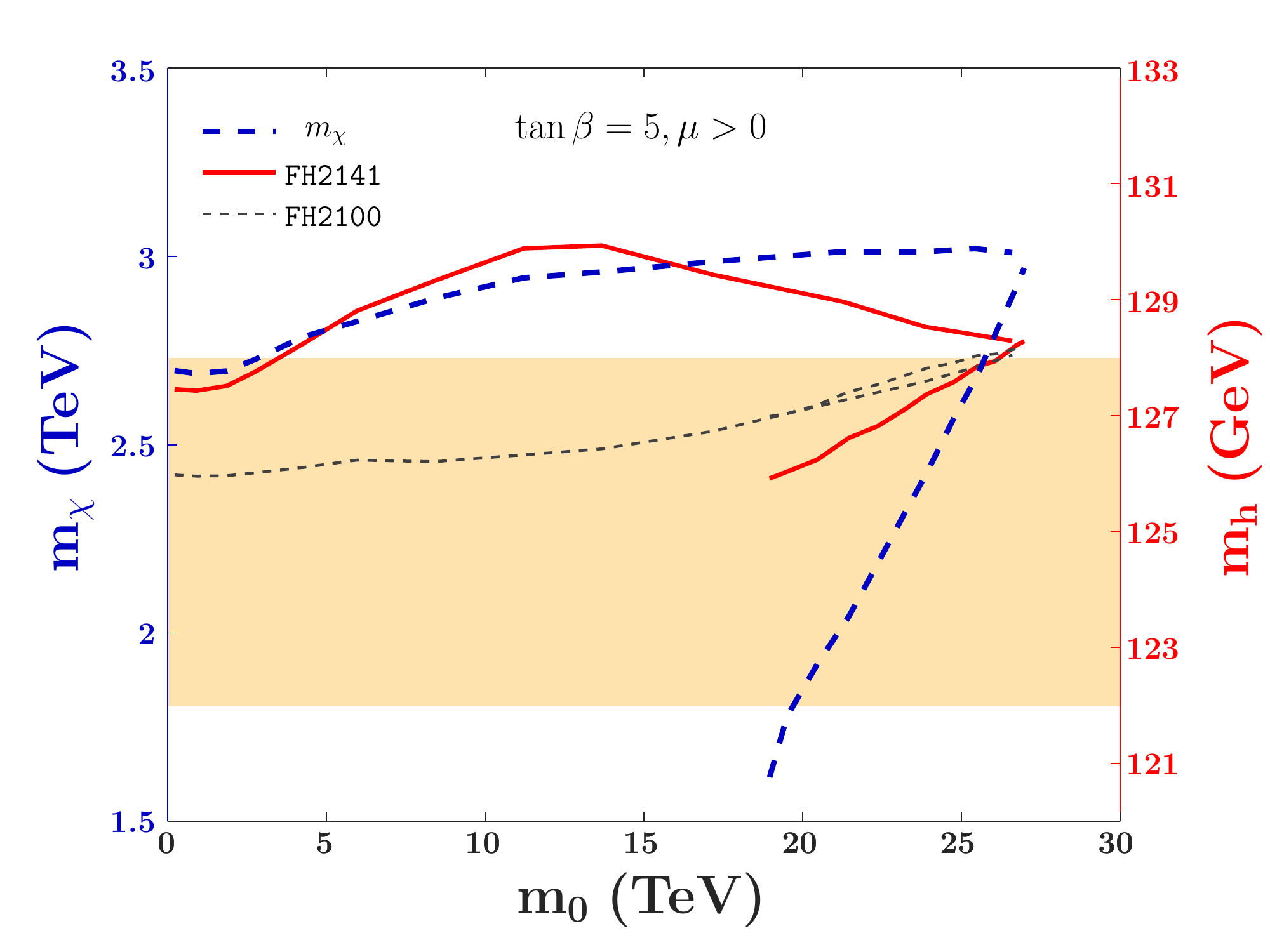}}
\centerline{
\includegraphics[height=6cm]{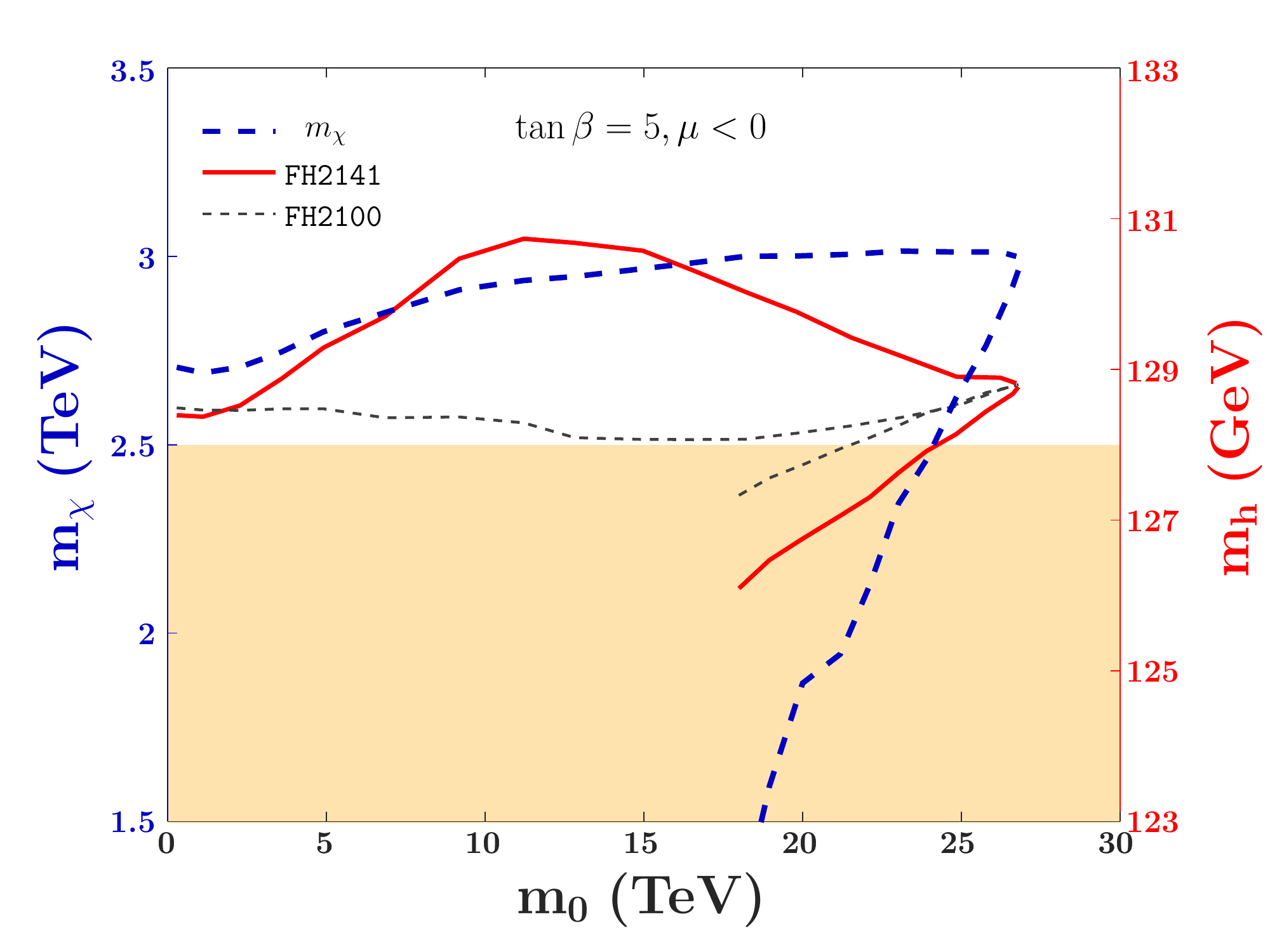}
\includegraphics[height=6cm]{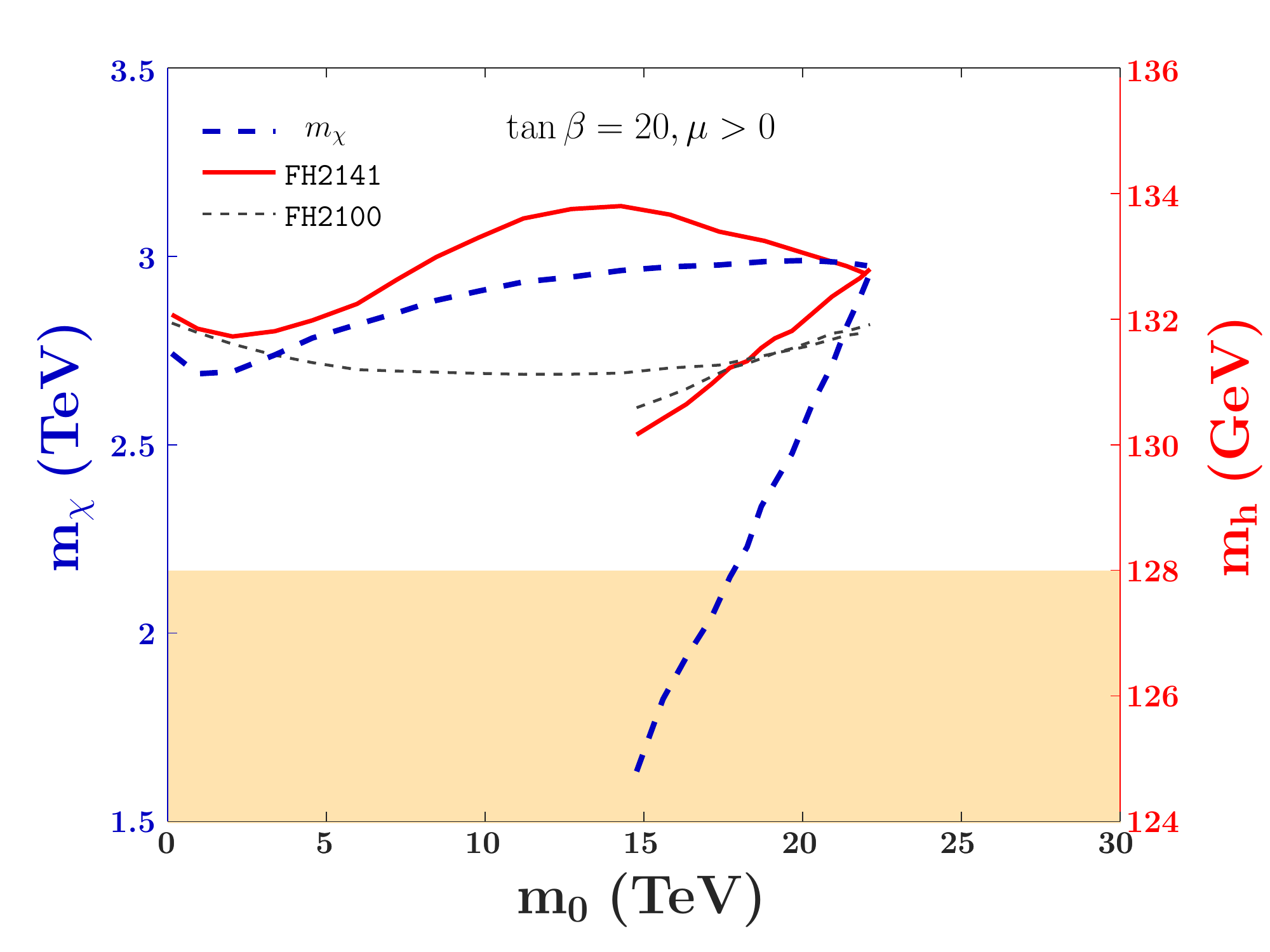}
}
\caption{
  \label{fig:mAMSBprofiles}
  The profiles of the mAMSB~strips for \mbox{$\tan \beta = 3.5$}, \mbox{$\mu > 0$} (upper left panel),
for \mbox{$\tan \beta = 5$}, \mbox{$\mu > 0$} (upper right panel), for \mbox{$\tan \beta = 5$}, \mbox{$\mu < 0$} (lower left panel),
and for \mbox{$\tan \beta = 20$}, \mbox{$\mu > 0$} (lower right panel).
The lower horizontal axes show~$m_{0}$,
the blue dashed lines show~$\mneu1$, to be read from the left vertical
axes.
The ``allowed'' range for~$\Mh$ is indicated by {the} horizontal
light orange shaded region for~\mbox{$\Mh \in [122, 128] \gev$}.
Calculated values for~$\Mh$: \FHnew\ (red) and \FHold\ (dashed black),
to be read from the right vertical axes.
}
\end{figure}

We see in the upper left panel of Fig.~\ref{fig:mAMSBprofiles} that,
for~\mbox{$\tan \beta = 3.5$} and~\mbox{$\mu > 0$}, the Higgs
mass~$\Mh$ calculated with \FHnew\ (red lines) is consistent with the
experimental value all along both strips, within the theoretical
uncertainties.  We do not show the results from \FHold\ in this case,
as they were not reliable for~\mbox{$\tan \beta = 3.5$}.  On the other
hand, calculations of~$\Mh$ with \FHnew\ are significantly higher than
the experimental value along the wino-like~strips in the other panels,
which are for larger values of~$\tb$.  In contrast,
\FHold\ calculations of~$\Mh$ were significantly lower along the
wino-like~strips, and compatible with experiment for~\mbox{$\tan \beta
  = 5$} and~\mbox{$\mu > 0$}.  In the cases of the
Higgsino-like~strips, \FHnew\ calculations of~$\Mh$ are compatible
with experiment along that for~\mbox{$\tan \beta = 5$} and~\mbox{$\mu
  > 0$} and most of the corresponding strip for~\mbox{$\tan \beta =
  5$} and~\mbox{$\mu < 0$}, though not for the Higgsino-like~strip
for~\mbox{$\tan \beta = 20$} and~\mbox{$\mu > 0$}.  \FHold\ gave
generally larger values of~$\Mh$ along these Higgsino-like~strips,
which are compatible with experiment only for the strip
for~\mbox{$\tan \beta = 5$} and~\mbox{$\mu > 0$} and part of the strip
for~\mbox{$\tan \beta = 5$} and~\mbox{$\mu < 0$}.

In the mAMSB, as seen in Fig.~\ref{fig:mAMSBprofiles}, in general
\FHnew\ yields values of $\Mh$ along the Higgsino strips that are more
consistent with the experimental measurement than the ones with
\FHold. On the other hand, the values of $\Mh$ along the wino strip
are generally larger for \FHnew\ than for \FHold, and in poorer
agreement with experiment.  Hence, in this case the improvements in
\FHnew\ yield a preference for a quite different region of the model
parameter space.


\subsection{The pMSSM11\label{sec:pmssm11}}

In contrast to the above models in which soft SUSY breaking is assumed
to originate from some specific theoretical mechanism, we now study a
model in which {the SUSY parameters} are constrained by purely
phenomenological considerations. In general, such
phenomenological~MSSM~(pMSSM)~\cite{pMSSM} models contain many more
parameters. Here we consider a variant of the~pMSSM with
$11$~parameters fixed at the electroweak-scale, the~pMSSM11, as
analyzed in \citere{MCpMSSM11} using the available experimental
constraints including many from the first LHC~run at~$13$~TeV.  The
model parameters are {three} independent gaugino
masses,~$M_{1,2,3}$, a common mass for the first-and second-generation
squarks,~$\msq$, a mass for the third-generation squarks,~$\msqt$,
that is allowed to be different, a common mass,~$\msl{}$, for the
first-and second-generation sleptons, a mass for the stau,~$\msl3$,
that is also allowed to be different,\footnote{Note that we assume
  equal soft SUSY-breaking parameters for {the superpartners of
    the} left- and right-handed fermions of the same flavor.} a
single trilinear mixing parameter,~$A$, the Higgs mixing
parameter~$\mu$, the pseudoscalar Higgs mass,~$\MA$, and the ratio of
Higgs vevs,~$\tb$.  These parameters are all fixed at a
renormalization scale~\mbox{$\msusy \equiv \sqrt{\mstop1\, \mstop2}$},
where~$\mstop1$,~$\mstop2$ are the masses of the two stop mass
eigenstates. This is also the scale at which electroweak vacuum
conditions are imposed. As in all the models we study, the sign of the
mixing parameter~$\mu$ may be either positive or negative.

\begin{figure}[hbtp!]
\centerline{
\includegraphics[height=5.5cm]{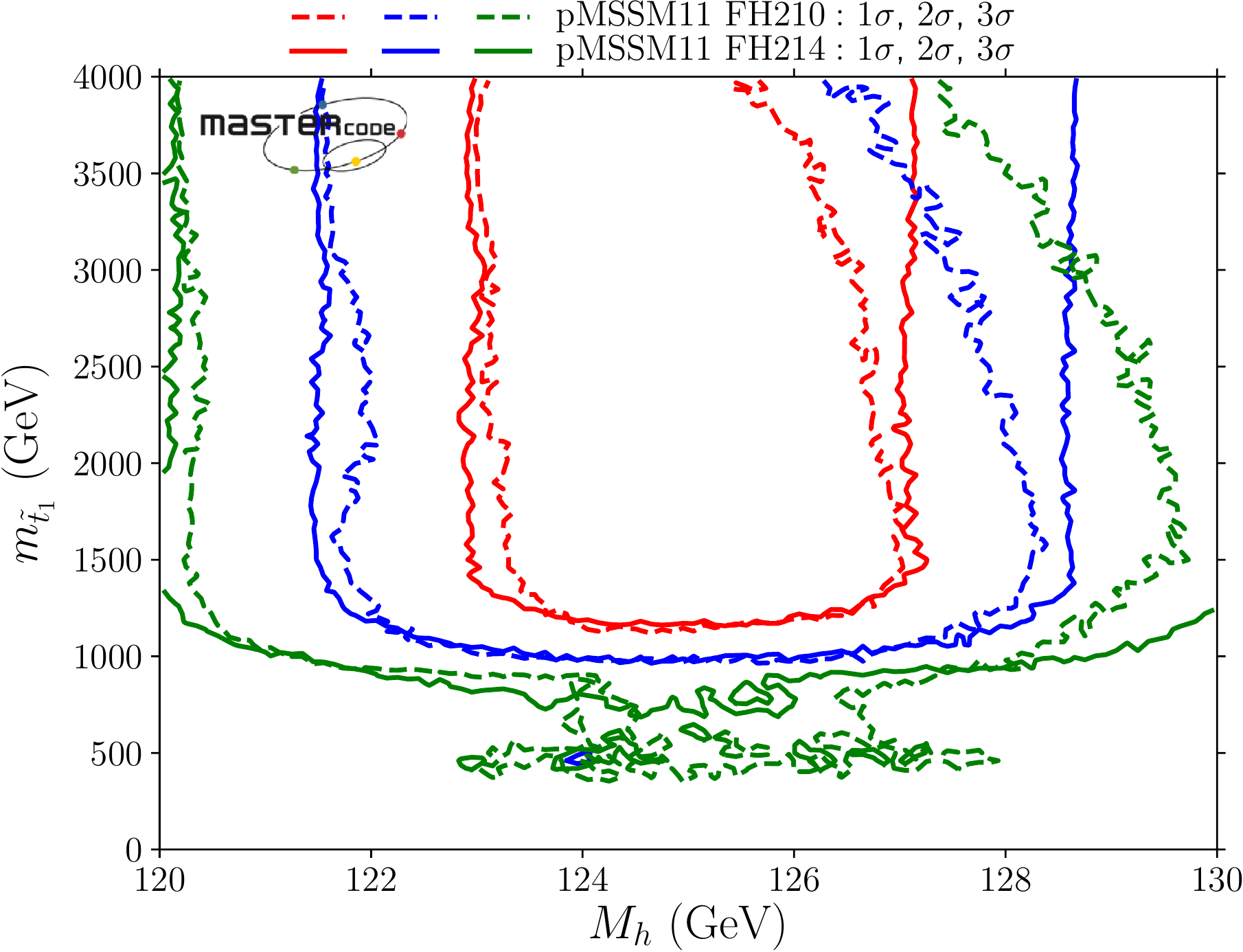}
\includegraphics[height=5.5cm]{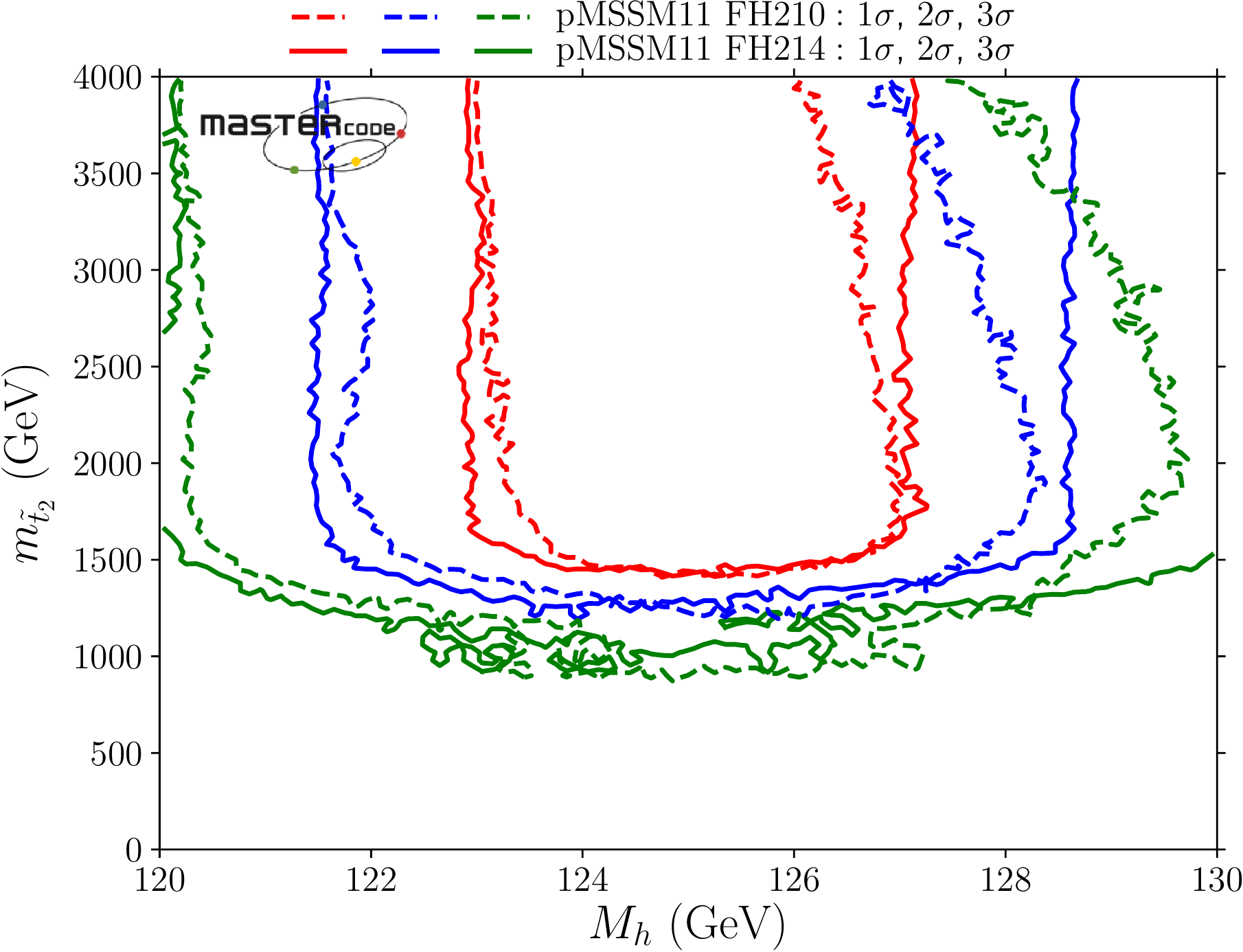}}
\centerline{
\includegraphics[height=5.5cm]{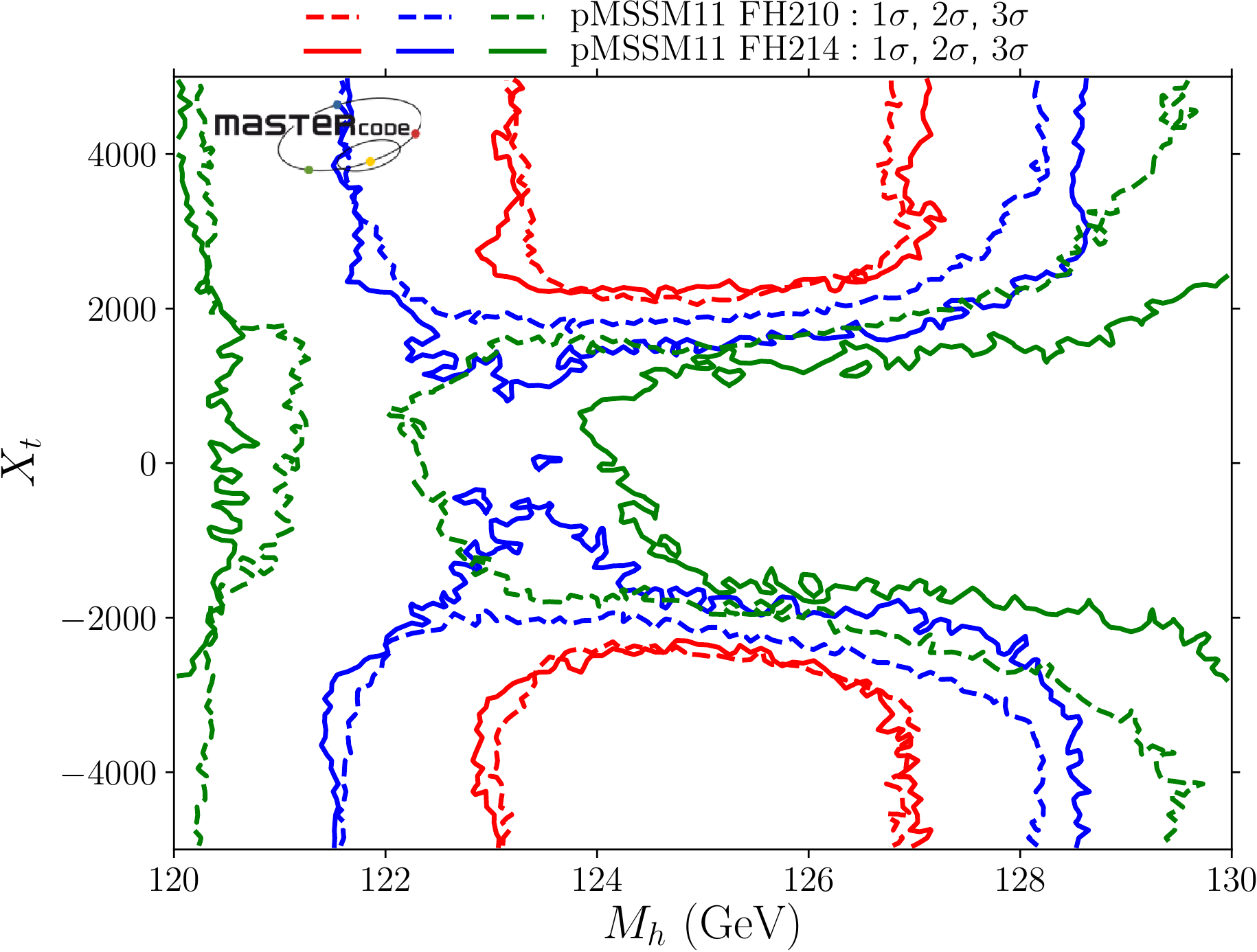}
\includegraphics[height=5.5cm]{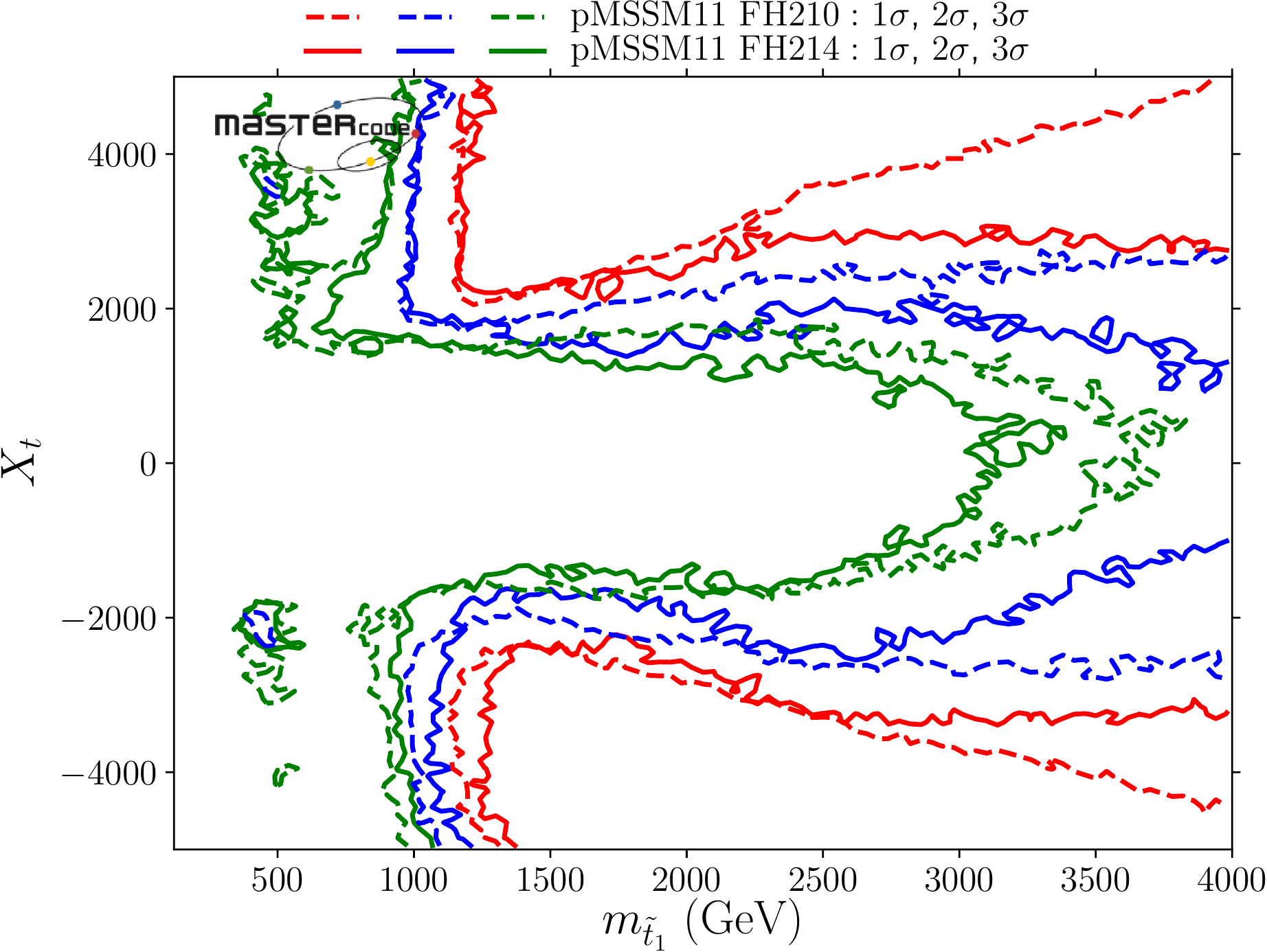}
}
\centerline{
\includegraphics[height=5.5cm]{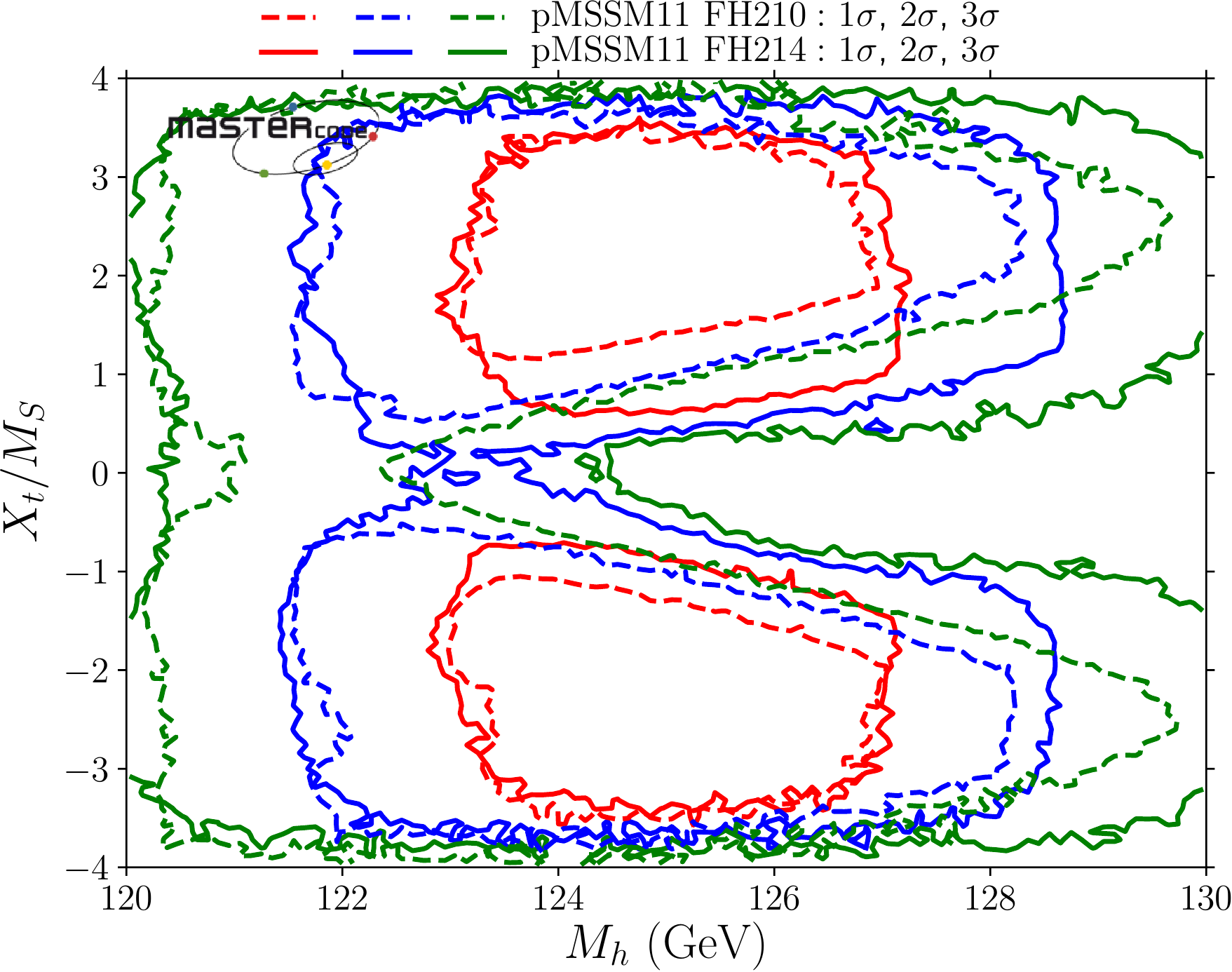}
\includegraphics[height=5.5cm]{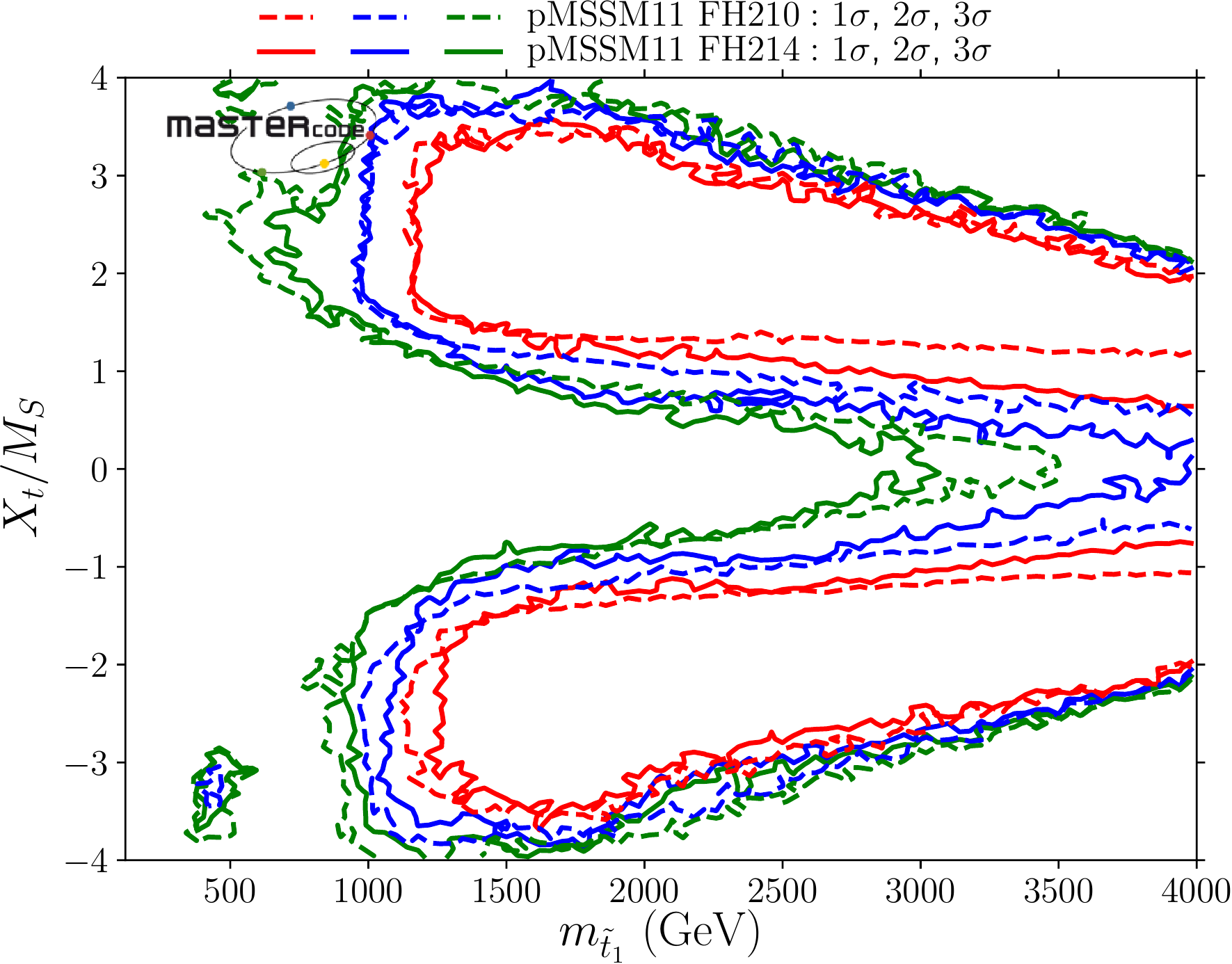}
}
\caption{
\label{fig:pMSSM11}
  Explorations of the sensitivities of~$\Mh$ to~$m_{\tilde t_{1,2}}$
  in preferred regions of the~pMSSM11 \hfill found}
\vspace{-2.2ex}
\caption*{in \protect\citere{MCpMSSM11}, displayed in the \mbox{$(\Mh,
    m_{\tilde t_{1}})$}~plane (top left panel), the \mbox{$(\Mh,
    m_{\tilde t_{2}})$~plane} (top right panel), the \mbox{$(\Mh,
    X_t)$}~plane (middle left panel), the \mbox{$(m_{\tilde t_{1}},
    X_t)$}~plane (middle right panel), the \mbox{$(\Mh,
    X_t/M_S)$}~plane (bottom left panel), and the \mbox{$(m_{\tilde
      t_{1}}, X_t/M_S)$}~plane (bottom right panel).  In each panel
  the red, blue and green contours outline the regions favored at the
  $68\%$~CL (1-$\sigma$), $95\%$~CL (2-$\sigma$) and $99.7\%$~CL
  (3-$\sigma$), respectively, and the solid (dashed) lines are those
  found in a global analysis of all relevant data using
  \FHnew\ (\FHold) to calculate $\Mh$.}
\end{figure}

The flexibility of the~pMSSM11~model allows, in principle, many
different mass hierarchies to be explored, and hence different aspects
of the~$\Mh$~calculation. In particular, since the Higgs mass is most
sensitive, in general, to the stop masses, we explore in
Fig.~\ref{fig:pMSSM11} what stop masses and mixing are compatible with
the measured Higgs mass, without being constrained by any preconceived
theoretical ideas such as those arising in the models discussed in the
previous sections. In each panel of Fig.~\ref{fig:pMSSM11}, the
regions favored at the~$68\%$~CL (1-$\sigma$), $95\%$~CL (2-$\sigma$)
and~$99.7\%$~CL (3-$\sigma$) are enclosed by red, blue and green
contours, respectively, which are shown solid (dashed) if \FHnew\ (\FHold) is used to calculate the~$\chi^2$~contribution\footnote{We recall here
  that the $\Mh$ contribution to the global likelihood is modeled using a Gaussian
  distribution with $\mu = 125.09$ GeV and $\sigma_{\mathrm{exp}} = 0.24$ GeV and $\sigma_{\mathrm{theo-SUSY}} = 1.5$ GeV.
  For further details on the likelihood, including a discussion of the other constraints, we refer the
reader to Ref.~\cite{MCpMSSM11}.} the LHC~measurement of~$\Mh$ to a frequentist global analysis of the
pMSSM11~parameter space~\cite{MCpMSSM11}.\footnote{It should be kept
  in mind that the set of pMSSM11~points used here was originally
  obtained in \citere{MCpMSSM11} using {\tt FeynHiggs~2.11.3}. Slight
  shifts in the contours shown below can be expected if the full fit
  would be done using either \FHold\ or \FHnew.}

We see in the top left panel of Fig.~\ref{fig:pMSSM11} that the
experimental value of~$\Mh$ can be accommodated by values
of~\mbox{$m_{\tilde t_{1}} \gtrsim 500$}~GeV ($1000$~GeV) ($1300$~GeV)
at the~$99.7$~($95$)~($68$)\%~CL, whether \FHnew\ or \FHold\ is used
to calculate~$\Mh$.  The most significant difference is a tendency for
\FHold\ to disfavor larger values of~$\Mh$ when~$m_{\tilde t_{1}}$ is
large, a tendency that is absent when \FHnew\ is used. We note also
that the
likelihood function is quite flat for~\mbox{$m_{\tilde t_{1}} \gtrsim
  1500 \gev$}, and for this reason we do not quote a best-fit point.
The upper right panel shows that values of~\mbox{$m_{\tilde t_{2}}
  \gtrsim 1$}~($1.3$)~($1.5$)~TeV are favored at
the~$99.7$~($95$)~($68$)\%~CL,
again with little difference between the results with \FHnew\ and \FHold.
Again, \FHold\ tends to disfavor larger values
of~$\Mh$ when~$m_{\tilde t_{2}}$ is large, but not \FHnew.

The middle panels of Fig.~\ref{fig:pMSSM11} explore the sensitivities
of the~$\Mh$~calculation to the stop mixing parameter~\mbox{$X_t
  \equiv A_t - \mu \cot \beta$} in the two versions of {\tt
  FeynHiggs~}\footnote{Note that here we use the sign convention
  for~$A_t$ of {\tt FeynHiggs}.}.  We see that they both favor values
of~\mbox{$\lvert X_t\rvert \gtrsim 2$}~TeV, though~\mbox{$X_t = 0$} is
allowed at the~$99.7\%$~CL.
However, we see in the middle right panel that this is possible only
for~\mbox{$m_{\tilde t_1} \gtrsim 3$}~($3.5$)~TeV when
\FHnew\ (\FHold) is used. This {behavior at~\mbox{$X_t = 0$}} may
be the origin of the {often}-repeated statement that the measured
value of~$\Mh$ requires a large stop mass. In fact, as already
mentioned above, the upper panels of Fig.~\ref{fig:pMSSM11} show
that~\mbox{$\Mh \simeq 125$}~GeV is quite compatible
with~\mbox{$m_{\tilde t_{1}} \sim 1.2$}~TeV, and the middle right
panel shows that this {is} possible if~\mbox{$\lvert X_t\rvert
  \sim 2$}~TeV.  The bottom plots of \reffi{fig:pMSSM11} show the same
results as in the middle row, but with~$X_t/M_S$ on the vertical
axes. In particular, in the lower right plot it can clearly be seen
that the correct Higgs-boson mass prediction requires {\it either}
large mixing in the stop sector, {\it or} large scalar top
masses. Here small mixing can more easily be reached with \FHnew.

Fig.~\ref{fig:pMSSM112} contains one-dimensional plots of the
global~$\chi^2$-likelihood~functions for~$m_{\tilde t_{1}}$ (left
panel) and~{$X_t/M_S$} (right panel), shown as solid (dashed)
lines as found using \FHnew\ (\FHold) to calculate
the~$\chi^2$~contribution from the LHC~measurement of~$\Mh$.  Here we
see again that the global minima are at~\mbox{$m_{\tilde t_{1}} \sim
  1.5$}~TeV and~{\mbox{$|X_t/M_S|\sim 2$}}, with little difference
between \FHnew\ and \FHold.  The~$\chi^2$~function for~$m_{\tilde
  t_{1}}$ rises very mildly as~$m_{\tilde t_{1}}$ approaches~$4$~TeV,
and the exact location of the minimum value cannot be regarded as
significant. We note also the appearance of a secondary minimum
with~\mbox{$\Delta \chi^2 < 3$} when~\mbox{$m_{\tilde t_{1}} \simeq
  500$}~GeV.  The~$\chi^2$~function for~{$X_t/M_S$} exhibits no
significant sign preference, but disfavors~\mbox{$X_t = 0$}
by~\mbox{$\Delta \chi^2 \simeq 4$} if \FHnew\ is used, compared
to~\mbox{$\Delta \chi^2 \simeq 8$} with \FHold.

\begin{figure}[tb!]
\centerline{
\hspace{-0.8cm}
\includegraphics[height=6.5cm]{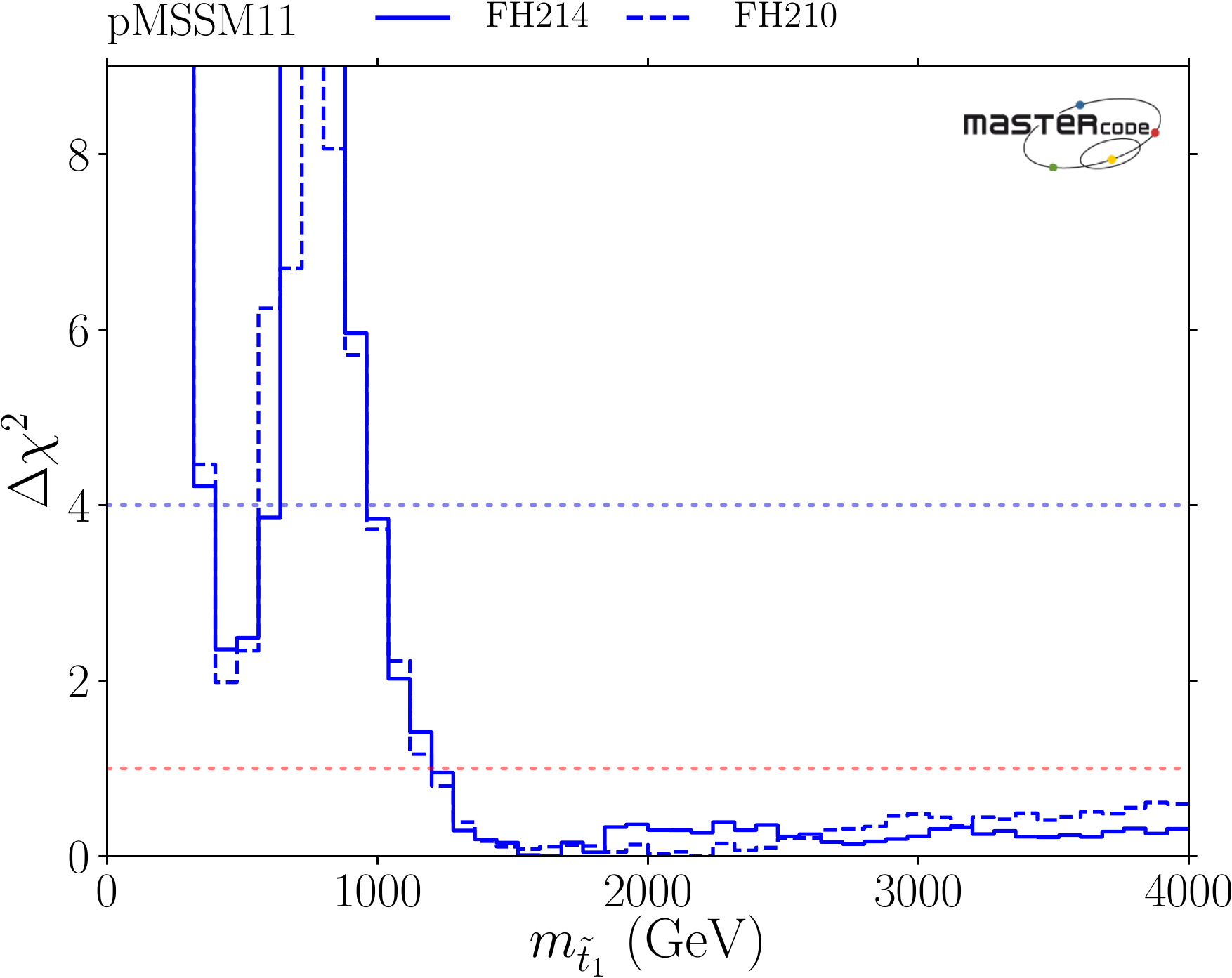}
\includegraphics[height=6.5cm]{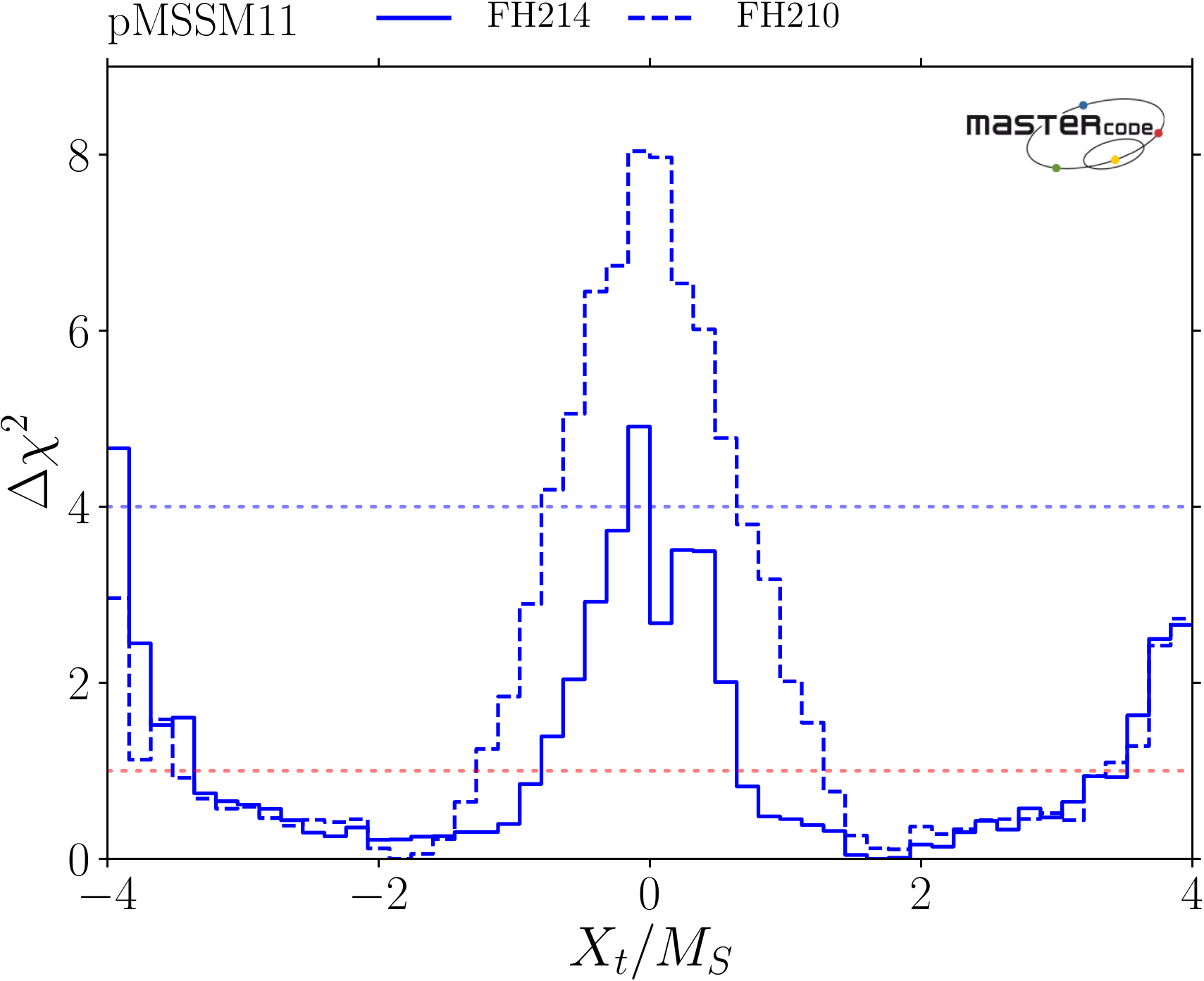}
}
\caption{
  \label{fig:pMSSM112}
The one-dimensional global $\Delta \chi^2$~functions for~$m_{\tilde
  t_{1}}$ (left panel) and~{$X_t/M_S$} (right panel), as found
using \FHnew\ (\FHold) to calculate the $\chi^2$~contribution from the
LHC~measurement of~$\Mh$, shown as solid (dashed) lines.}
\end{figure}

Overall, in the~pMSSM11 we see no clear trend towards lower or higher
values of~$M_h$ when going from \FHold\ to \FHnew. {Al}though individual
parameter choices may yield different values for~$\Mh$, the
experimental constraints on the~pMSSM11 favor regions in the
parameter space where marginalization to minimize~$\chi^2$ yields
milder effects on the light $\cp$-even Higgs-boson mass.


\section{Conclusions\label{sec:concl}}

We have investigated the physics implications of improved Higgs-boson
mass {predictions} in the~MSSM, comparing results from \FHnew\ and
\FHold. The main differences, as discussed in this paper, are~$3$-loop
RG~effects and~$2$-loop threshold corrections that can accommodate
three separate mass scales:~$m_{\tilde q}$, $\mgluino$ and an
electroweakino mass scale, as well as an improved {treatment of
  \DRbar\ input parameters in the scalar top sector} avoiding problems
with the conversion to on-shell parameters, that yields significant
improvements for large SUSY-breaking scales. These changes
reflect the progress
made over the last~$\simord 5$~years in ``hybrid'' Higgs-mass
calculations in the~MSSM.

The examples presented in this paper illustrate how the preferred
ranges of the parameter space of the~MSSM can change when \FHnew\ is
used to calculate~$\Mh$, as compared to when \FHold\ is used.  The
first representative model is the~CMSSM.  As is well known, in
the~CMSSM reproducing the correct CDM~density of neutralinos, despite
the rising lower limits on sparticle masses from the~LHC, tends to
favor narrow strips of parameter space that extend to large~$m_{1/2}$
and/or~$m_0$.The improvements in
\FHnew\ can play important roles in these parameter regions. Examples of these high-mass~strips
include some where stop~coannihilation is important, and others where
the focus-point~mechanism is operative. In both these cases, using
\FHnew\ rather than \FHold\ changes significantly the parts of the
strips that are consistent with the experimental measurement
of~$\Mh$. This reflects the different dependences on~$m_{1/2}$ of the
\FHnew\ and \FHold\ calculations of~$\Mh$.

We have also studied sub-GUT models, in which the soft SUSY-breaking
masses are assumed to be universal at some scale~\Min\ below the
conventional grand unification scale~\MGUT\ assumed in the~CMSSM.
Both the stop-coannihilation and focus-point~mechanisms may be
operational in different regions of the sub-GUT parameter
space. Depending on the choice of~\Min, the forms of the DM~strips can
be very different from those allowed in the~CMSSM, with the
possibility of two (or more) DM~strips with different values of~$m_0$
for the same value of~$m_{1/2}$. In general, along the
lower-$m_0$~strips the agreement between \FHnew\ calculations of~$\Mh$
and experiment is better than that for the \FHold\ calculations.

As a third case we investigated the~mAMSB, where two different classes
of DM~strips occur: one where the~LSP may be mainly a~wino, or one
where it may have a large Higgsino~component. Both of these types of
dark-matter~{strips} extend to relatively large values of~$m_0$,
with an LSP~mass~$\simord 3 \tev$ or~$\gsimord 1 \tev$,
respectively. Calculations of~$\Mh$ using \FHnew\ favor the
Higgsino~region, whereas calculations using \FHold\ favored the
wino~region.

In the case of the~pMSSM11, we find little change in the regions of
parameter space favored by~$\Mh$, which can be ascribed to the fact
that there is no big mass hierarchy.  The predictions from both
\FHnew\ and \FHold\ are consistent with~\mbox{$\Mh \simeq 125 \gev$}
and~\mbox{$m_{\tilde t_{1}} \sim 1.3 \tev$} at the~$68\%$~CL, and they
both allow~\mbox{$m_{\tilde t_{1}} \sim 500 \gev$} with~\mbox{$\Delta
  \chi^2 \sim 3$}. Both versions of {\tt FeynHiggs} disfavor small
stop mixing,~\mbox{$X_t = 0$}, by~\mbox{$\Delta \chi^2 \sim 4\ (8)$}
in the case of \FHnew\ (\FHold), with~{\mbox{$|X_t/M_S| \sim 2$}}
being favored.  Obtaining the correct prediction for the Higgs-boson
mass requires {\it either} {large mixing in the scalar top sector
  (with~\mbox{$|X_t/M_S| \sim 2$}),} {\it or} large scalar top masses,
though smaller values of~{$|X_t/M_S|$} can be reached more easily
with \FHnew.  We find no clear preference towards lower or higher
values of~$M_h$ when going from \FHold\ to \FHnew\ in the~pMSSM11. The
experimental constraints yield parameter combinations with mild
effects on the light $\cp$-even Higgs-boson mass after marginalization
to minimize~$\chi^2$.

In conclusion, we comment that in this paper we have limited ourselves
to exploratory studies, and have not attempted to make global fits to
the parameters of any of the SUSY models we have discussed. However,
we find an overall tendency towards better compatibility with the
experimental data when employing the updated Higgs-boson mass
calculations. Performing new fits with updated calculations of~$M_h$
would clearly be an interesting next step, and we hope that the
studies described here will give some insight into the results to be
expected from such more complete investigations.


\subsection*{Acknowledgments}

\noindent
The work of JE was supported in part by the United Kingdom STFC Grant
ST/P000258/1, and in part by the Estonian Research Council via a
Mobilitas Pluss grant.  The work of SH was supported in part by the
MEINCOP (Spain) under contract FPA2016-78022-P, in part by the Spanish
Agencia Estatal de Investigaci\'on (AEI), in part by the EU Fondo
Europeo de Desarrollo Regional (FEDER) through the project
FPA2016-78645-P, in part by the ``Spanish Red Consolider MultiDark''
FPA2017-90566-REDC, and in part by the AEI through the grant IFT
Centro de Excelencia Severo Ochoa SEV-2016-0597.  The work of KAO was
supported in part by DOE grant DE-SC0011842 at the University of
Minnesota.  The work of SP was supported by the ANR grant
``HiggsAutomator'' (ANR-15-CE31-0002).
The work of HR was partially funded by the Danish National
Research Foundation, grant number DNRF90.
The work of JZ was supported by
 KAKENHI Grant Number JP26104009.


\end{document}

%% file: paperdef.tex
\usepackage[T1]{fontenc}
\usepackage[utf8]{inputenc}

\usepackage{amsmath,amsfonts,amssymb}
\usepackage[mathscr]{euscript}
\usepackage{booktabs}
\usepackage{graphicx, subfigure, ulem, hhline}
\usepackage[table]{xcolor}
\usepackage{siunitx}
\usepackage{soul}
\usepackage{wasysym}              
\usepackage{hyphenat}             

\usepackage{etoolbox,xstring,xspace,calc,xifthen}

\let\theparentequation\theequation
\patchcmd{\theparentequation}{equation}{parentequation}{}{}

\newcommand*{\nextParentEquation}[1][]{
  \refstepcounter{parentequation}
  \setcounter{equation}{0}
  \ifx\\#1\\\relax\else\parentlabel{#1}\fi
}

\makeatletter
\renewcommand\@makefntext[1]{\leftskip=.8em\hskip-.62em\@makefnmark#1}
\makeatother
\usepackage{setspace}

\input{loadhyperref.tex}

\usepackage[all]{hypcap}
\usepackage{footnotebackref}

\usepackage[format=plain, margin=0.5cm, font=footnotesize, labelfont=bf]{caption}
\DeclareCaptionSubType*[arabic]{figure}
\DeclareCaptionSubType*[arabic]{table}
\let\theparentequation\theequation
\patchcmd{\theparentequation}{equation}{parentequation}{}{}

\apptocmd{\thebibliography}{\normalsize}{}{}
\let\OLDthebibliography\thebibliography
\renewcommand\thebibliography[1]{
  \OLDthebibliography{#1}
  \setlength{\parskip}{1pt}
  \setlength{\itemsep}{1pt plus 0.3ex}
}

\newcommand{\simord}{\mathord{\sim}\,}
\newcommand{\simeqord}{\mathord{\simeq}\,}

\newcommand{\gsimord}{\mathord{\gtrsim}\,}

\newcommand{\gsim}{\gtrsim}
\newcommand{\tev}{\ensuremath{~\mathrm{TeV}}}
\newcommand{\gev}{\ensuremath{~\mathrm{GeV}}}
\newcommand{\mev}{\ensuremath{~\mathrm{MeV}}}
\def\order#1{\ensuremath{{\cal O}(#1)}}

\def\ga{\gamma}

\def\la{\lambda}

\def\beq{\begin{equation}}
\def\eeq{\end{equation}}

\let\sz\ss

\def\ss{\scriptscriptstyle}
\def\ga{\mathrel{\raise.3ex\hbox{$>$\kern-.75em\lower1ex\hbox{$\sim$}}}}
\def\la{\mathrel{\raise.3ex\hbox{$<$\kern-.75em\lower1ex\hbox{$\sim$}}}}
\def\gyr{{\rm \, G\kern-0.125em yr}}

\def\gappeq{\mathrel{\rlap {\raise.5ex\hbox{$>$}}
{\lower.5ex\hbox{$\sim$}}}}
\def\lappeq{\mathrel{\rlap{\raise.5ex\hbox{$<$}}
{\lower.5ex\hbox{$\sim$}}}}
\def\Toprel#1\over#2{\mathrel{\mathop{#2}\limits^{#1}}}

\def\sl{{\widetilde \ell}_{\scriptscriptstyle\rm R}}

\def\MGUT{\ensuremath{M_{\rm GUT}}}
\def\Min{\ensuremath{M_{\rm in}}}

\def\m12{m_{1\!/2}}

\newcommand{\mstop}[1]{m_{\tilde t_#1}}

\def\mgluino{m_{\tilde{g\,}\!}}
\def\mew{m_{\tilde\chi}}

\def\mt{m_{t}}

\def\msqt{m_{\widetilde{q}_3}}
\newcommand{\msl}[1]{m_{\tilde \ell_{#1}}}

\newcommand{\FH}{{\tt FeynHiggs}}
\newcommand{\FHold}{{\tt FeynHiggs~2.10.0}}
\newcommand{\FHnew}{{\tt FeynHiggs~2.14.1}}

\newcommand{\ETslash}{\ensuremath{/ \hspace{-.7em} E_T}}
\usepackage{xcolor}

\newcommand{\msq}{\ensuremath{m_{\tilde q}}}

\newcommand{\mste}[1]{\ensuremath{m_{\tilde t_{#1}}}}

\newcommand{\msusy}{M_{\rm SUSY}}

\newcommand{\Amp}[4][\mathcal{A}]{#1^{\mbox{\tiny #2}}_{\mbox{\tiny #3}}\ifthenelse{\isempty{#4}}{}{{\left[#4\right]}}}

\newcommand{\IE}{\textit{i.\,e.}}
\newcommand{\EG}{\textit{e.\,g.}}

\newcommand{\Mh}{\ensuremath{M_h}}
\newcommand{\MA}{\ensuremath{M_A}}
\newcommand{\MHp}{\ensuremath{M_{H^\pm}}}
\newcommand{\MZ}{\ensuremath{M_Z}}

\newcommand{\tb}{\ensuremath{\tan\beta}}

\newcommand{\cha}[1]{\tilde \chi^\pm_{#1}}

\newcommand{\mcha}[1]{m_{\tilde \chi^\pm_{#1}}}
\newcommand{\neu}[1]{\tilde \chi^0_{#1}}
\newcommand{\mneu}[1]{m_{\tilde \chi^0_{#1}}}
\def\order#1{\ensuremath{{\cal O}(#1)}}
\newcommand{\DRbar}{\ensuremath{\overline{\mathrm{DR}}}}
\newcommand{\MSbar}{\ensuremath{\overline{\mathrm{MS}}}}
\newcommand{\al}{\alpha}
\newcommand{\als}{\alpha_s}
\newcommand{\alt}{\alpha_t}
\newcommand{\alb}{\alpha_b}
\newcommand{\fh}{{\tt FeynHiggs}}
\newcommand{\mgrav}{\ensuremath{m_{3/2}}}

\newcommand{\cp}{\ensuremath{{\cal CP}}}

\newcommand{\reffi}[1]{Fig.~\ref{#1}}

\newcommand{\refse}[1]{Sect.~\ref{#1}}

\newcommand{\citere}[1]{Ref.~\cite{#1}}
\newcommand{\citeres}[1]{Refs.~\cite{#1}}

\makeatletter
\DeclareRobustCommand\em{%
  \@nomath\em \ifdim \fontdimen\@ne\font >\z@\scshape
  \else \slshape \fi}
\makeatother

\renewcommand{\emph}[1]{{\em #1}}

\addtokomafont{disposition}{\rmfamily}
\BeforeTOCHead[toc]{{\pdfbookmark[1]{\contentsname}{toc}}}

\usepackage{diagbox}

\usepackage{array}

%% file: loadhyperref.tex
\RequirePackage[numbers,sort&compress]{natbib}
\def\NAT@spacechar{\,}
\RequirePackage{color}
\RequirePackage[
colorlinks=true
,urlcolor=blue
,anchorcolor=blue
,citecolor=blue
,filecolor=blue
,linkcolor=blue
,menucolor=blue
,linktoc=all
,unicode
,backref=page
]{hyperref}
\usepackage{hypernat}

\renewcommand*{\backref}[1]{}
\renewcommand*{\backrefalt}[4]{%
  \ifcase #1%
  \or \![p\,#2]%
  \else \![pp\,#2]%
  \fi%
}

\newcommand\note[2][]{%
\if!#1!%
\stepcounter{footnote}\footnotetext{#2}%
\else%
{\renewcommand\thefootnote{#1}%
\footnotetext{#2}}%
\fi}

\makeatletter
\newdimen\p@ \p@=1pt
\newdimen\z@ \z@=0pt

\skip\@mpfootins = \skip\footins
\renewcommand{\@dotsep}{10000}

\numberwithin{equation}{section}

\renewcommand\section{\addtocontents{toc}{\protect\vspace{-5\p@}}%
                      \@startsection{section}{1}{\z@}%
                                   {-3.5ex \@plus -1.3ex \@minus -.7ex}%
                                   {2.3ex \@plus.4ex \@minus .4ex}%
                                   {\normalfont\large\bfseries}}
\renewcommand\subsection{\@startsection{subsection}{2}{\z@}%
                                   {-2.3ex\@plus -1ex \@minus -.5ex}%
                                   {1.2ex \@plus .3ex \@minus .3ex}%
                                   {\normalfont\normalsize\bfseries}}
\renewcommand\subsubsection{\@startsection{subsubsection}{3}{\z@}%
                                   {-2.3ex\@plus -1ex \@minus -.5ex}%
                                   {1ex \@plus .2ex \@minus .2ex}%
                                   {\normalfont\normalsize\bfseries}}
\renewcommand\paragraph{\@startsection{paragraph}{4}{\z@}%
                                   {1.75ex \@plus1ex \@minus.2ex}%
                                   {-1em}%
                                   {\normalfont\normalsize\bfseries}}
\renewcommand\subparagraph{\@startsection{subparagraph}{5}{\parindent}%
                                   {1.75ex \@plus1ex \@minus .2ex}%
                                   {-1em}%
                                   {\normalfont\normalsize\bfseries}}

\def\fnum@figure{\textbf{\figurename\nobreakspace\thefigure}}
\def\fnum@table{\textbf{\tablename\nobreakspace\thetable}}

\long\def\@makecaption#1#2{%
  \vskip\abovecaptionskip
  \sbox\@tempboxa{\small #1. #2}%
  \ifdim \wd\@tempboxa >\hsize
    \small #1. #2\par
  \else
    \global \@minipagefalse
    \hb@xt@\hsize{\hfil\box\@tempboxa\hfil}%
  \fi
  \vskip\belowcaptionskip}
\makeatother

\renewenvironment{thebibliography}[1]{%
\begin{oldthebibliography}{#1}%
\small%
\raggedright%
\setlength{\itemsep}{5pt plus 0.2ex minus 0.05ex}%
}%
{%
\end{oldthebibliography}%
}

\newif\ifbackrefshowonlyfirst
\backrefshowonlyfirsttrue
\makeatletter
\let\BR@direct@old@hyper@natlinkstart\hyper@natlinkstart
\renewcommand*{\hyper@natlinkstart}{\phantomsection\BR@direct@old@hyper@natlinkstart}
\let\BR@direct@oldBR@citex\BR@citex
\renewcommand*{\BR@citex}{\phantomsection\BR@direct@oldBR@citex}%

\long\def\hyper@page@BR@direct@ref#1#2#3{\hyperlink{#3}{#1}}

\ifx\backrefxxx\hyper@page@backref
    \let\backrefxxx\hyper@page@BR@direct@ref
    \ifbackrefshowonlyfirst
    \fi
\else
    \ifbackrefshowonlyfirst
    \fi
\fi

\patchcmd{\Hy@backout}{Doc-Start}{\@currentHref}{}{\errmessage{I can't seem to patch backref}}
\makeatother